\documentclass[fleqn,usenatbib]{mnras}

\usepackage{newtxtext,newtxmath}

\usepackage[T1]{fontenc}

\DeclareRobustCommand{\VAN}[3]{#2}
\let\VANthebibliography\thebibliography
\def\thebibliography{\DeclareRobustCommand{\VAN}[3]{##3}\VANthebibliography}

\usepackage{graphicx}	
\usepackage{amsmath}	
\usepackage{subcaption}
\usepackage{comment}
\usepackage[T1]{fontenc}
\usepackage{multirow}

\defcitealias{Mingo2019}{M19}

\title[Accretion mode vs radio morphology in LoTSS-Deep]{Accretion mode versus radio morphology in the LOFAR Deep Fields}

\author[B. Mingo et al.]{
B. Mingo$^{1}$\thanks{E-mail: bmingo@extragalactic.info (BM)},
J. H. Croston$^{1}$,
P. N. Best$^{2}$,
K. J. Duncan$^{2}$,
M. J. Hardcastle$^{3}$,
R. Kondapally$^{2}$,
I. Prandoni$^{4}$, \newauthor
J. Sabater$^{2,5}$,
T. W. Shimwell$^{6,7}$,
W. L. Williams$^{7}$,
R. D. Baldi$^{4,8}$,
M. Bonato$^{4,9,10}$,
M. Bondi$^{4}$,
P. Dabhade$^{11}$, \newauthor
G. G\"urkan$^{12}$,
J. Ineson$^{8}$,
M. Magliocchetti$^{13}$,
G. Miley$^{7}$,
J. C. S. Pierce$^{3}$,
and H. J. A. R\"ottgering$^{7}$
\\
$^{1}$School of Physical Sciences, The Open University, Walton Hall, Milton Keynes, MK7 6AA, UK\\
$^{2}$Institute for Astronomy, University of Edinburgh, Royal Observatory, Blackford Hill, Edinburgh, EH9 3HJ, UK\\
$^{3}$Centre for Astrophysics Research, University of Hertfordshire, College Lane, Hatfield AL10 9AB, UK\\
$^{4}$INAF-IRA, Via Gobetti 101, I-40129, Bologna, Italy\\
$^{5}$STFC UK Astronomy Technology Centre, Royal Observatory, Blackford Hill, Edinburgh, EH9 3HJ, UK\\
$^{6}$ASTRON, Netherlands Institute for Radio Astronomy, Oude Hoogeveensedijk 4, 7991 PD, Dwingeloo, The Netherlands\\
$^{7}$Leiden Observatory, Leiden University, PO Box 9513, NL-2300 RA Leiden, the Netherlands\\
$^{8}$School of Physics and Astronomy, University of Southampton, Southampton, SO17 1BJ, UK\\
$^{9}$Italian ALMA Regional Centre, Via Gobetti 101, I-40129, Bologna, Italy\\
$^{10}$INAF-Osservatorio Astronomico di Padova, Vicolo dell'Osservatorio 5, I-35122, Padova, Italy\\
$^{11}$Observatoire de Paris, LERMA, Coll\`ege de France, CNRS, PSL University, Sorbonne University, 75014, Paris, France\\
$^{12}$Th\"{u}ringer Landessternwarte (TLS), Sternwarte 5, D-07778 Tautenburg, Germany\\
$^{13}$INAF - IAPS, Via Fosso del Cavaliere 100, 00133, Roma, Italy\\
}

\date{Accepted XXX. Received YYY; in original form ZZZ}

\pubyear{2021}

\begin{document}
\label{firstpage}
\pagerange{\pageref{firstpage}--\pageref{lastpage}}
\maketitle

\begin{abstract}
Radio-loud active galaxies have two accretion modes [radiatively inefficient (RI) and radiatively efficient (RE)], with distinct optical and infrared signatures, and two jet dynamical behaviours, which in arcsec- to arcmin-resolution radio surveys manifest primarily as centre- or edge-brightened structures [Fanaroff-Riley (FR) class I and II]. The nature of the relationship between accretion mode and radio morphology (FR class) has been the subject of long debate. We present a comprehensive investigation of this relationship for a sample of 286 well-resolved radio galaxies in the LOFAR Two-metre Sky Survey Deep Fields (LoTSS-Deep) first data release, for which robust morphological and accretion mode classifications have been made. We find that two-thirds of luminous FRII radio galaxies are RI, and identify no significant differences in the visual appearance or source dynamic range (peak/mean surface brightness) of the RI and RE FRIIs, demonstrating  that both RI and RE systems can produce FRII structures. We also find a significant population of low-luminosity FRIIs (predominantly RI), supporting our earlier conclusion that FRII radio structures can be produced at all radio luminosities. We demonstrate that in the luminosity range where both morphologies are present, the probability of producing FRI or FRII radio morphology is directly linked to stellar mass, while across all morphologies and luminosities, RE accretion occurs in systems with high specific star formation rate, presumably because this traces fuel availability. In summary, the relationship between accretion mode and radio morphology is very indirect, with host-galaxy environment controlling these two key parameters in different ways.

\end{abstract}





\section{Introduction}\label{sect:Intro}

Jets in active galactic nuclei (AGN) are highly collimated beams of plasma launched as part of the accretion process from the vicinity of the central supermassive black hole, and can reach distances of up to a few Mpc \citep[e.g][]{Blandford2019}. The large-scale radio morphology of AGN is a consequence of the interaction between the jet and the surrounding environment, and provides fundamental insight on the underlying physics regulating this process and on the nature of the jet itself \citep[see e.g. the recent review by][]{HardcastleCroston2020}. The morphology of (extended) radio-loud AGN on kpc to Mpc scales has been the subject of investigation for many decades \citep[e.g.][]{Miley1980}, and typically falls into the two categories identified by \citet{FR1974} -- centre-brightened and edge-brightened structures (FRI and FRII). There is now extensive observational evidence that this dichotomy in kpc to Mpc structures is a consequence of jet deceleration and decollimation in FRI jets, which occurs on $\sim$ kpc scales \citep[e.g.][]{Bicknell1994,Laing2002,LaingBridle2014}. The long-standing scenario in which this behaviour is controlled by a combination of jet power and environmental density on small scales \citep{Bicknell1994,Ledlow1996} has found support in recent LOFAR work \citep[][hereafter M19]{Mingo2019}.

As well as morphology, radio galaxies have traditionally been classified according to two different sets of emission line properties \citep[see e.g.][]{Hine1979,Best2012}, with those displaying strong optical emission lines called high-excitation radio galaxies (HERGs, or HEGs) and those where the lines are faint or absent being labelled as low-excitation radio galaxies (LERGs, or LEGs). Sometimes they are also referred to as weak- or strong-line radio galaxies \citep[WLRGs and SLRGs, see e.g.][]{Tadhunter2016}. These two classes have been linked to a dichotomy in the accretion properties of the AGN central engine, motivated by the behaviour of X-ray binaries \citep[see e.g. the review by][]{Fender2014}. LERGs are associated with low-rate, radiatively-inefficient accretion \citep[RI;][]{Narayan1995}, effective at producing jets but generating little radiation beyond the radio. HERGs are instead linked to high-rate, radiatively-efficient accretion \citep[RE;][]{Shakura1973}, generating large amounts of radiation but only sometimes producing jets.

In the literature, the FRI and II morphological classes have been frequently associated with the RI and RE accretion states, respectively, \citep[e.g.][]{Jackson1997}. Such a connection is supported by the fact that there are few (though not zero) known examples of FRI HERGs (see e.g. Gurkan et al., subm.), but is contradicted by the existence of a significant population of FRII LERGs \citep[e.g.][]{Hine1979,Willott1999,Chiaberge2000,Hardcastle2004,Baldi2010,Croston2018}. One possible explanation for the existence of FRII LERGs in a model where the central engine determines the large-scale radio morphology is that they are systems where the central engine has recently switched off, or suffered a significant change in accretion rate, but this information has not reached the hotspots yet \citep[e.g.][]{Tadhunter2016}. This model predicts observable differences in the large-scale radio structures of FRII LERGs and HERGs, with the former frequently presenting a more relaxed, `fat double' structure \citep{Owen1989}. Additional support for a model in which the central engine controls the jet physics comes from the relationship between blazar peak frequency and luminosity (the `blazar envelope'), which may be explained by intrinsic differences in jet structure on sub-parcsec scales that are linked to accretion mode rather than jet power \citep[e.g.][]{Meyer2011,Keenan2021}. 

New facilities and deeper radio surveys are now demonstrating that the faint radio-loud AGN population encompasses more diverse behaviour than previously observed, which must be incorporated into physical models, e.g. small sources with very weak jets, and low-luminosity FRIIs \citep[e.g.][]{Best2012,Fernandes2015,Baldi2015,Capetti2017FRII,Miraghaei2017,Mingo2019,Jimenez-Gallardo2019,Macconi2020,Pierce2020,Grandi2021,Vardoulaki2021}. Environmental studies have also demonstrated links between environment and both radio morphology and accretion mode \citep{Gendre2013,Ineson2015,Croston2019,Vardoulaki2021}. In the high accretion regime there is increasing evidence of the presence of jets along with radiatively-driven winds in sources often classified as radio-quiet AGN \citep[e.g.][]{Jarvis2019,Fawcett2020}, though there seems to be evidence for an inverse correlation between the strength of the wind and that of the jet \citep{Mehdipour2019}. Many of these radio sources seem to be young or unlikely to grow to large sizes, so it is yet unclear how they tie in with the evolved FRI/II classes, and what impact they might have beyond their host galaxies \citep[e.g.][]{Chhetri2020,Patil2020}. We are also achieving greater insight on the bright compact steep spectrum (CSS) and gigahertz peaked spectrum (GPS) populations, with radio powers comparable to those of extended radio galaxies, but which are far too numerous to always evolve into FRIs and FRIIs \citep[see e.g. the recent review by][]{ODea2021}.

In this work we exploit the newly-released LOFAR Two Metre Sky Survey Deep Fields dataset \citep[LoTSS-Deep;][]{Tasse2021,Sabater2021,Kondapally2021,Duncan2021} to examine the relationship between radio morphology and accretion mode for a large and well-resolved sample with high-quality optical and infrared data, and to investigate in detail the role of host galaxy stellar mass, fuel availability, and jet power as controlling influences. The LoTSS-Deep survey covers an area of $\sim25$ square degrees with an average noise level of $\sim25$ $\mu$Jy beam$^{-1}$ and the same 6-arcsec resolution as the LOFAR Two Metre Sky Survey first data release \citep[ LoTSS-DR1;][]{Shimwell2019,Williams2019,Duncan2019}, resulting in a source density of up to $\sim5000$ sources per square degree, a factor of $\sim7$ higher than  LoTSS-DR1, in a sky area $\sim17$ times smaller. The increased depth and outstanding multiwavelength coverage of LoTSS-Deep allows us to reliably classify sources up to redshift $(z)\sim2.5$, for the first time allowing us to examine this relationship for faint and moderately-bright sources well beyond our local Universe. LoTSS-Deep is representative of the data quality and source density expected for the low-frequency Square Kilometre Array (SKA) surveys \citep{SKA2009,Prandoni2015}, which means that investigating the demographics and different physical mechanisms driving these populations will be a key step towards their large-scale identification and classification in the SKA surveys, and to understand their evolution.

In Section \ref{sec:Analysis} we describe the data from the three LOFAR deep fields that are the subject of this study (ELAIS-N1, Bo\"otes, and the Lockman Hole), our radio source morphological classification results following the method developed by \citetalias{Mingo2019}, and the host galaxy spectral energy distribution (SED) classifications obtained by Best et al. (subm.). We then analyse the properties of our sources in Section \ref{sec:Results}, comparing our results with those obtained using  LoTSS-DR1. In Section \ref{sec:FR_accretion_link} we then discuss the relationship between jet power, accretion mode, and radio galaxy morphology, while in Section \ref{sec:Mstar_SFR} we explore the role of the host galaxy in regulating both accretion and FR class. Section \ref{sec:Conclusions} summarises our conclusions. 

For this paper we have used a concordance cosmology with $H_{0} = 70$ km s$^{-1}$ Mpc$^{-1}$, $\Omega_{m}= 0.3$ and $\Omega_{\Lambda}= 0.7$.


\section{Data and Analysis}\label{sec:Analysis}


\subsection{The LOFAR deep fields}\label{sec:DF_data}

\begin{table}
\caption{LOFAR deep fields coverage. Columns show (l to r): field name, sky area with available multi-wavelength coverage, RMS noise in the central region of the field at 150 MHz, total combined exposure, and our initial sample size (sources with catalogued sizes $>8$ arcsec) prior to morphological classification. For details on each individual field see \citet{Sabater2021,Tasse2021}.}
\label{tab:DF}
\centering
\begin{tabular}{c c c c c}
\hline 
Field & Sky area & RMS & Exposure & N sources \\ 
& [sq deg] & [$\mu$Jy/beam] & [h] &\\
\hline 
ELAIS-N1 & 6.74 & 19 & 164 & 8251 \\ 
Bo\"otes & 8.63 & 30 & 80 & 3886 \\
Lockman & 10.28 & 22 & 112 & 8078 \\\hline
\end{tabular}
\end{table}

The three fields covered in this paper -- ELAIS-N1 \citep{Oliver2000}, the Lockman Hole \citep{Lockman1986}, and Bo\"otes \citep{Jannuzi1999} -- were chosen as targets for the LoTSS-Deep surveys because of their excellent multiwavelength coverage \citep[see][and references therein]{Kondapally2021}, as well as their location in the sky -- high in declination to allow high elevation observing and mitigating additional ionospheric thickness and beam elongation effects, whilst allowing for good uv-coverage. The ultimate aim of the deepest LOFAR surveys is to achieve a central noise level of $\sim10$ $\mu$Jy/beam, ten times deeper than the target depth of the all-northern-sky LoTSS survey \citep{Shimwell2017,Shimwell2019}, over a sky area of 30-50 square degrees, and with a resolution of 6 arcsec. As described by \citet{Sabater2021}, this coverage and depth is sufficient to probe radio source populations beyond $z\sim1$ with sufficient statistics so as to obtain representative samples of sources. Fig. 1 by \citet{Tasse2021} provides a direct comparison of the resolution and sensitivity for the current and planned LoTSS-Deep surveys, in comparison to other existing and planned surveys.

The LOFAR and multiwavelength data for the three deep fields are slightly different, with the Lockman Hole having the largest sky area with multiwavelength coverage, and ELAIS-N1 the deepest radio images (lowest RMS). Details of the observation strategy, data processing and calibration, and compilation of the multiwavelength catalogues for the three LOFAR deep fields are described by \citet{Tasse2021}, \citet{Sabater2021}, \citet{Kondapally2021}, and \citet{Duncan2021}. Table \ref{tab:DF} provides a summary of the broad characteristics of each field in terms of their LOFAR coverage. Any implications of the differences between the three fields for our results are noted in Sections \ref{sec:Classifications} and \ref{sec:Results}. We note that we have observed no field-dependent effects at a significant level in the results presented in this work.

Similar to the value-added cataloguing of LoTSS-DR1 \citep{Williams2019}, the basic sizes and shapes of our extended sources have been catalogued using \textsc{PyBSDF} \citep{PyBDSF2015} and for the largest sources \textsc{PyBDSF} was complemented by Zooniverse component association. Host identification was achieved through a combination of maximum likelihood and Zooniverse visual identification \citep{Kondapally2021}. We used the value-added and component catalogues for each deep field for our analysis, in a similar manner as that described by \citetalias{Mingo2019}. As our science aims require morphological classification, we imposed an initial cut at a catalogued size of 8 arcsec to discard any sources that would be too small to classify, but did not impose an initial flux density cut. Table \ref{tab:DF} shows the initial number of sources for each field that we have used for our analysis. The total numbers listed represent $\sim26$ per cent of all the sources in ELAIS-N1 and the Lockman Hole, and $\sim20$ per cent of the sources in Bo\"otes.

Thanks to the excellent multiwavelength coverage available in our deep fields we were able to rely on spectral energy distribution (SED) fitting to separate galaxies from AGN, to derive star formation rates and stellar masses for each source, and to classify AGN based on their radio loudness and accretion mode. Details of this process are given in Section \ref{sec:SED_fitting}.


\subsection{Morphological classification}\label{sec:Classifications}

Our aim in classifying the extended sources from LoTSS-Deep was to obtain clean samples of FRIs and FRIIs. To do so we used a two-fold approach, similar to that described by \citetalias{Mingo2019}, in which the sources were first automatically classified (as described in Section \ref{sec:Auto_class} below) and then further refinements to the samples were carried out via visual inspection (Section \ref{sec:Eye_class}).

While we had enough information from the SED fitting results (described in Section \ref{sec:SED_fitting}) to pre-filter star-forming galaxies, we decided to automatically classify and visually inspect the radio sources first, in order to provide additional validation for the SED results. This independent estimate on the potential level of contamination is also useful for comparison with other automatic classification methods, and with samples which do not have the multiwavelength information needed to easily filter out star-forming galaxies.


\subsubsection{Automatic classification}\label{sec:Auto_class}

\begin{table*}
\caption{Automatic classification statistics. See Section \ref{sec:Auto_class} for details and \citetalias{Mingo2019} for a full description of the individual classes. The second set of columns includes only the sources with well-defined redshifts, and $z\leq2.5$, used throughout the rest of this work. The last column includes in square brackets the percentage of sources remaining for each class after the $z$ cut. As detailed in the main text, we have focused primarily on the (large) FRI and FRII classes for the rest of the analysis.}
\label{tab:auto_class}
\centering
\begin{tabular}{c@{\hspace{2.5\tabcolsep}}c@{\hspace{0.7\tabcolsep}}c@{\hspace{0.7\tabcolsep}}c@{\hspace{0.7\tabcolsep}}c@{\hspace{2.5\tabcolsep}}c@{\hspace{0.7\tabcolsep}}c@{\hspace{0.7\tabcolsep}}c@{\hspace{0.7\tabcolsep}}c}
\hline 
        &\multicolumn{4}{c}{All sources}&\multicolumn{4}{c}{$z\leq2.5$}\\
		Morphology&EN1&Bo\"otes&Lockman&total&EN1&Bo\"otes&Lockman&total [\%]\\\hline
		FRI&109&74&123&306&108&73&120&301 [98\%]\\
		FRII&55&48&58&161&49&47&51&147 [91\%]\\
		Hybrid&36&30&50&116&36&29&47&112 [97\%]\\
		Small FRI&181&134&258&573&174&125&238&537 [94\%]\\
		Small FRII&18&25&24&67&17&21&24&62 [93\%]\\
		Small hybrid&30&35&24&89&30&31&24&85 [96\%]\\
		Unresolved&185&218&338&741&168&206&304&678 [91\%]\\
		Too faint&7637&3322&7203&18162&6540&2987&6165&15692 [86\%]\\\hline
\end{tabular}
\end{table*}

We used the morphological code \textsc{LoMorph}\footnote{\url{https://github.com/bmingo/LoMorph/}}, described in detail by \citetalias{Mingo2019}, to automatically classify all the resolved sources in each field. The morphological categories are defined by the same process as in our earlier work, to which we refer the reader for details. Sources whose peaks of emission are closer to the centre (optical host position) than to the edges were classified as FRIs, and those with the brightest emission closer to the edges of the source than the host position as FRIIs. Sources with different classifications on each side were classified as `Hybrids' and excluded from FRI/II comparisons; however, we know from \citetalias{Mingo2019} that a combination of data resolution and source projection can produce this appearance, and that true hybrid sources are very rare \citep{Harwood2020}.

We carried out some minor improvements to the masking of unassociated components in \textsc{LoMorph} to minimise problems caused by the higher source density present in LoTSS-Deep compared to the main LoTSS survey \citep{Shimwell2019} -- these changes are described in full in the online repository. As for \citetalias{Mingo2019}, we found that the best noise threshold compromise was again achieved at $4 \times$ the local RMS noise. The local RMS was estimated directly from the \textsc{PyBSDF} RMS image, as it provides good noise estimates for these deeper datasets, even considering the high source density \citep{Tasse2021,Sabater2021}. 

We retained the `too faint' filter in \textsc{LoMorph} requiring that at least five pixels in each source be above the $4\times$RMS threshold, and that its total flux (post-RMS filtering) be above 1 mJy. We visually inspected samples of `too faint' sources, and verified that they did not contain enough information to be reliably classified, despite the increased image depth. In contrast, sources in the `unresolved' class were sufficiently bright, but with too small an angular size to identify individual structures.

We also separated the sources into large and small classes, for which the morphological classifications are expected to have different reliability, using the resolution criteria of \citetalias{Mingo2019}, where the small sources have sizes $\le$27 arcseconds, or sizes below 40 arcseconds but a combined distance between the peaks of emission (d1+d2) smaller than 20 arcseconds. 

The automatic classification results are summarised for the three fields in Table \ref{tab:auto_class}. The results shown in Table \ref{tab:auto_class} are prior to any sample cleaning, so they include misclassified sources and star-forming galaxies. Sample cleaning is described in the following Section.


\subsubsection{Visual inspection adjustments}\label{sec:Eye_class}

\begin{table*}
\caption{Visual classification statistics. See the main text and \citetalias{Mingo2019} for a detailed description of the individual classes. For each field the columns list, from left to right, the number of straight (lobed/tailed) FRIs, narrow-angle tails, wide-angle tails, FRIIs, double-doubles, `core-dominated' sources, `fuzzy blobs', hybrid candidates, star-forming galaxies, and sources with technical issues. The top four rows correspond to the sources automatically classified as FRIs, and the bottom four to those automatically classified as FRIIs. Numbers in bold show the sources for which the visual inspection matches the automatic classification}
\label{tab:eye_class}
\centering
\begin{tabular}{ccccccccccc}
\hline 
Field&Straight FRI&NAT&WAT&FRII&D-D&Core-D&Blob&Hybrid&SF galaxy&Bad\\\hline
EN1 - FRI &\textbf{27}&\textbf{9}&\textbf{22}&9&0&11&15&2&5&8\\
Bo\"otes - FRI &\textbf{28}&\textbf{3}&\textbf{10}&1&5&16&2&1&3&4\\
Lockman - FRI &\textbf{31}&\textbf{6}&\textbf{25}&16&3&13&14&2&7&3\\
Total - FRI &\textbf{86}&\textbf{18}&\textbf{57}&26&8&40&31&5&15&15\\\hline
EN1 - FRII &0&0&1&\textbf{39}&0&0&0&0&0&9\\
Bo\"otes - FRII &0&0&0&\textbf{42}&0&0&0&2&0&3\\
Lockman - FRII &0&1&3&\textbf{45}&1&0&0&1&0&0\\
Total - FRII &0&1&4&\textbf{126}&1&0&0&3&0&12\\\hline
\end{tabular}
\end{table*}

As discussed by \citetalias{Mingo2019}, the (large, $>27$ arcsec) FRI and FRII classes are the cleanest and most reliable, so we will focus almost exclusively on these for the remainder of this work. The FRI class in particular suffers from contamination from sources that fulfil the automatic classification criteria for FRIs, but are physically distinct from this population \citepalias{Mingo2019}. Given the higher source density and the necessity for the cleanest possible samples, we visually inspected both the FRIs and the FRIIs for the three LOFAR deep fields analysed in this work, using the same criteria described by \citetalias{Mingo2019}. The results are shown in Table \ref{tab:eye_class}, with a final, securely-classified sample of 161 FRIs and 126 FRIIs. Galleries of these two classes can be found in Sections \ref{sec:FRI} and \ref{sec:FR2}.

We did not incorporate into our final sample the 26 visually-identified FRIIs that were automatically classified as FRIs by \textsc{LoMorph} \citepalias[due to issues with deconvolution, noise, and projection effects, as described in][]{Mingo2019}, nor the five visually-identified FRIs automatically classified as FRIIs. The double automated and visual test guarantees that we have the cleanest, most unambiguous sample obtainable with our sample, directly comparable to the results of \citetalias{Mingo2019}, and useful for future work on automated classification. The impact of prioritising reliability over completeness is discussed where relevant in Sections \ref{sec:Results}, \ref{sec:FR_accretion_link}, and \ref{sec:Mstar_SFR}.

As in \citetalias{Mingo2019} we subdivided the FRIs into straight (lobed/tailed), narrow-angle tails \citep[NATs,][]{Rudnick1976,ODea1985NAT}, and wide-angle tails \citep[WATs,][]{Owen1976}. Following our \citetalias{Mingo2019} definitions, we classified sources with very bright cores surrounded by a drop and subsequent rise in brightness as `core-dominated', and those with bright cores surrounded by halo-like extended emission not resembling lobes or tails as `fuzzy blobs'. We classified as `Bad' any sources with technical issues: suspected bad host IDs, bad region associations (including suspected blends), or problems with masking or flood-filling. We again kept the double-double sources \citep{Schoenmakers2000,Mahatma2019}, mostly classified as FRIs by LoMorph (Table \ref{tab:auto_class}), separate from the FRIIs, since they cannot be treated on the same terms as sources where only one activity cycle is visible. For the same reason, we tried to avoid including visually-obvious remnants in either of our clean FRI/II classes.

We also visually inspected the optical images for all sources to filter out obvious nearby galaxies where the radio emission is arising from star formation processes, rather than an AGN. These sources are labelled as `SF galaxy' in Table \ref{tab:eye_class}.

The percentage of sources correctly classified by \textsc{LoMorph} (compared to the visual inspection) is slightly worse than that we obtained for LoTSS, due to not pre-filtering star-forming galaxies and a higher incidence of deblending complications in the deeper images. ELAIS-N1, being the deepest field, was particularly problematic in terms of visually disentangling blends and overlapping sources in the Galaxy Zoo, as described by \citet{Kondapally2021}, but the higher sky density compared to  LoTSS-DR1 also caused some overlaps and complications in the other two fields, which were not completely resolved by the \textsc{LoMorph} masking modifications. While the small sky areas involved mean that the three fields are not entirely comparable due to cosmic variance, it is interesting to note that the classification statistics are generally better for the two shallower fields compared to ELAIS-N1. This may point to future challenges with applying automated classification methods to increasingly deep surveys at this imaging resolution.

In terms of morphology, it is interesting to see a much higher fraction of FRIIs compared to \citetalias{Mingo2019}: FRIIs constitute 44 per cent of all sources, versus 25 per cent in  LoTSS-DR1, although the fraction changes slightly for each field. This is a direct consequence of having both deeper radio data and deeper multiwavelength coverage. We discuss this in more detail in Section \ref{sec:zLumSiz}. Within the FRIs we also found a slightly higher fraction of bent sources (NATs and WATs) compared to our earlier work (47 vs 37 per cent). This is likely due to the increased depth of the radio data, making the tails of these sources easier to find.


\subsection{SED classification}\label{sec:SED_fitting}

\begin{table}
\caption{Table of SED classifications produced by Best et al. (subm.). The first two columns refer to the labels output by the fitting and consensus process, with values 1=True, 0=False, and -1=undetermined (when the classification could not be determined within the uncertainty margin). The last column indicates the type of source selected based on combinations of values from the two labels.}
\label{tab:SED_classes}
\centering
\begin{tabular}{c c c}
\hline 
AGN&Radio AGN&Sources selected\\\hline
0&0&Star-forming galaxy\\
1&0&`Radio-quiet' AGN\\
0&1&LERG (`Jet-mode' radio AGN)\\
1&1&HERG (`Quasar mode' radio AGN)\\
-1&(Any)&No secure classification\\
(Any)&-1&No secure classification\\\hline
\end{tabular}
\end{table}

Spectral energy distribution (SED) fitting works by fitting a series of model templates to the broad-band photometry and/or spectra of astronomical sources (galaxies, in this case). Since the focus of our work is on radio AGN, our goal in using the SED fitting information was threefold: i) to identify which of our sources fell into different accretion regimes (RI/RE -- HERG/LERG); ii) to eliminate contaminants from our sample (star-forming galaxies with no radio jets); iii) to constrain the stellar masses and star formation rates in the host galaxies of our radio sources.

The three LOFAR Deep Fields datasets covered in this work have deep multiwavelength coverage, spanning the far-IR \citep[\textit{Herschel;}][]{Herschel2010}, mid-IR \citep[\textit{Spitzer, WISE}][]{SpitzerIRAC2004,WISE2011,AllWISE2014}, near-IR \citep[including UKIDSS -- UK Infrared Telescope Deep Sky Survey -- extragalactic survey for Lockman Hole and ELAIS-N1;][]{UKIDSS2007}, optical \citep[Pan-STARRS, Subaru HSC-SSP -- Hyper Suprime-Cam Subaru Strategic Program -- and NDWFS -- NOAO Deep Wide-Field Survey;][]{PanSTARRS2010,Aihara2018,Jannuzi1999}, and UV \citep[\textit{GALEX;}][]{GALEX2007}. \textit{Herschel} observations were deblended using XID+ \citep{Hurley2017}, to avoid issues with contamination caused by companion late-type galaxies blending with the point-spread function (PSF) of early-type radio galaxy hosts \citep{Drouart2014,Falkendal2019}. Full details on these data and how they were combined into the value-added catalogues are given by \citet{Kondapally2021} and McCheyne et al. (subm.). Photometric redshifts were determined by \citet{Duncan2021}, following a similar procedure to that used for LoTSS \citep{Duncan2019}, and fed into the SED fitting codes (along with spectroscopic redshifts where available).

The SED-fitting procedure and resulting classifications are described in detail by Best et al. (subm.). The broadband spectra were individually fitted with each of \textsc{magphys} \citep[Multi-wavelength Analysis of Galaxy Physical Properties,][]{magphys2008}, \textsc{cigale} \citep[Code Investigating Galaxy Evolution,][]{cigale2005,cigale2020}, \textsc{agnfitter} \citep{agnfitter2016}, and \textsc{bagpipes} \citep[Bayesian Analysis of Galaxies for Physical Inference,][]{bagpipes2018,bagpipes2019}. The information from the different fits was then used to determine whether the source was classified as an (optical) AGN (understood as radiatively-efficient sources, with detectable AGN emission in the FIR through X-rays) or not. Both \textsc{cigale} and \textsc{agnfitter} include AGN templates, which allowed Best et al. to calculate an `AGN fraction' with each code per source, and estimate the relative goodness of fit of these compared with \textsc{magphys} and \textsc{bagpipes}.

If the source was not classified as an AGN, then \textsc{magphys} and \textsc{bagpipes} generally provided consistent star formation rates (SFRs) to +/- 0.2 dex. \textsc{cigale} generally provided values which were systematically the same, but with larger scatter. This might be due to the less complete range of galaxy SEDs included. \textsc{agnfitter} performed slightly worse, presenting a systematic offset and more scatter. In the vast majority of these cases (see Best et al. for details of the exceptions), Best et al. adopted the logarithmic median of the \textsc{bagpipes} and \textsc{magphys} values as the consensus measurement, to maximise accuracy.

If the source was classified as an AGN, the SFRs derived by \textsc{cigale} were used instead. These were mostly consistent with the \textsc{bagpipes/magphys} values, except for a subset of sources which had significant AGN contribution in the optical or the far-IR, where the lack of AGN templates in \textsc{bagpipes/magphys} caused inconsistencies, making the \textsc{cigale} values the safest option. Again, \textsc{agnfitter} tended to have greater scatter (though less so than for the non-AGN sources).

This approach established a consensus estimate of physical galaxy properties, such as star formation rate and stellar mass, for every source. Best et al. also determined which sources had a radio excess over the emission expected purely from star formation \citep[see e.g.][]{Read2018,Smith2021} at a $\sim3 \sigma$ level (0.7 dex), similarly to \citet{Hardcastle2019}.

Sources were then assigned a series of labels depending on the result of the SED classification, as detailed in Table \ref{tab:SED_classes}. It is worth noting that the RI/RE (LERG/HERG) classifications derived from this SED-fitting approach are primarily based on the overall continuum shape (although contributions of emission lines within given filters are included in the SED models), and are therefore not strictly comparable to classifications based purely on line intensity ratios and line breaks, such as those used by \citet{Best2012}. However, for a small proportion of the sources (5.1, 21.1, and 4.7 per cent of all sources in, respectively, the ELAIS-N1, Bo\"otes and Lockman Hole fields) spectroscopic information is also available and folded in to the classifications.

As the SED classifications are most reliable below $z\sim2.5$ (Best et al., subm.) we restricted all our analysis to this redshift range. As shown in Table \ref{tab:auto_class}, this had minimal impact on the sample size.


\section{Results}\label{sec:Results}


\subsection{Redshift, luminosity and size}\label{sec:zLumSiz}

\begin{figure}
   \resizebox{\hsize}{!}{\includegraphics[trim={0.5cm 0cm 1.5cm 1cm},clip]{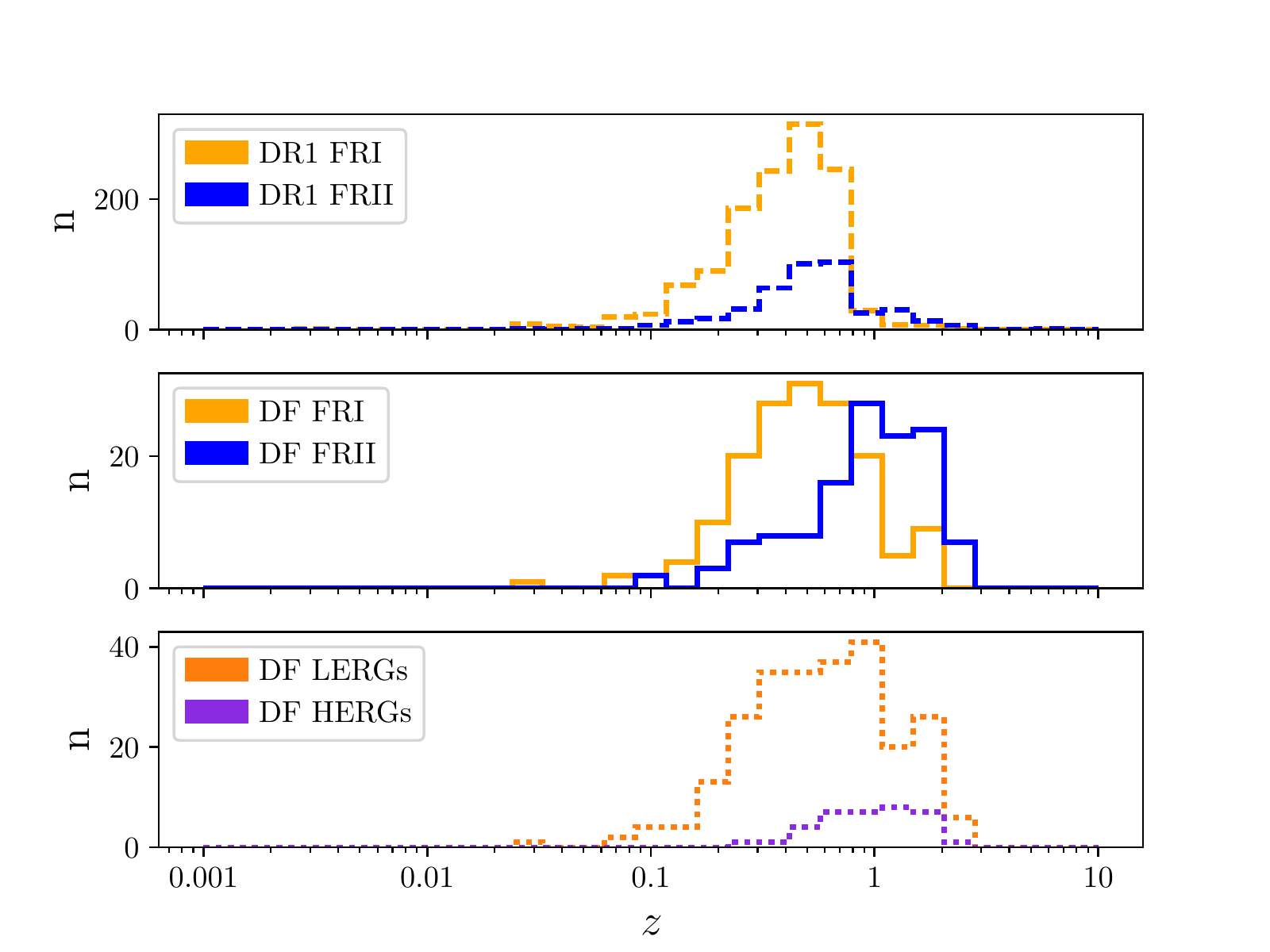}}
     \caption{\textbf{Top two panels:} redshift distribution comparison between the three combined deep fields and the  LoTSS-DR1 sources from \citetalias{Mingo2019}. Following the same colour convention, the histograms for the FRIs are in yellow and those for the FRIIs in blue. \textbf{Bottom panel:} redshift distribution for the LERG (orange) and HERG (purple) sources in LoTSS-Deep. Note that our LoTSS-Deep sample is limited to $z\leq2.5$, as mentioned in Section \ref{sec:SED_fitting}. See Section \ref{sec:AGN_SF} for detailed statistics.}
     \label{fig:DF_Hetdex_z_histo}
\end{figure}

Fig. \ref{fig:DF_Hetdex_z_histo} shows the $z$ distribution of FRIs and FRIIs for the three combined LOFAR deep fields in comparison with the  LoTSS-DR1 results presented by \citetalias{Mingo2019}. As mentioned in Section \ref{sec:Eye_class} it is striking that the FRI/II proportions are substantially different from what was found in the  LoTSS-DR1 morphological analysis. However, the distribution of sources up to $z\sim0.8$ is not substantially different from that of the  LoTSS-DR1 FRIs and FRIIs, which is reassuring. It is mostly the deeper multiwavelength data that enable host identification to higher redshift in our LoTSS-Deep sample. LoTSS-Deep has host identifications for $>97$ per cent of sources \citep{Kondapally2021}, a much larger fraction than for  LoTSS-DR1 \citep[73 per cent; see][]{Williams2019}. The differences in redshift distribution between the FRIs and FRIIs are also more apparent, with the latter peaking at higher $z$ values. As we pointed out in our earlier work \citepalias{Mingo2019} this is a combination of selection and evolution \citep[see also e.g.][]{Willott2001,Wang2008,Donoso2009,Gendre2010,Kapinska2012,Williams2015, Williams2018a}. 

Fig. \ref{fig:DF_Hetdex_z_histo} also shows the overall redshift distribution of the LERGs and HERGs in our sample. LERGs are the dominant population, as also discussed by Kondapally et al. (subm.), and overall HERGs tend to be found at slightly higher redshifts (the median values are $z=0.602$ and $z=0.922$ for LERGs and HERGs, respectively, though there is large scatter). We note that the absence of low-$z$ HERGs in our sample is likely due to the small sky area covered (LERGs are the dominant population at low-$z$). A detailed breakdown of the accretion statistics for each morphological class is presented in Section \ref{sec:AGN_SF}, and the implications are discussed in detail in Section \ref{sec:FR_accretion_link}. 

\begin{figure}
  \resizebox{\hsize}{!}{\includegraphics[trim={0.5cm 0cm 1.5cm 1cm},clip]{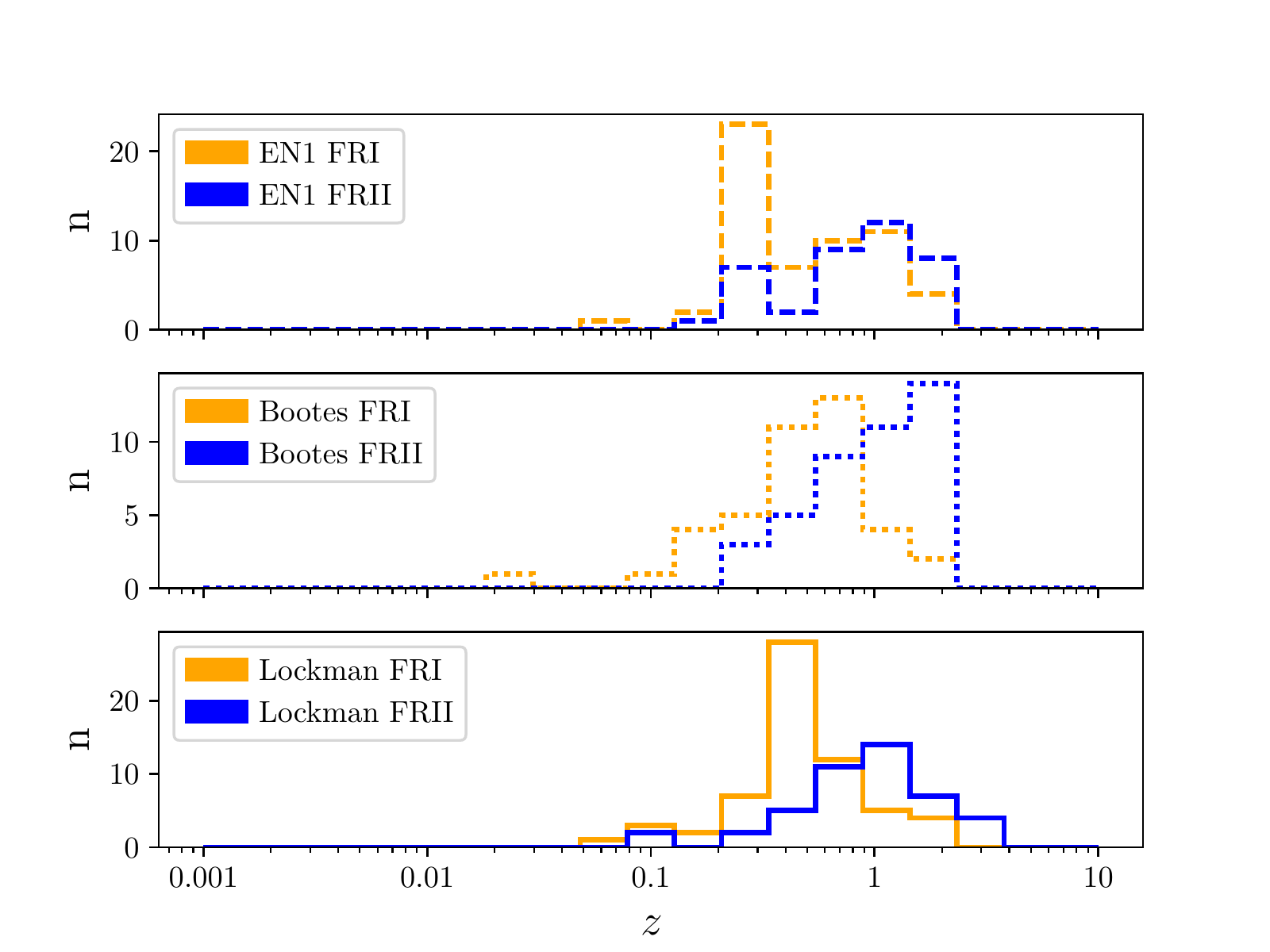}}
  \caption{Redshift distribution histograms for the three deep fields. Colours as in Fig. \ref{fig:DF_Hetdex_z_histo}.}
  \label{fig:DF_z_histo}
\end{figure}

Fig. \ref{fig:DF_z_histo} illustrates the distribution of redshifts across the three fields. Given the different noise levels in the three fields (Table \ref{tab:DF}), this plot helps us understand how different flux density and surface brightness limits preferentially include or exclude different populations. The fact that the highest overlap between the FRI and FRII distributions happens in the deepest field (EN1), and the lowest overlap in the shallowest (Bo\"otes) hints at the fact that FRIs at high redshift do exist, and in large numbers, but they are the first victims of surface brightness limits. However, we are working with small number statistics, and there might be small differences in terms of host identification due to the different optical catalogues used in the three fields \citep[see][for details]{Kondapally2021}. There are some low-significance differences in the $z$ distributions of FRIs and FRIIs for the three fields, likely caused by a combination of effects (binning, underlying differences in the multiwavelength catalogues, radio survey depth, etc.), but these have no impact on our conclusions. We checked whether our main results showed any differences across the three fields and found no evidence of discrepancies at a statistically significant level. As such, all the analysis and results described in the following Sections were carried out using the combined dataset.

\begin{figure*}
\centering
  \includegraphics[width=17cm, trim={0cm 0cm 0cm 0cm},clip]{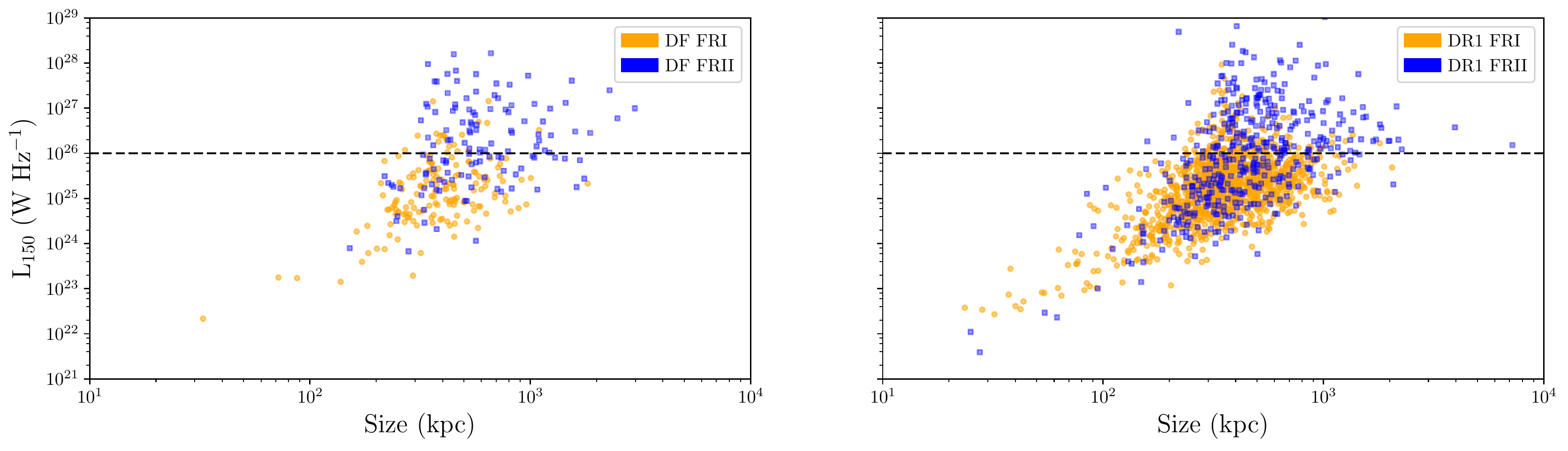}
  \caption{Luminosity vs size distribution of the FRIs (yellow circles) and FRIIs (blue squares) for the combined deep fields (left) and the  LoTSS-DR1 sources from \citetalias{Mingo2019} (right). The dashed line shows the `traditional' boundary between FRIs and FRIIs, at $L_{150}\sim10^{26}$ W Hz$^{-1}$ \citep{FR1974,Ledlow1996}.}
  \label{fig:DF_Hetdex_L150_size}
\end{figure*}

The luminosity and size distribution of our LoTSS-Deep sources is very similar to that of the morphological catalogue for  LoTSS-DR1 (Fig. \ref{fig:DF_Hetdex_L150_size}). The median sizes for the FRI and FRII in our LoTSS-Deep sample are $380\pm20$ and $540\pm50$ kpc, respectively, and the median 150-MHz luminosities $1.5\pm1.7\times10^{25}$ and $1.2\pm2.7\times10^{26}$ W Hz$^{-1}$. The distributions are roughly consistent with the \citetalias{Mingo2019} DR1 sample -- in both samples the dominant population is large, mature systems (partly due to our angular size cut), though we note that in our LoTSS-Deep sample we find no FRII sources with sizes below 100 kpc. These sources are quite rare, and the combination of angular size and surface brightness limits only allows us to identify them at low $z$. Our DR1 sample covered a much larger sky area, making it easier to catch some examples of rarer populations.

We have again drawn a horizontal line to represent the canonical FR luminosity break at L$_{150} \sim 10^{26}$ W Hz$^{-1}$ \citep{FR1974,Ledlow1996}. The number of FRIs above this line is rather small: 19 across the three fields, representing a fraction of 13 per cent, identical to that found for  LoTSS-DR1. We again found a large number of sources with FRII morphology below the canonical line, which we will refer to as FRII-Low throughout this work, after \citetalias{Mingo2019} (similarly, we will refer to the luminous FRIIs above the line as FRII-High). There are 50 FRII-Low across the three deep fields (respectively 21, 14, and 15 in EN1, Bo\"otes, and Lockman), representing a fraction of $\sim40\pm9$ per cent, lower than that found in  LoTSS-DR1 ($\sim51$ per cent), but consistent within the errors. Notably, we only found 9 of these sources at very low luminosities (L$_{150} < 10^{25}$ W Hz$^{-1}$), or 7 per cent of all FRIIs, which is a much smaller fraction than that found in  LoTSS-DR1 (21 per cent), though still consistent within the errors. These differences are likely driven by our ability to detect more distant, high luminosity FRIIs in LoTSS-Deep, thanks to the lower flux density limit and better host identification, thus lowering the overall fraction of low-luminosity sources. The properties of our FRII-Low are discussed in detail in Section \ref{sec:FR2L}.

In terms of size distribution we also found 21 sources (19 are FRIIs, of which 5 are FRII-Low) with sizes greater than 1 Mpc, and thus giant radio galaxies \citep[GRG, see e.g.][and references therein]{Ishwara-Chandra1999,schoenmakers2000b,Machalski2001,Machalski2008,Dabhade2017,Dabhade2020,Dabhade2020b}, a larger proportion than we found in our earlier work, indicating that this population greatly benefits from the increased survey depth. 

As in our LoTSS-DR1 work, it is crucial to emphasise that the apparent correlation between size and luminosity in our sample is primarily driven by selection effects \citep[see also the discussion by][]{Turner2015,Hardcastle2019}. The top-left corner of both plots in Fig. \ref{fig:DF_Hetdex_L150_size} is empty because our angular size cut at 27 arcsec eliminates the small, bright sources. The bottom-right corner of both plots could contain physically large yet faint sources, falling below the surface brightness limits of either sample. Similarly to \citetalias{Mingo2019}, this makes our sample selection quite different from that of several older surveys, which contain more compact, luminous sources -- we return to the implications of this point in later sections.


\subsection{Accretion mode vs radio morphology}\label{sec:AGN_SF}

As introduced in Section \ref{sec:SED_fitting} we used the SED classifications from Best et al. (subm.) to classify our sources according to their accretion properties (see Table \ref{tab:SED_classes}). In Table \ref{tab:accr_mode} we summarise the SED-based AGN/SF classifications for each of our morphological classes. We removed the FRI source with an `undefined' SED classification from the sample for any further analysis, leaving the final number of FRIs at 160. Other than for this source, our results show that the sample cleaning via visual inspection (Section \ref{sec:Eye_class}) was highly consistent with the SED-fitting filters, and that the automatic classification for the large FRI and FRII categories is robust even in the absence of any pre-selection.

\begin{table*}
\caption{SED-fitting classifications by morphological class (see Section \ref{sec:Auto_class} and Table \ref{tab:auto_class} for details on the individual classes). The second column (n) shows the number of sources in each category; the other columns show the number [percentage] of sources in each category. The first set of rows shows the classification results for the source classes that are the main focus of this work \citepalias[FRII-High are FRIIs with $L_{150}>1\times10^{26}$ W Hz$^{-1}$; FRII-Low are those below the boundary, as introduced by][]{Mingo2019}. Our morphological classifications are summarised in Tables \ref{tab:auto_class} and \ref{tab:eye_class}, and the SED labels (Best et al., subm.) can be found in Table \ref{tab:SED_classes}. Note that the FRI source with `undefined' classification was excluded from the rest of our analysis.}\label{tab:accr_mode}
    \centering
    \begin{tabular}{ccccccc}\hline
         Morphology&n&SF&HERG&LERG&RQ AGN&undefined\\\hline 
         Straight FRI&86&0 [0\%]&5 [6\%]&81 [94\%]&0 [0\%]&0 [0\%]\\
         NAT&18&0 [0\%]&1 [5.5\%]&16 [89\%]&0 [0\%]&1 [5.5\%]\\
         WAT&57&0 [0\%]&1 [2\%]&56 [98\%]&0 [0\%]&0 [0\%]\\
         \textbf{Total FRI}&161&0 [0\%]&7 [4\%]&153 [95\%]&0 [0\%]&1 [<1\%]\\
         FRII-High&69&0 [0\%]&24 [35\%]&45 [65\%]&0 [0\%]&0 [0\%]\\
         FRII-Low&57&0 [0\%]&5 [9\%]&52 [91\%]&0 [0\%]&0 [0\%]\\
         \textbf{Total FRII}&126&0 [0\%]&29 [23\%]&97 [77\%]&0 [0\%]&0 [0\%]\\\hline
         Core-D&40&0 [0\%]&10 [25\%]&30 [75\%]&0 [0\%]&0 [0\%]\\
         Blobs&31&1 [3\%]&6 [19.5\%]&24 [77.5\%]&0 [0\%]&0 [0\%]\\\hline
         Hybrid&111&3 [3\%]&13 [12\%]&95 [85\%]&0 [0\%]&0 [0\%]\\
         Small FRI&536&52 [10\%]&54 [10\%]&407 [76\%]&4 [1\%]&19 [3\%]\\
         Small FRII&62&3 [5\%]&6 [9.5\%]&52 [84\%]&0 [0\%]&1 [1.5\%]\\
         Small Hybrid&85&5 [6\%]&7 [8\%]&71 [84\%]&0 [0\%]&2 [2\%]\\
         Unresolved&677&249 [37\%]&35 [5\%]&326 [48\%]&17 [3\%]&50 [7\%]\\
         Too faint&15688&11663 [74\%]&211 [1.5\%]&2809 [18\%]&813 [5\%]&192 [1.5\%]\\\hline
    \end{tabular}
\end{table*}

In terms of accretion mode we found the overwhelming majority (95 per cent) of FRI sources to be LERGs, a percentage that is in broad agreement with those found in earlier surveys \citep[e.g.][]{Best2012,Best2014,Capetti2017FRI}, and with no significant variations in the fractions of straight or bent sources. Of the 7 FRI HERGs at least 4 have been confirmed as such by optical, X-ray, and/or IR surveys, including two sources identified as broadline QSOs in the Sloan Digital Sky Survey as of data release 16 \citep[SDSS DR16,][]{Blanton2017,Ahumada2020}. 

The FRII-Low are also mostly LERGs (91 per cent). This is consistent with recent results by \citet{Miraghaei2017} and \citet{Grandi2021}, who show that at low luminosities and fluxes FRIIs are predominantly LERGs (80 and 89 per cent, respectively at $z<0.1$ and $z<0.15$). Most importantly, our sample covers extra parameter space in terms of luminosity and redshift compared to FIRST \citep{FIRST1995,FIRST2015}, as LoTSS-Deep is $\sim50-70$ times more sensitive.

Surprisingly, we found that nearly two thirds (65 per cent) of the FRII-High are LERGs. This is not consistent with the fractions derived from older surveys, though the LOFAR work of \citet{Williams2018a} already showed a larger fraction of LERGs at high radio luminosities compared to existing surveys. We discuss these results and their implications in more depth in Section \ref{sec:FR_accretion_link}. 

The core-dominated sources and fuzzy blobs are also LERG-dominated, though the percentage of HERGs in these populations (25 and 19.5 per cent, respectively) is only slightly smaller than that of the FRII-High. It is thus unlikely that these classes are predominantly populated by QSOs, though a fraction of them are clearly luminous AGN at other wavelengths. As suggested in \citetalias{Mingo2019} they are likely to contain a mixture of populations, including restarting sources.

Among the other morphological classes obtained with \textsc{LoMorph}, only the (large - as defined in Section \ref{sec:Auto_class}) hybrids contain a noticeable fraction of HERGs (12 per cent), consistent with the likely scenario that at least some of them might be FRII sources in projection \citep{Harwood2020}.


\subsection{Mid-IR diagnostics}\label{sec:MIR}

We investigated how the accretion and morphology classifications relate to AGN sample selections at other wavelengths, so as to understand how reliably LOFAR sources can be classified when the sophisticated accretion mode classification method employed here is unfeasible. 

Mid-infrared colour-colour plots are commonly used as a diagnostic tool for radiatively efficient AGN activity and sample selections \citep[see e.g.][]{Mateos2012,Stern2012,Donley2012,Assef2013,Secrest2015,Chhetri2020}, though their usefulness for selecting AGN is limited to very bright radiatively efficient sources (quasars and bright Seyfert galaxies), where the torus component dominates, and often fainter AGN are missed \citep[see e.g.][]{Rovilos2014,Gurkan2014,Gurkan2018a,Gurkan2019,Mingo2016,RetanaMontenegro2020}. Given that all three fields have \textit{Spitzer} IRAC \citep{Spitzer2004,SpitzerIRAC2004} and \textit{WISE} \citep{Wright2010,WISE2011} coverage, these are the most useful diagnostics to test with our sample (see also the discussion by Best et al., subm.).

\begin{figure}
  \resizebox{\hsize}{!}{\includegraphics[trim={0cm 0cm 1.2cm 1cm},clip]{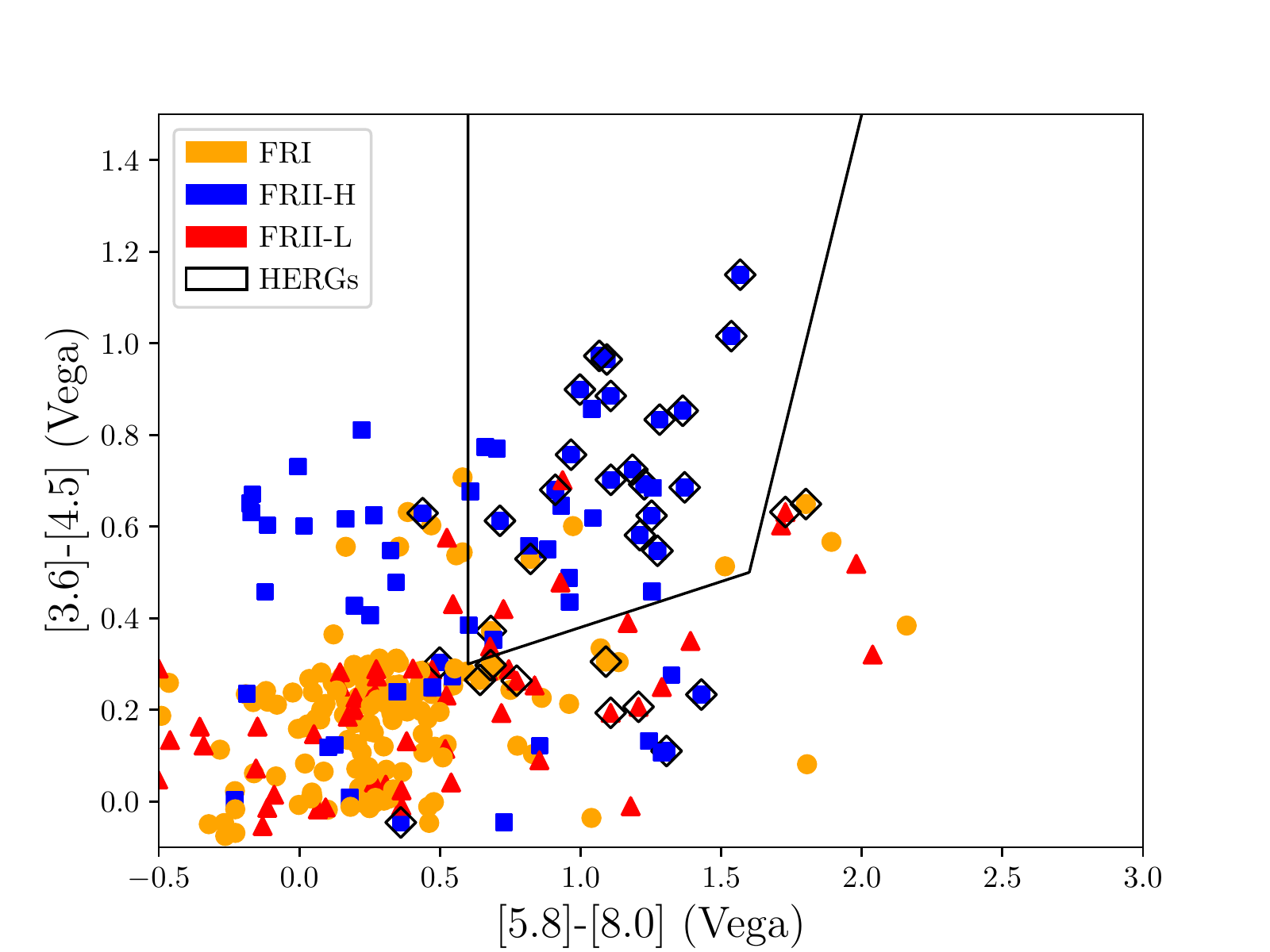}}
  \caption{\textit{Spitzer} IRAC colour-colour diagnostic plot for the FRI (yellow circles), FRII-High (blue squares), and FRII-Low (red triangles) in the combined LOFAR deep fields, including the AGN selection wedge from \citet{Stern2005} and \citet{Assef2013}. HERGs are outlined with black diamonds, all the other sources are LERGs.}
  \label{fig:IRAC}
\end{figure}

\begin{figure}
  \resizebox{\hsize}{!}{\includegraphics[trim={0cm 0cm 1.2cm 1cm},clip]{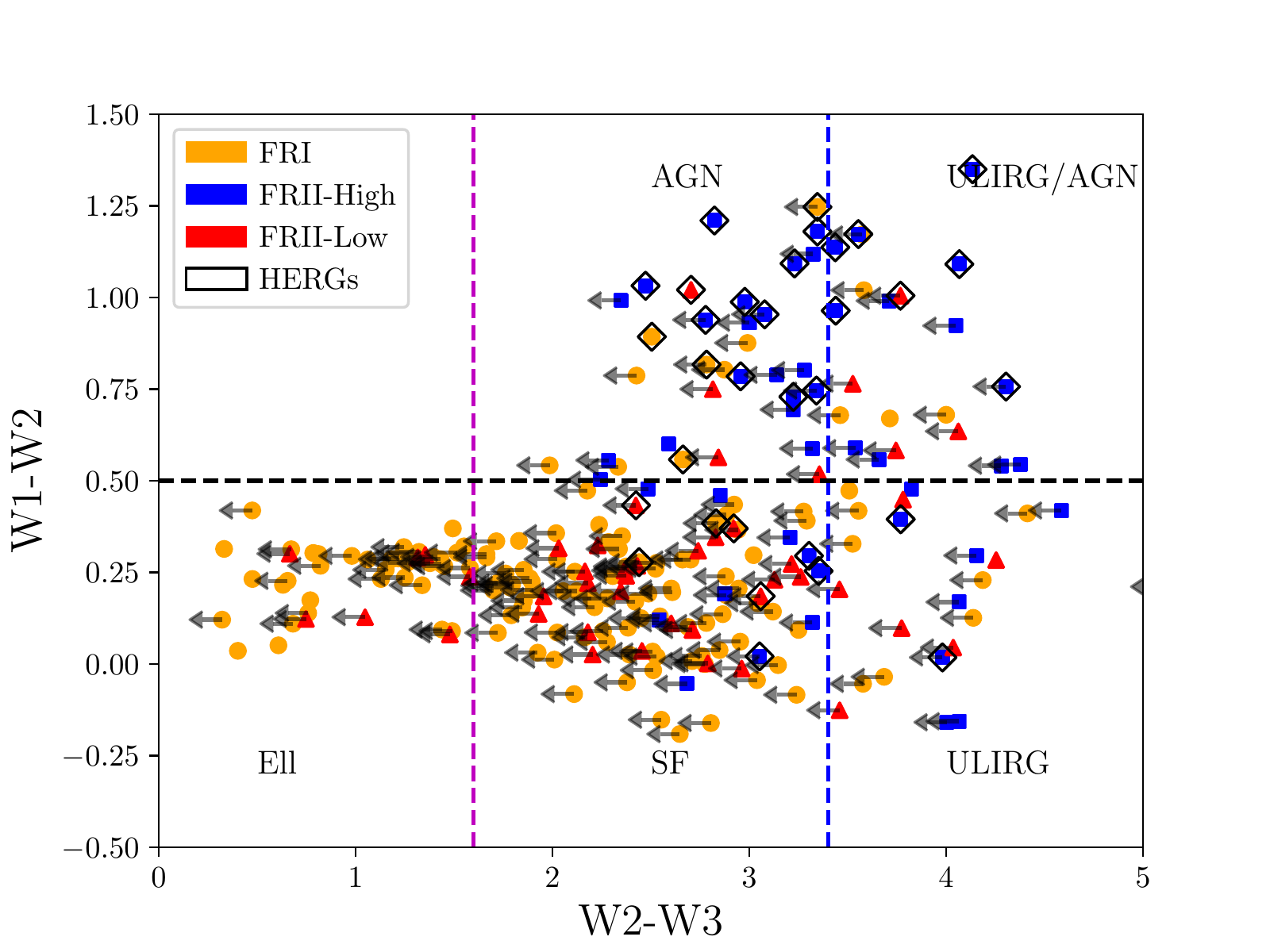}}
  \caption{\textit{WISE} colour-colour diagnostic plot. Symbols and colours as in fig. \ref{fig:IRAC}. Rough class divisions as in \citet{Mingo2016,Mingo2019}. W3 upper limits are represented with grey arrows.}
  \label{fig:WISE}
\end{figure}

Fig. \ref{fig:IRAC} shows the \textit{Spitzer} IRAC colour-colour distribution of our sources, including the AGN selection wedge of \citet{Stern2005} and \citet{Assef2013}, which is less restrictive than that of \citet{Donley2012}. Nearly all sources ($>97$ per cent) were detected by IRAC. The selection wedge encompasses 61 per cent of the HERGs, but 47 per cent of the sources inside it are LERGs. This is not unexpected, as both \citet{Assef2013} and \citet{Donley2012} highlight the likely presence of contaminants in this area for data covering a large redshift range.

We matched our sample in \textsc{topcat} \citep{Taylor2005} with the AllWISE catalogue of \citet{AllWISE2014}, using a 2 arcsec matching radius \citep[following the SDSS DR12 matching criteria of][which should yield a false positive rate of just 2 per cent]{Paris2017} around the optical host positions for each source. The detection fraction was lower for \textit{WISE} than IRAC (93/75/82 per cent of FRI/FRII-High/FRII-Low), and particularly bad for the FRII-High LERGs, of which only 64 per cent had \textit{WISE} counterparts. Fig. \ref{fig:WISE} illustrates these results. We have used the rough divisions of \citet{Mingo2016} and \citet{Mingo2019}, based on the synthetic SED results of \citet{Wright2010} and \citet{Lake2012}, to categorise the sources. The `AGN' box encompasses 44 per cent of the \textit{WISE}-detected HERGs, while 53 per cent of the sources in the box are LERGs. A wedge like that of \citet{Mateos2012} would still only contain $\sim65$ per cent of the HERGs.

These results confirm the fact \citep[e.g.][]{Gurkan2014,Mingo2016} that neither diagnostic plot offers a reliable accretion mode diagnostic or selection tool for radio AGN, and both -- but especially \textit{WISE}, being shallower -- must be used with caution, particularly for samples with a wide redshift range. 

Luminous FRII LERGs, which constitute nearly 2/3 of all our FRII-High and are not being picked up by shallower radio surveys, are not well detected by \textit{WISE} (and likely other surveys selecting AGN at wavelengths other than radio). We discuss the implications of this result in more detail in Section \ref{sec:FR_accretion_link}.

\section{Is there a link between accretion mode and FR class?}\label{sec:FR_accretion_link}

\begin{table*}
	\caption{Breakdown of LERG/HERG fractions and comparison with other samples. Agresti-Coull confidence intervals (last column) were calculated at the 2-sigma level, and shown here multiplied by 100 to compare with the percentages for each sample. The first group of rows shows the results for all the sources in our sample, across the combined Deep Fields and various luminosity intervals. The second set of rows shows the results broken down by morphology. The third group shows the results from the $z<1$ FRIIs in the 3CRR sample \citep{Laing1983,Hardcastle2007,Hardcastle2009,Mingo2014}. The last group of rows shows a comparison with the results of \citet{Williams2018a}, with three subsets of our sources matched in terms of luminosity and redshift distribution.}\label{tab:AC}
	\centering
	\begin{tabular}{cccccc}
		Sample&Total&LERG&HERG&\% LERG&AC error\\\hline
		All DF FRI/FRII&286&250&36&87.4&3.8\\\hline
		All DF, $L_{150}>10^{26}$&88&62&26&72.6&9.3\\
		All DF, $L_{150}\leq10^{26}$&198&188&10&95.0&2.8\\
		All DF, $10^{25}<L_{150}\leq10^{26}$&117&108&9&92.3&4.3\\
		All DF, $L_{150}\leq10^{25}$&81&80&1&98.8&(-)\\\hline
		All FRI&160&153&7&95.6&2.7\\
		All FRII&126&96&29&76.2&3.7\\
		FRII-High&69&45&24&65.2&10.9\\
		FRII-Low&57&52&5&91.2&5.8\\\hline
		All 3CRR FRII ($z\leq1$)&136&18&118&13.2&6.0\\
		3CRR FRII ($z\leq1$), $L_{178}\leq2\times10^{28}$&62&13&49&21.0&10.5\\\hline
		Bo\"otes sample from \citet{Williams2018a}&47&24&23&51.1&13.9\\
		All DF, broad cut ($L_{150}>5.65\times10^{25}, z\leq2$)&111&82&29&73.9&8.0\\
		DF FRII, matched&78&52&26&66.7&10.2\\
		All DF, strict cut ($L_{150}>3.16\times10^{26}, 0.5<z\leq2$)&42&26&16&61.9&13.9\\\hline
	\end{tabular}
\end{table*}

\begin{figure}
  \resizebox{\hsize}{!}{\includegraphics[trim={0cm 0cm 1.2cm 1cm},clip]{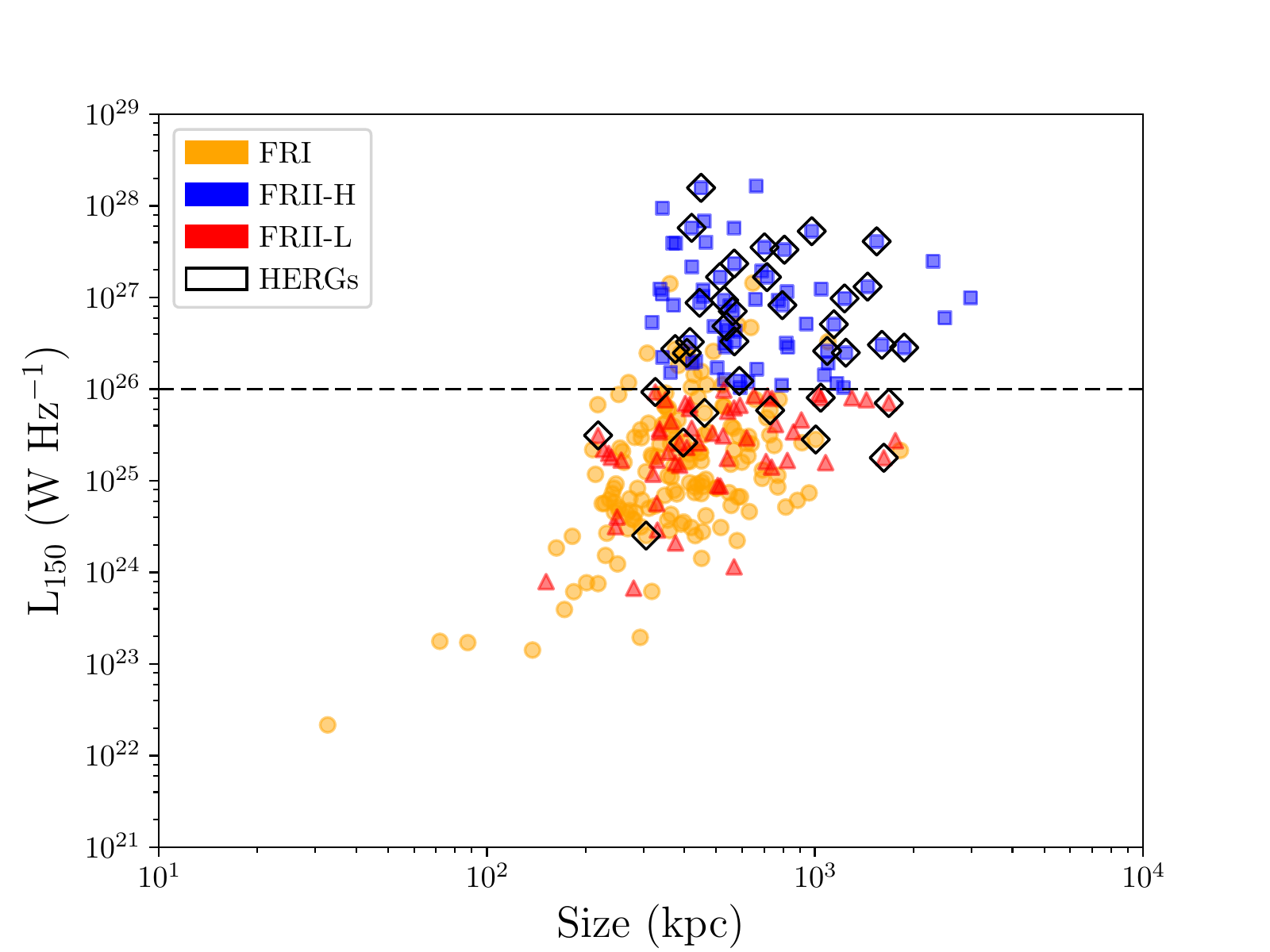}}
  \caption{Luminosity versus size distribution of FRI, FRII-Low, and FRII-High, including accretion mode. Colours and symbols as in Fig. \ref{fig:IRAC}. The dashed line shows the `traditional' boundary between FRI and FRII, at $L_{150}\sim10^{26}$ W Hz$^{-1}$ \citep{FR1974,Ledlow1996}.}
  \label{fig:L150_size_HERG_LERG}
\end{figure}

\begin{figure*}
\centering
  \includegraphics[width=0.49\textwidth, trim={0cm 0cm 1.2cm 1cm},clip]{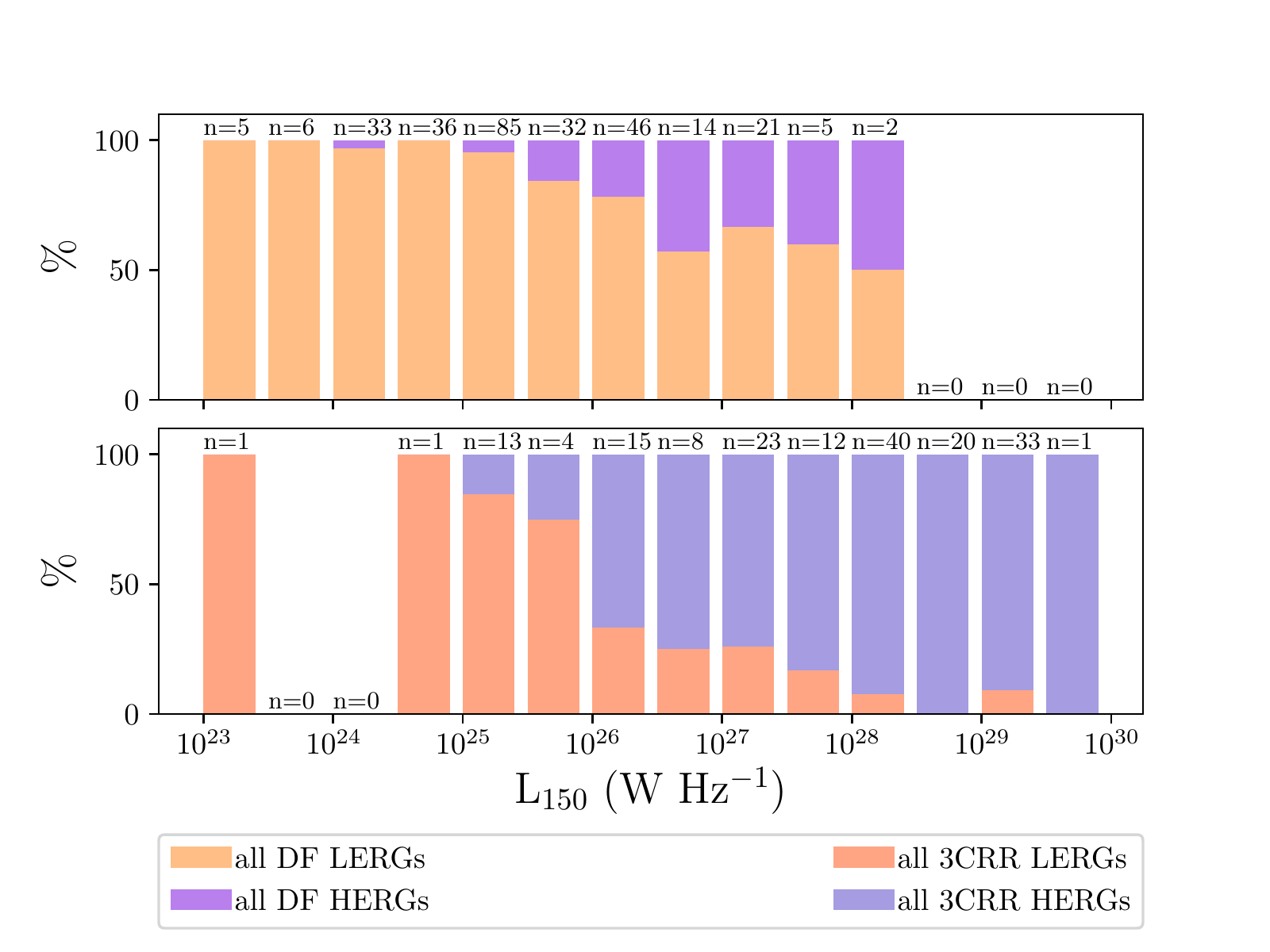}
  \includegraphics[width=0.49\textwidth, trim={0cm 0cm 1.2cm 1cm},clip]{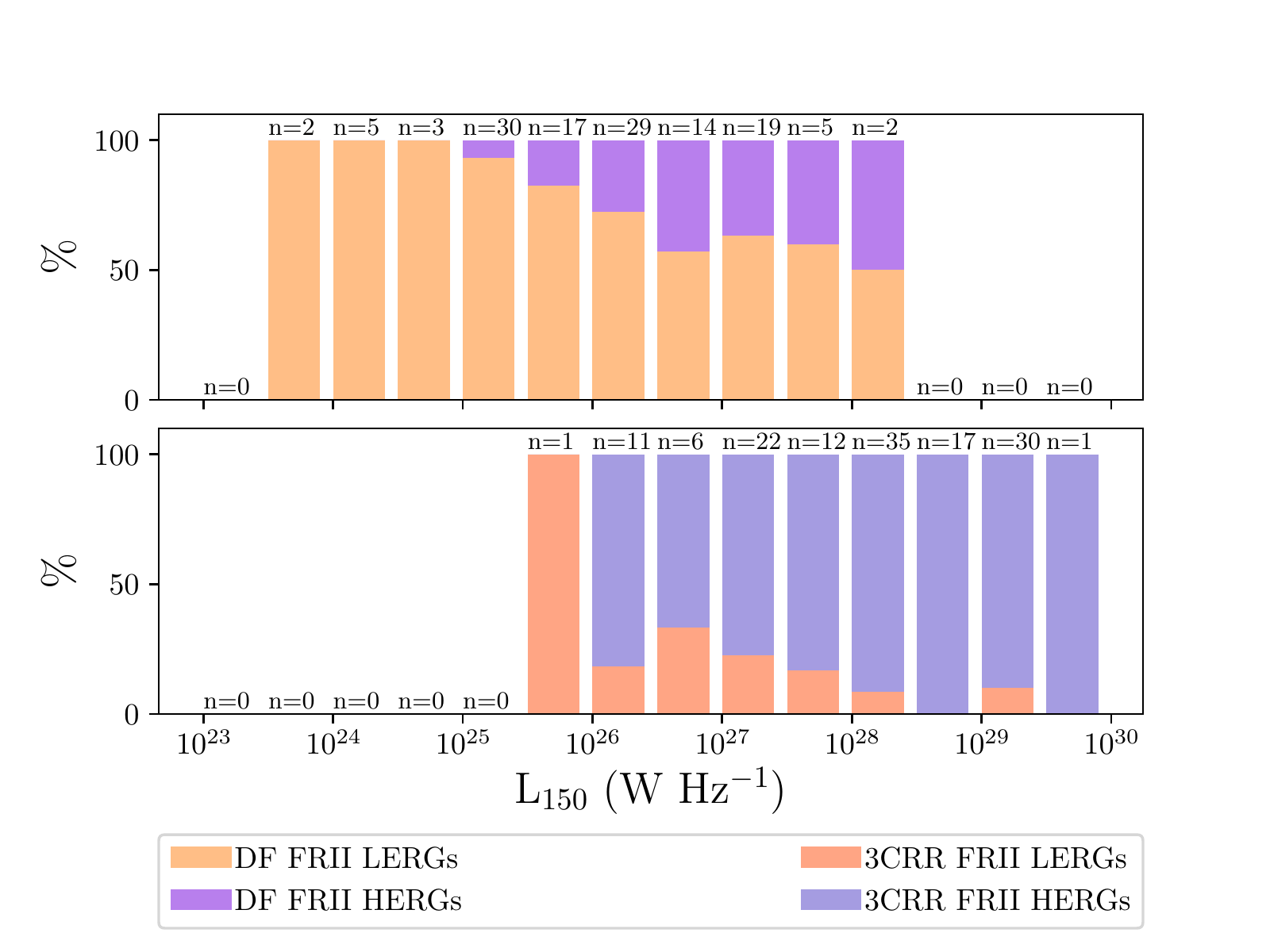}
  \caption{Left: stacked histograms showing the percentile fraction of LERGs and HERGs in our LoTSS-Deep dataset (top) and the 3CRR sample (bottom) as a function of luminosity, regardless of their FR classification. Right: same plot, restricted to FRII sources in both samples. In all panels LERGs are shown in shades of orange and HERGs in shades of purple. 3CRR Data obtained from M. Hardcastle's updated version of the \citet{Laing1983} 3CRR catalogue (available at \url{https://3crr.extragalactic.info/}). 3C 343.1 was excluded from the sample \citep{Arp2004}.}
  \label{fig:comb_LH_histo}
\end{figure*}

Section \ref{sec:AGN_SF} and Table \ref{tab:accr_mode} show that our samples are dominated by LERGs across all populations, though in different proportions depending on luminosity. The LERG fraction for different luminosity and FR class subsets of our sample is shown in Table \ref{tab:AC}. These results strongly argue that radio morphology on 100-kpc scales is not directly controlled by either accretion mode, given the considerable fraction of FRII LERGs, or by jet power, given the presence of the FRII-Low.

We do find some dependence of accretion mode on luminosity (see Table \ref{tab:AC}), but as we found in \citetalias{Mingo2019}, this is not necessarily linked to morphology. Fig. \ref{fig:L150_size_HERG_LERG} shows the luminosity vs size distribution of FRI, FRII-Low, and FRII-High, with black diamonds surrounding all sources classified as HERGs; this clearly shows that LERGs are the dominant population at low radio luminosities, regardless of morphology. HERGs are most likely to be found above the traditional FRI/II boundary, though even in this area of the parameter space they are a minority. As mentioned in Section \ref{sec:zLumSiz}, we note that the bottom-right corner of the parameter space is inaccessible due to surface-brightness limits \citep[see e.g.][]{Hardcastle2019}, so we might be missing further FRII-Low at large sizes.

It is clear that the HERG/LERG ratio in our sample is very different compared to older surveys \citep[e.g. 3CRR, 6C, 7C, 2Jy, see][and references therein]{Laing1983,Mingo2014,Fernandes2015}, particularly at higher luminosities \citep[LERGs have long been known to dominate at low luminosities out to at least $z\sim1$, e.g.][]{Best2014,Pracy2016}. This is illustrated in the left-hand panel of Fig. \ref{fig:comb_LH_histo}, which shows the luminosity dependence of the HERG/LERG fraction for our sample and all the sources in 3CRR, regardless of their FR classification. While at lower luminosities (FRI-dominated in 3CRR) the HERG/LERG ratios appear compatible, the high fraction of LERGs among the FRII-High sources in particular is different to that found for earlier surveys, as is clearly shown in the right-hand panel of Fig. \ref{fig:comb_LH_histo}. It is likely that our lower flux density limit drives the discrepancy with older, shallower surveys. For instance, as shown in Table \ref{tab:AC}, the LERG fraction for 3CRR FRIIs is only 13 per cent. However, the LOFAR Bo\"otes results of \citet{Williams2018a} already showed a trend in the direction that we see in our work. Splitting our sample in several ways to match that of \citet{Williams2018a}, as illustrated in the last section of Table \ref{tab:AC}, yields HERG/LERG percentages that are different for both samples, but consistent within the errors. Kondapally et al. (subm.) note that the small discrepancy between our Bo\"otes results and those of \citet{Williams2018a} is likely due to a combination of the lower flux density limit and the differences in the SED classification method.

It is very unlikely that these very bright radio sources could be associated with radiatively-efficient but optically-faint AGN, misclassified as LERGs via the SED fitting process. Even considering the large scatter, and the caveats on how jet power translates into radio luminosity \citep[e.g.][]{Hardcastle2013}, there is a broad correlation between the radiative and kinetic (jet) output of RE AGN \citep[see e.g.][]{Mingo2016}. If the FRII-High LERGs were RE they would be outliers in this correlation, and thus still intrinsically different, in terms of their underlying physics, from the luminous, radiatively efficient FRIIs observed in older surveys.

In the two subsections that follow we consider the radio morphologies of the FRI and FRII subsamples and their relation to accretion mode, including exploring the reasons for differences with shallower surveys, in more detail. Section~\ref{sec:Mstar_SFR} then builds on these results to incorporate host-galaxy parameters (stellar mass and star formation rate obtained from SED fitting) into our understanding of what controls radio morphology and accretion mode.


\subsection{FRI}\label{sec:FRI}

\begin{figure*}
	\centering
	\includegraphics[width=0.23\textwidth, trim=1.4cm 1cm 6.8cm 2.4cm, clip=true]{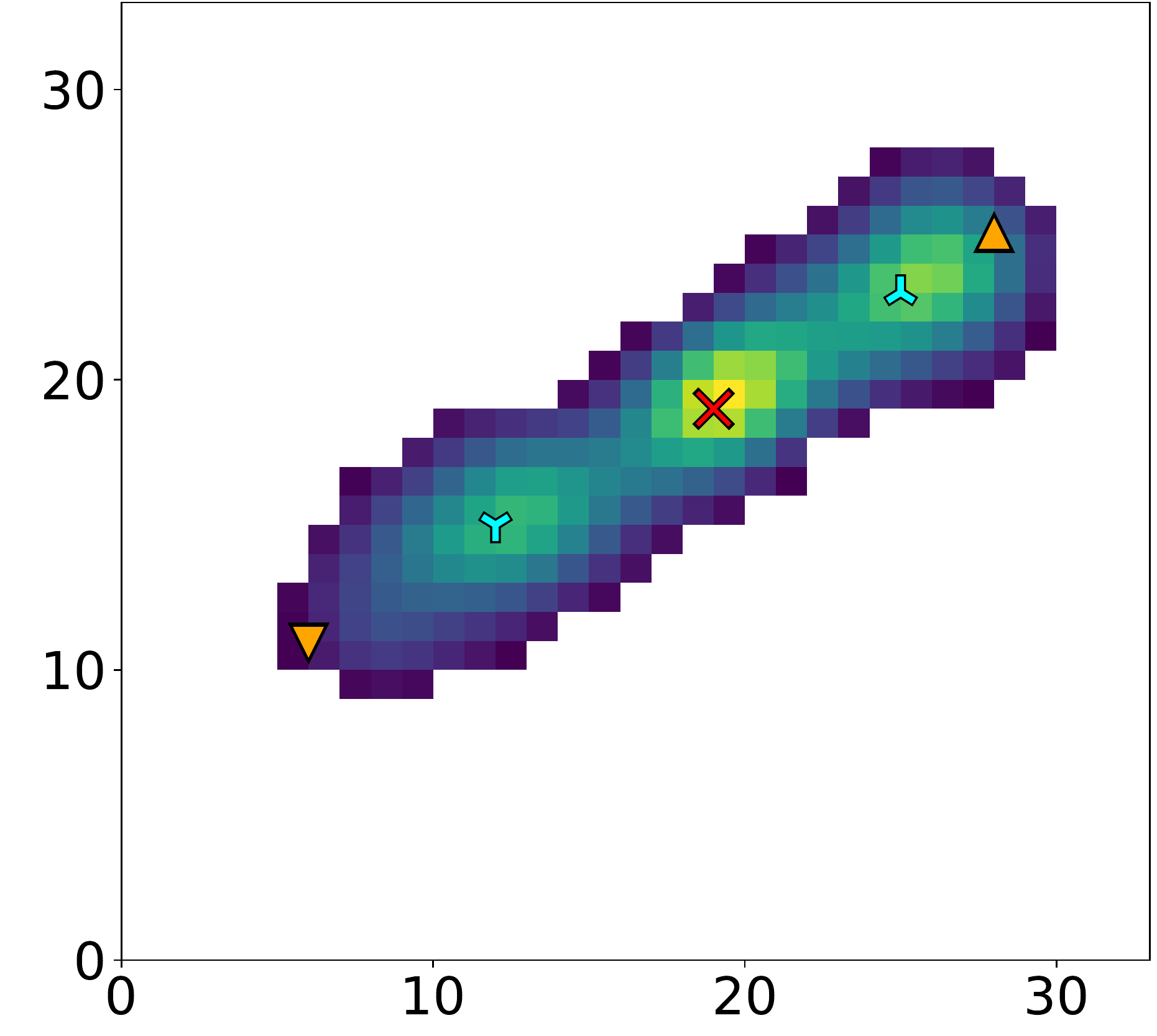} 
	\includegraphics[width=0.23\textwidth, trim=1.4cm 1cm 6.8cm 2.4cm, clip=true]{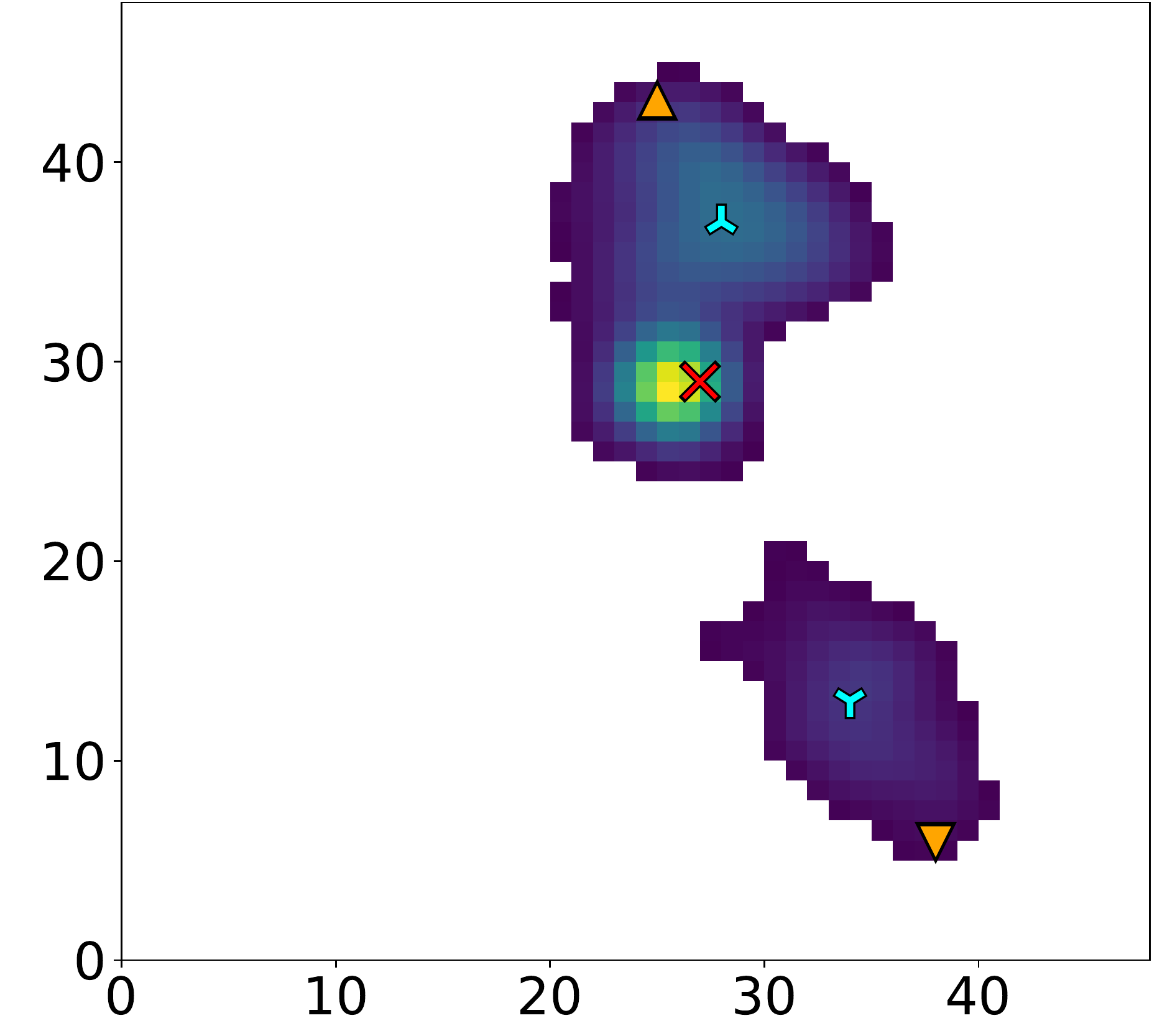} 
	\includegraphics[width=0.23\textwidth, trim=1.4cm 1cm 6.8cm 2.4cm, clip=true]{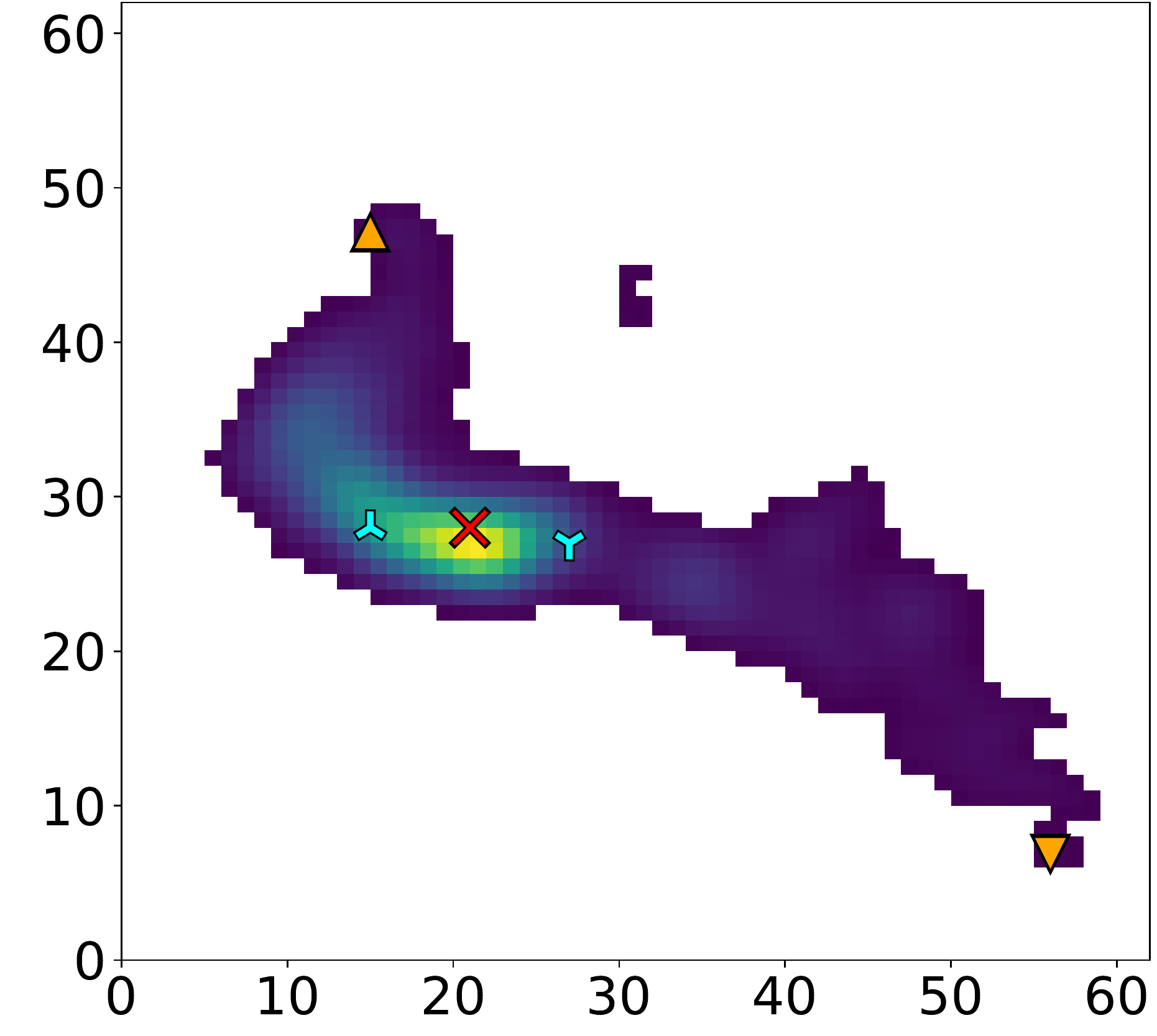} 
	\includegraphics[width=0.23\textwidth, trim=1.4cm 1cm 6.8cm 2.4cm, clip=true]{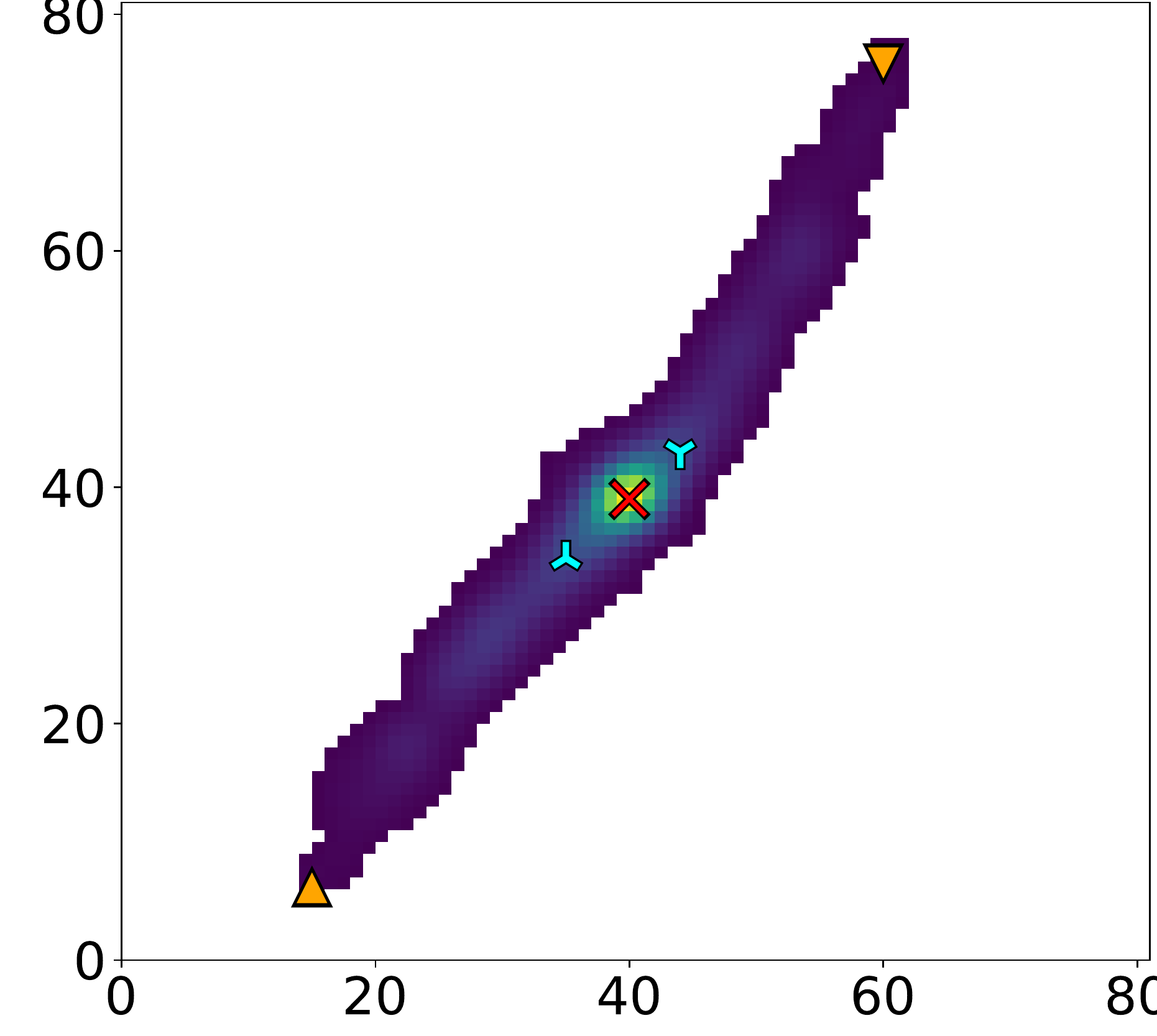} 
	\includegraphics[width=0.23\textwidth, trim=1.4cm 1cm 6.8cm 2.4cm, clip=true]{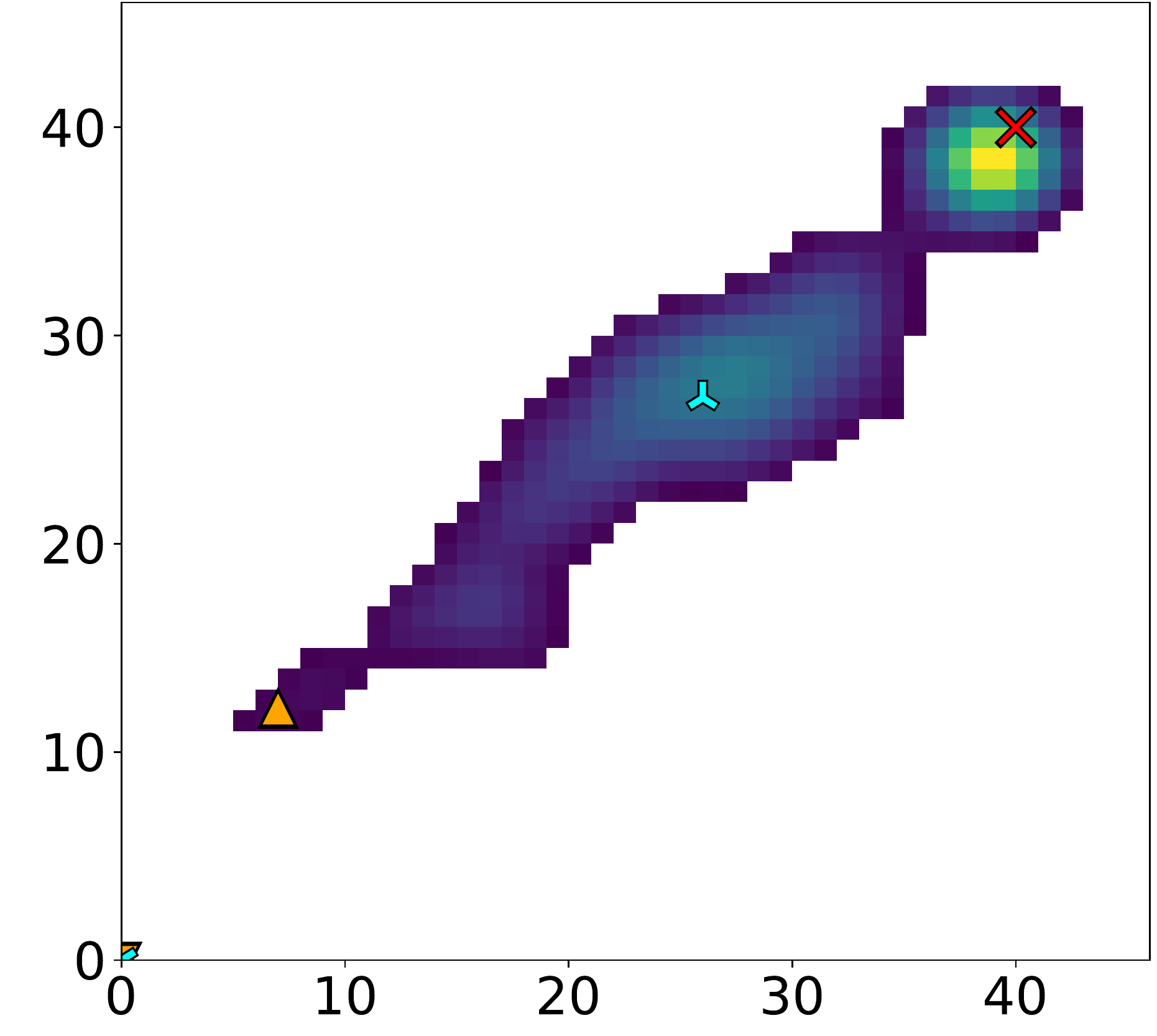} 
	\includegraphics[width=0.23\textwidth, trim=1.4cm 1cm 6.8cm 2.4cm, clip=true]{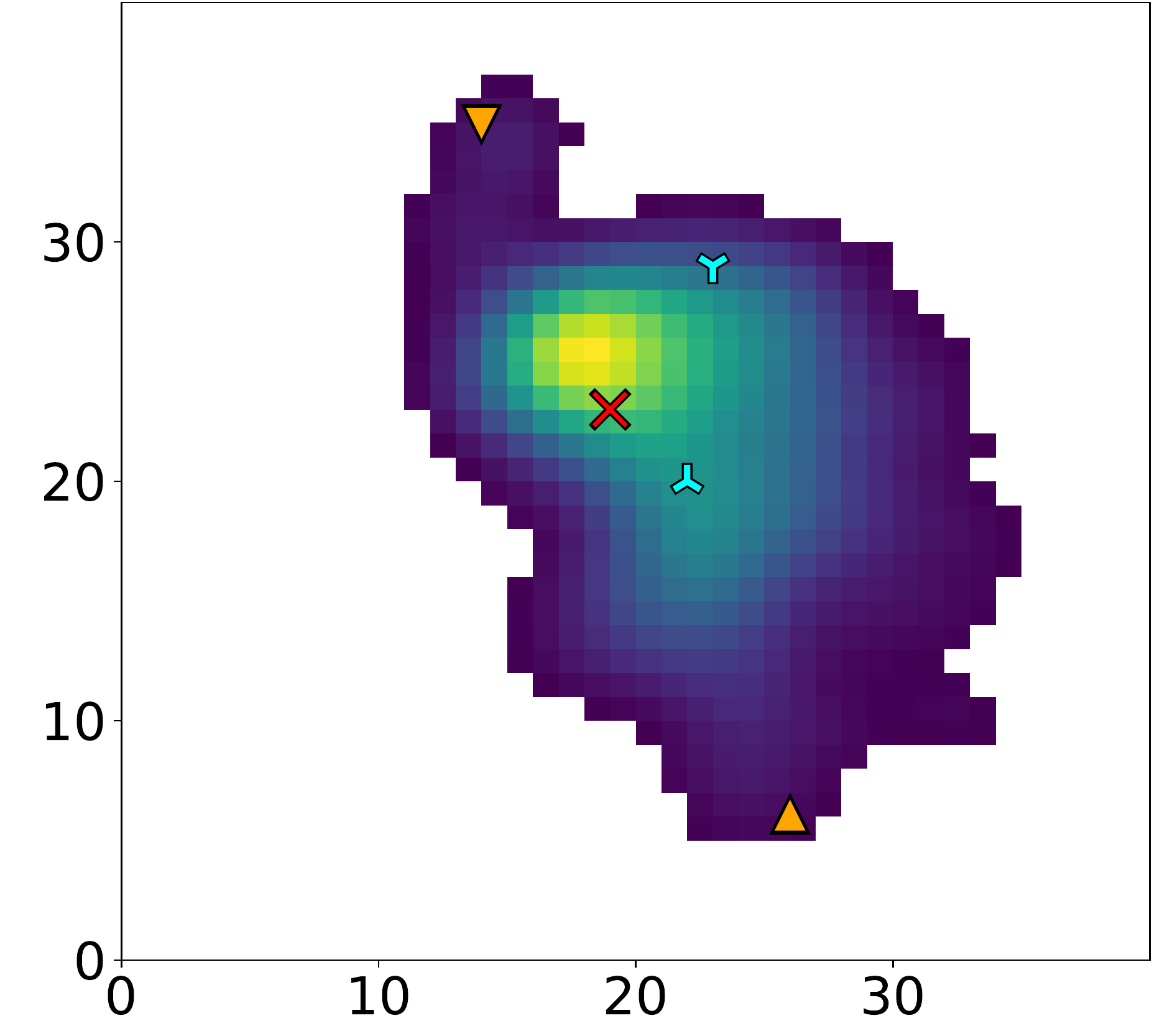} 
	\includegraphics[width=0.23\textwidth, trim=1.4cm 1cm 6.8cm 2.4cm, clip=true]{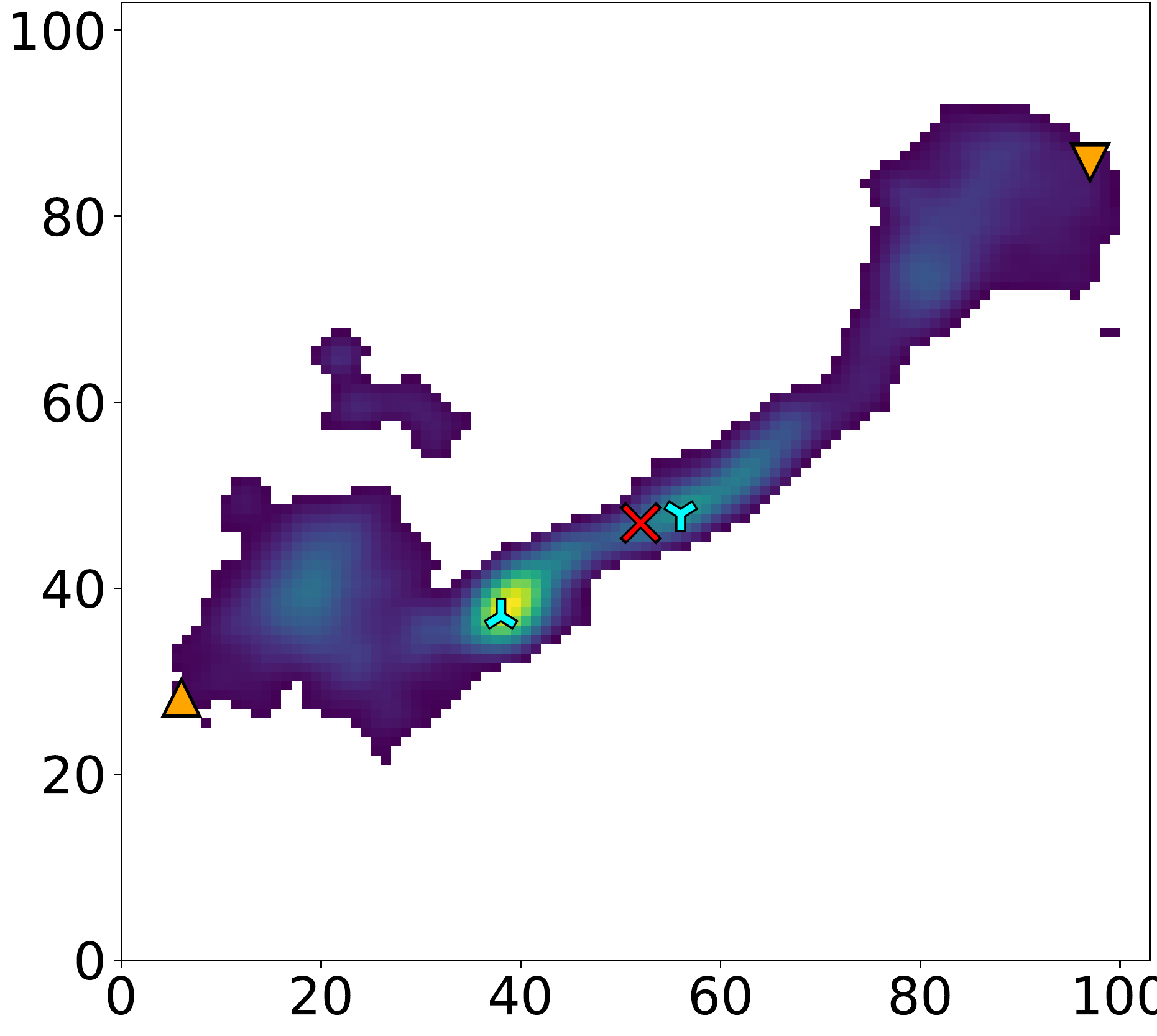}
	\includegraphics[width=0.23\textwidth, trim=1.4cm 1cm 6.8cm 2.4cm, clip=true]{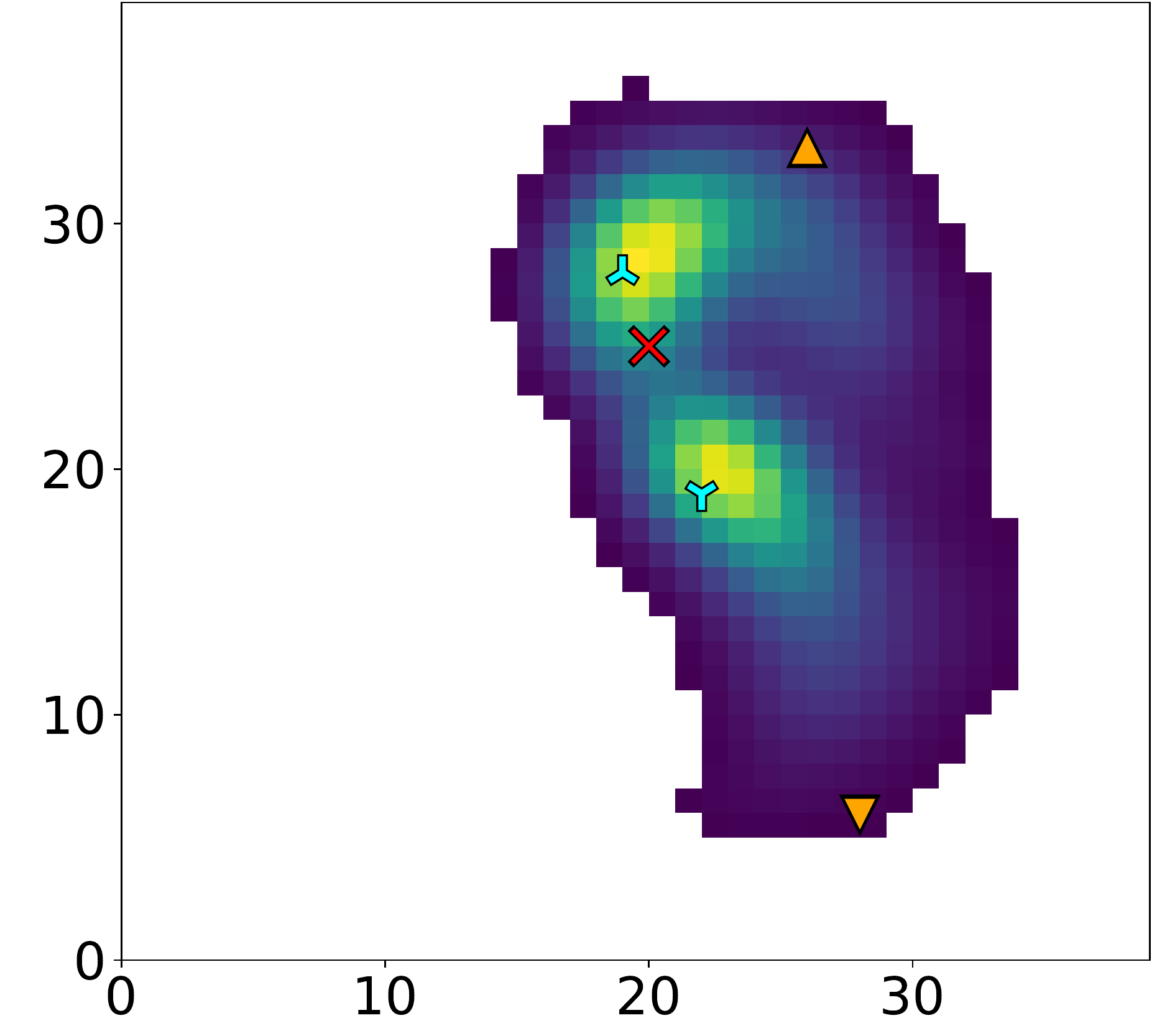}
	\caption{Gallery of FRI LERGs, showing two of each FRI class, in order: lobed, tailed, NAT and WAT. All images have been drawn at random within their respective classes, so as to be representative of the overall subsets. The plots are produced as output by \textsc{LoMorph} based on the radio maps filtered at the desired RMS level. The symbols indicate the positions of the optical host (red X), the first and second-brightest peak of emission beyond the core (d1 and d2 -- inverted and non-inverted cyan Y, respectively), and the maximum extent of the source in both directions (D1 and D2 -- up and down pointing orange triangles, respectively for the directions to the brightest and second-brightest peak). The scale is in pixel coordinates, with a scale of 1.5 arcsec per pixel. See \citetalias{Mingo2019} for more details.} 
	\label{fig:FRI_LERG_gallery}
\end{figure*}

\begin{figure*}
	\centering
	\includegraphics[width=0.23\textwidth, trim=1.4cm 1cm 6.8cm 2.4cm, clip=true]{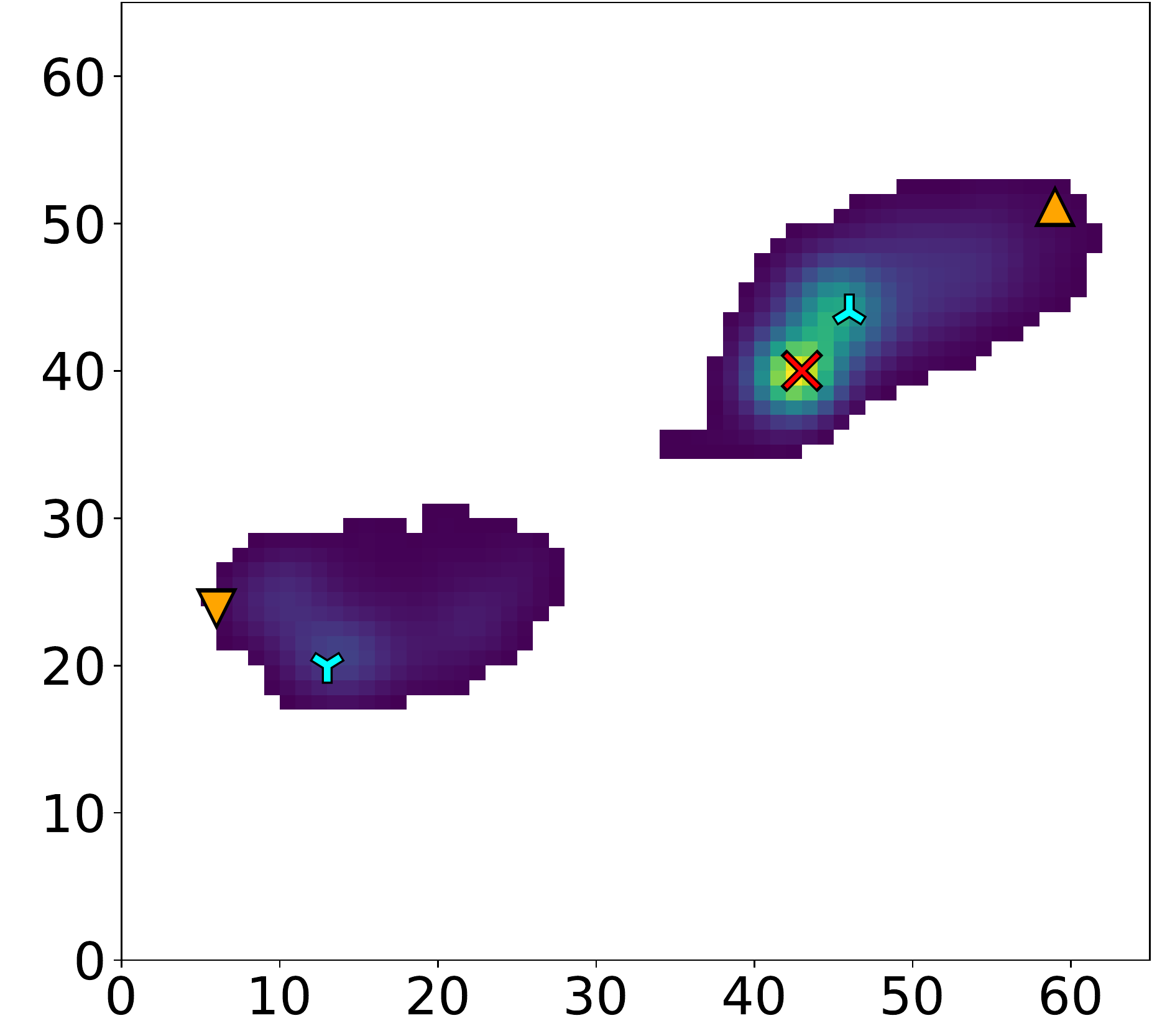} 
	\includegraphics[width=0.23\textwidth, trim=1.4cm 1cm 6.8cm 2.4cm, clip=true]{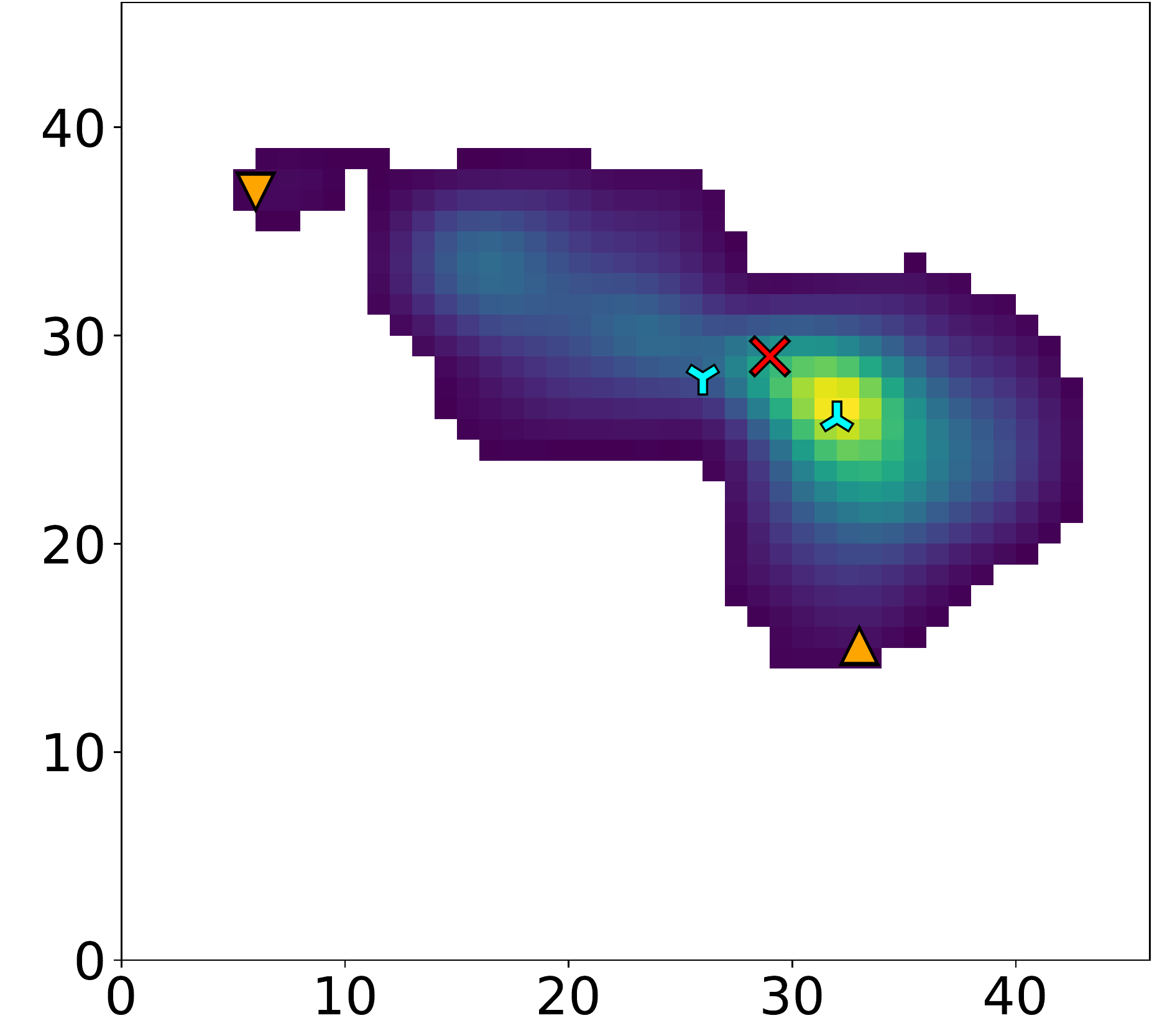} 
	\includegraphics[width=0.23\textwidth, trim=1.4cm 1cm 6.8cm 2.4cm, clip=true]{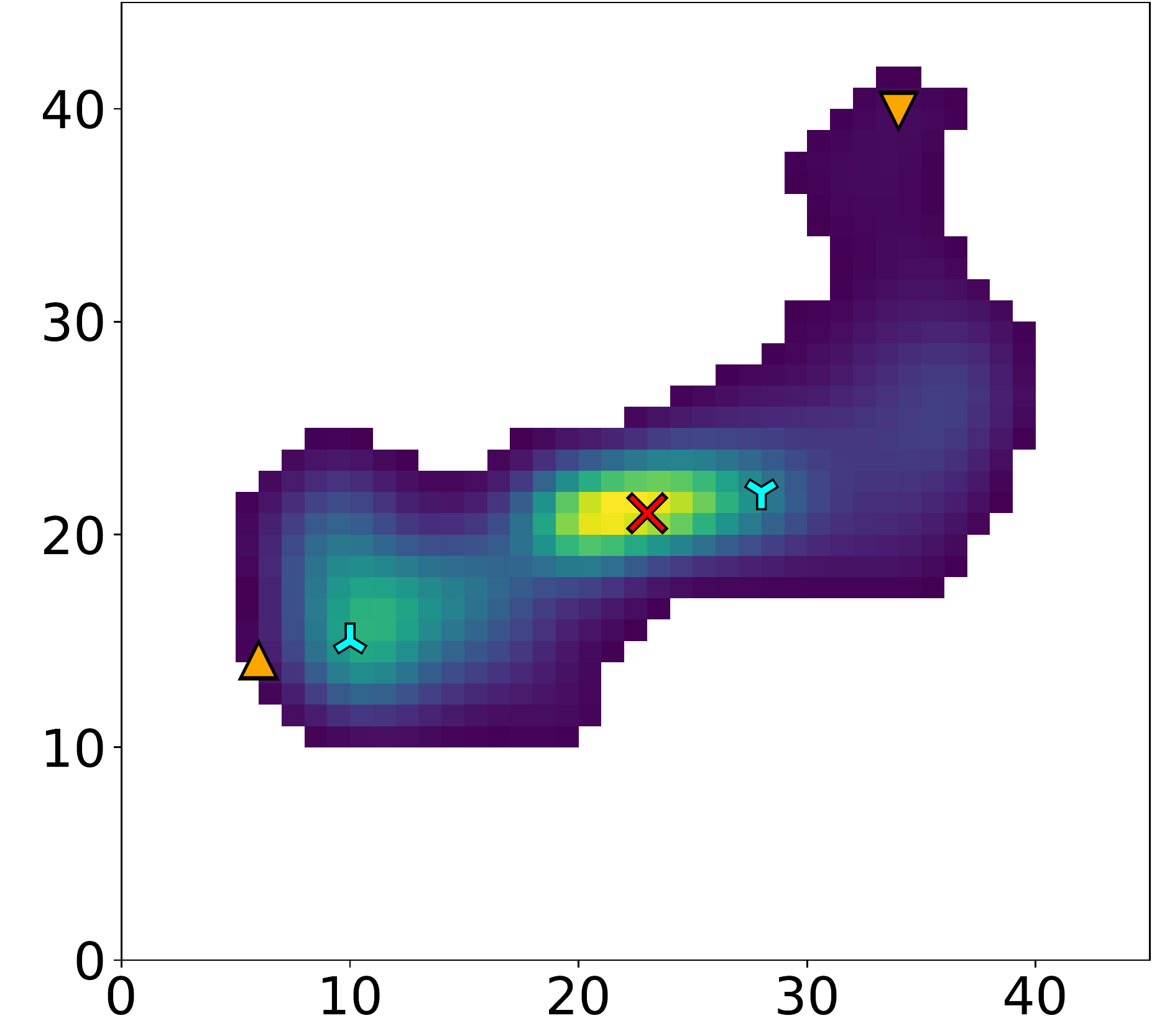} 
	\includegraphics[width=0.23\textwidth, trim=1.4cm 1cm 6.8cm 2.4cm, clip=true]{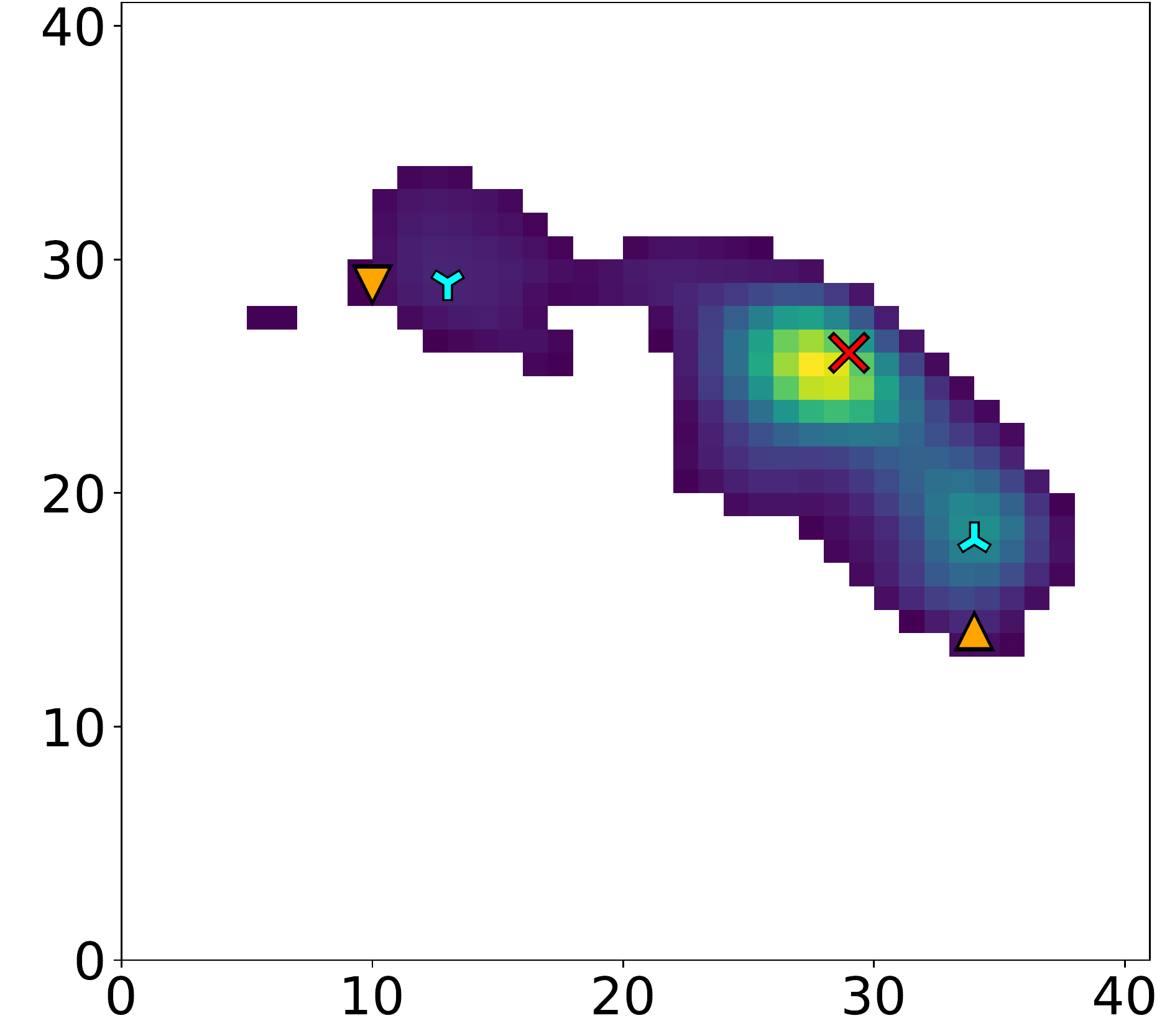} 
	\includegraphics[width=0.23\textwidth, trim=1.4cm 1cm 6.8cm 2.4cm, clip=true]{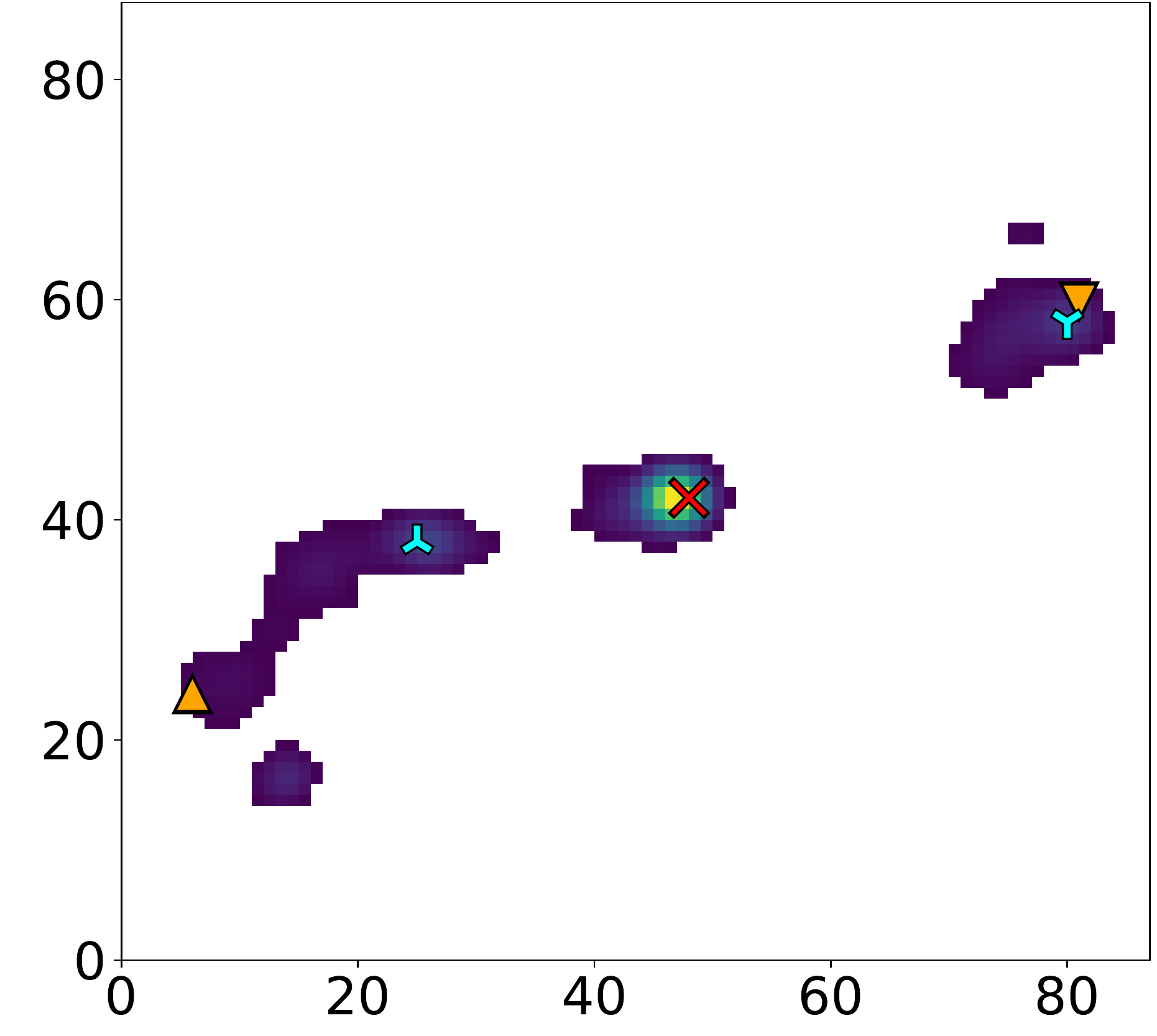} 
	\includegraphics[width=0.23\textwidth, trim=1.4cm 1cm 6.8cm 2.4cm, clip=true]{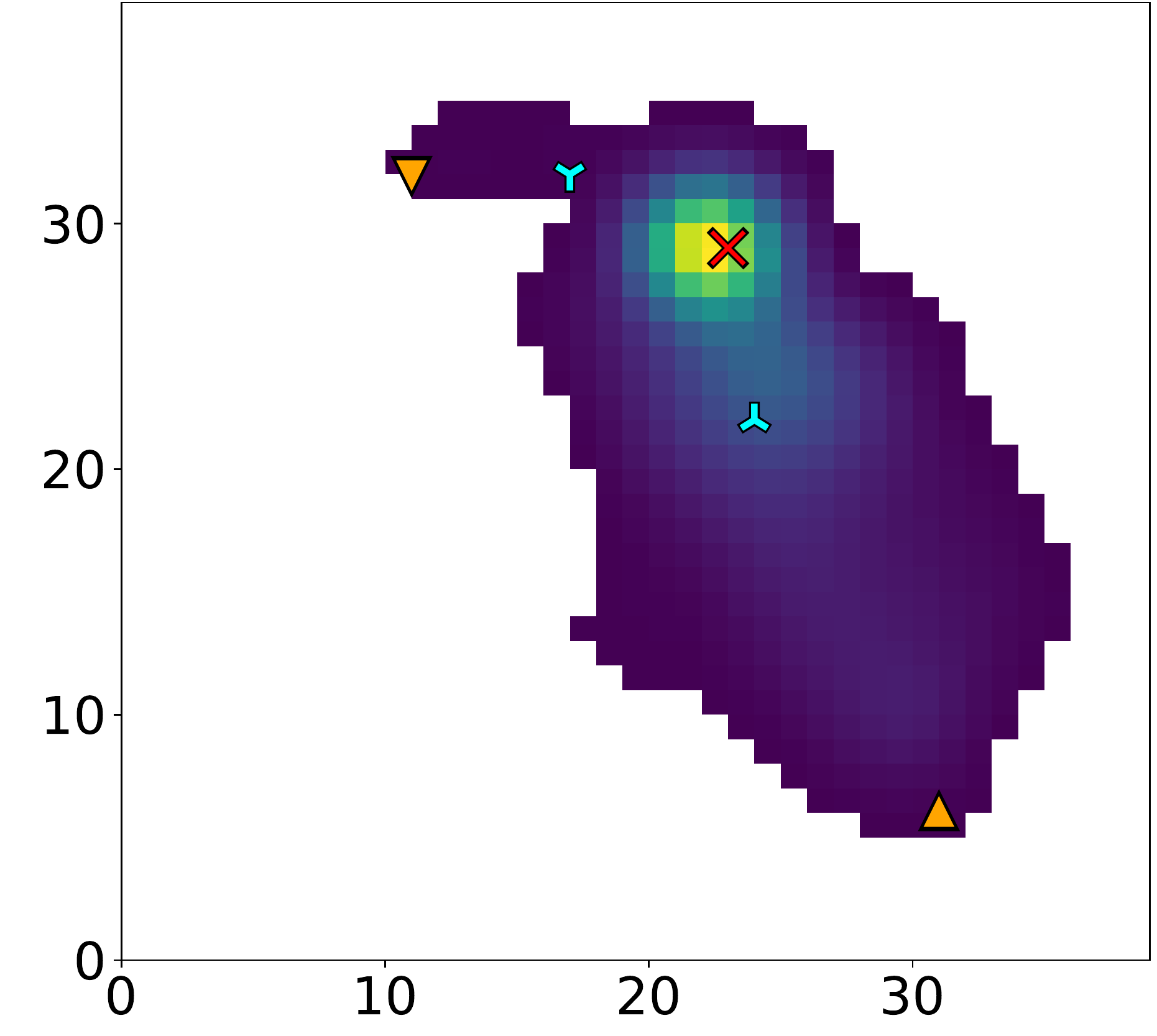} 
	\includegraphics[width=0.23\textwidth, trim=1.4cm 1cm 6.8cm 2.4cm, clip=true]{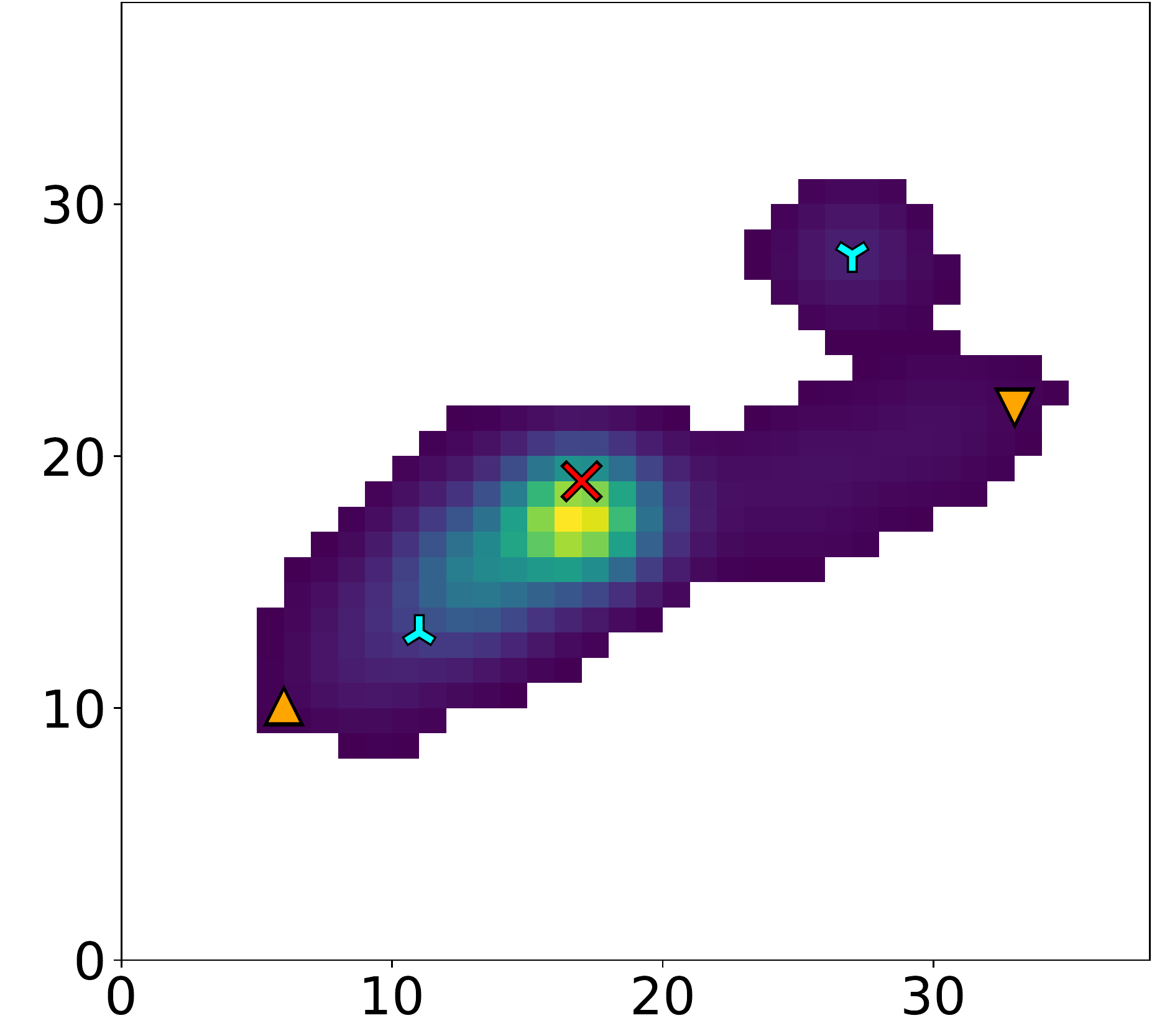}
	\caption{Gallery of all the FRI HERGs. Symbols and colours as in Fig. \ref{fig:FRI_LERG_gallery}. The scale is in pixel coordinates, with a scale of 1.5 arcsec per pixel.} 
	\label{fig:FRI_HERG_gallery}
\end{figure*}

As introduced in previous sections, our FRI sources are mostly consistent with results from previous surveys in terms of their size and luminosity distribution (Fig. \ref{fig:DF_Hetdex_L150_size}). In addition, their hosts are predominantly found in the red elliptical locus in mid-IR colour-colour plots (Figs. \ref{fig:IRAC} and \ref{fig:WISE}), and they are overwhelmingly LERGs (Figs. \ref{fig:L150_size_HERG_LERG} and \ref{fig:FRI_LERG_gallery}, Tables \ref{tab:accr_mode} and \ref{tab:AC}). 

The deeper radio and optical data in LoTSS-Deep have allowed us to identify FRI sources at higher $z$, particularly in the ELAIS-N1 field (Fig. \ref{fig:DF_z_histo}), but due to the smaller sky area limiting our sample size it is not possible to investigate any trends in terms of evolution of their accretion properties. This will become clearer as LOFAR samples grow, and with future surveys carried out with MeerKAT and the SKA.

As mentioned in Section \ref{sec:AGN_SF}, our results are consistent with the established picture in which FRI HERGs are very rare (e.g. \citeauthor{Best2012} \citeyear{Best2012}; \citeauthor{Best2014} \citeyear{Best2014}; \citeauthor{Miraghaei2017} \citeyear{Miraghaei2017}; \citeauthor{Capetti2017FRI} \citeyear{Capetti2017FRI}; Gurkan et al., subm.). Given the very small number of FRI HERGs in our sample (7 sources, all shown in Fig. \ref{fig:FRI_HERG_gallery}) it is not possible to draw firm conclusions as to what distinguishes these sources from the more numerous FRI LERGs shown in Fig. \ref{fig:FRI_LERG_gallery}, other than the fact the core seems to be the brightest structure in all but one of them, and that none of them seem to be WATs or NATs (although most do present some bends and twists). Table \ref{tab:AC} shows that the fraction of FRI HERGs must be below 8 per cent. The key as to why these FRIs are HERGs likely lies in the combination of their host masses and gas availability, as we will discuss in Section \ref{sec:Accr_mode}.


\subsection{FRII}\label{sec:FR2}

\begin{figure*}
	\centering
	\includegraphics[width=0.2\textwidth, trim=1.4cm 1cm 6.8cm 2.4cm, clip=true]{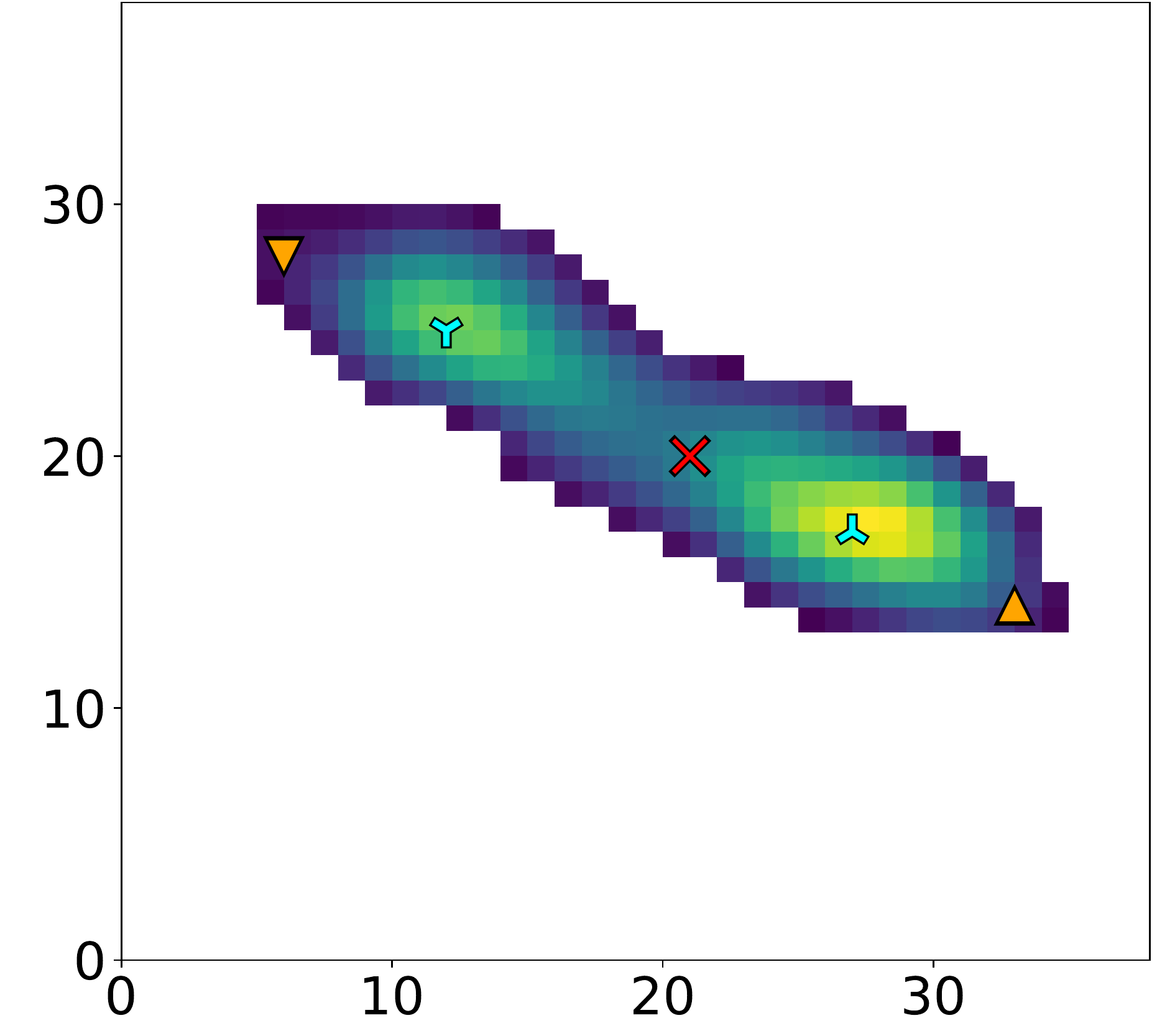} 
	\includegraphics[width=0.2\textwidth, trim=1.4cm 1cm 6.8cm 2.4cm, clip=true]{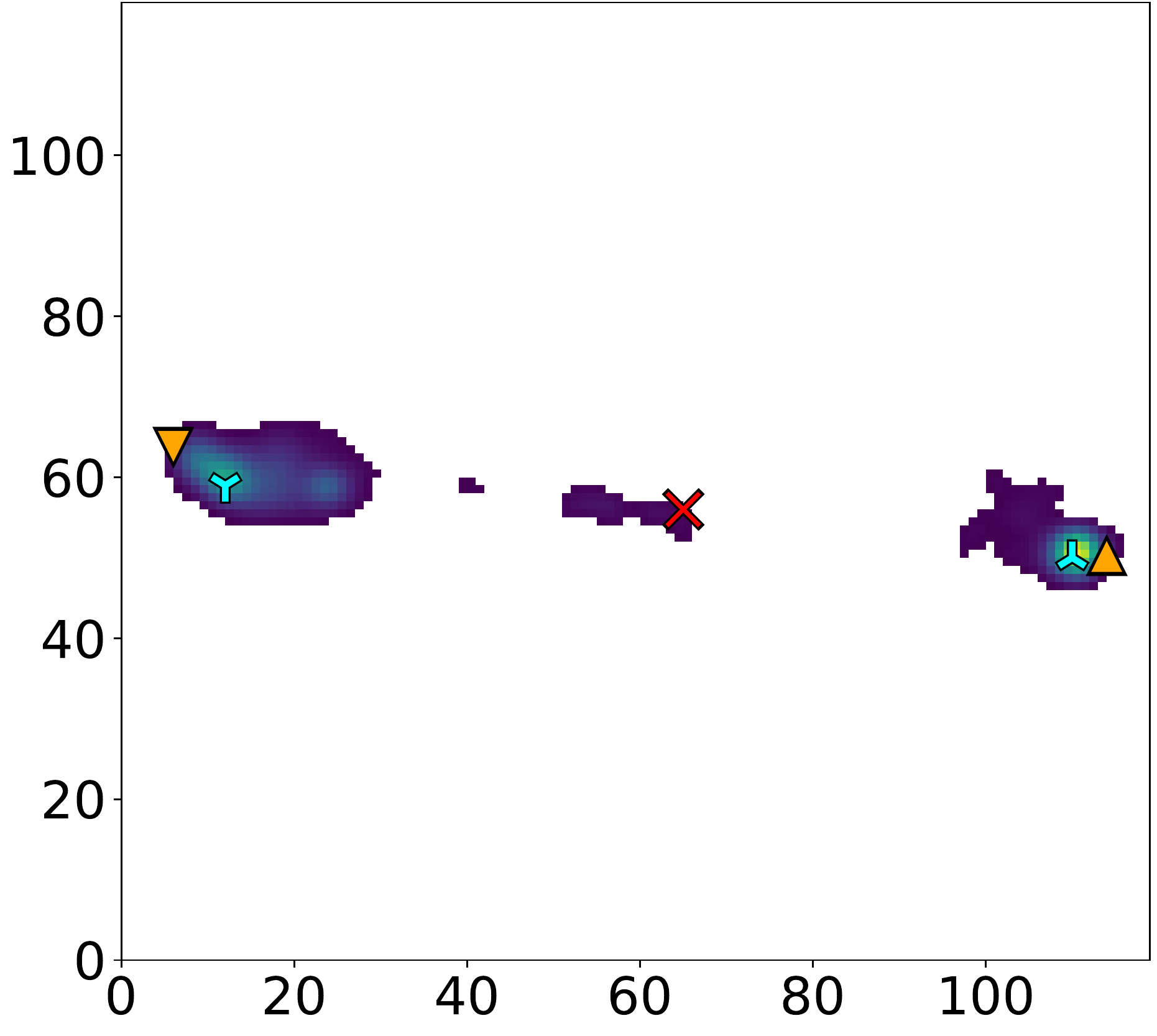} 
	\includegraphics[width=0.2\textwidth, trim=1.4cm 1cm 6.8cm 2.4cm, clip=true]{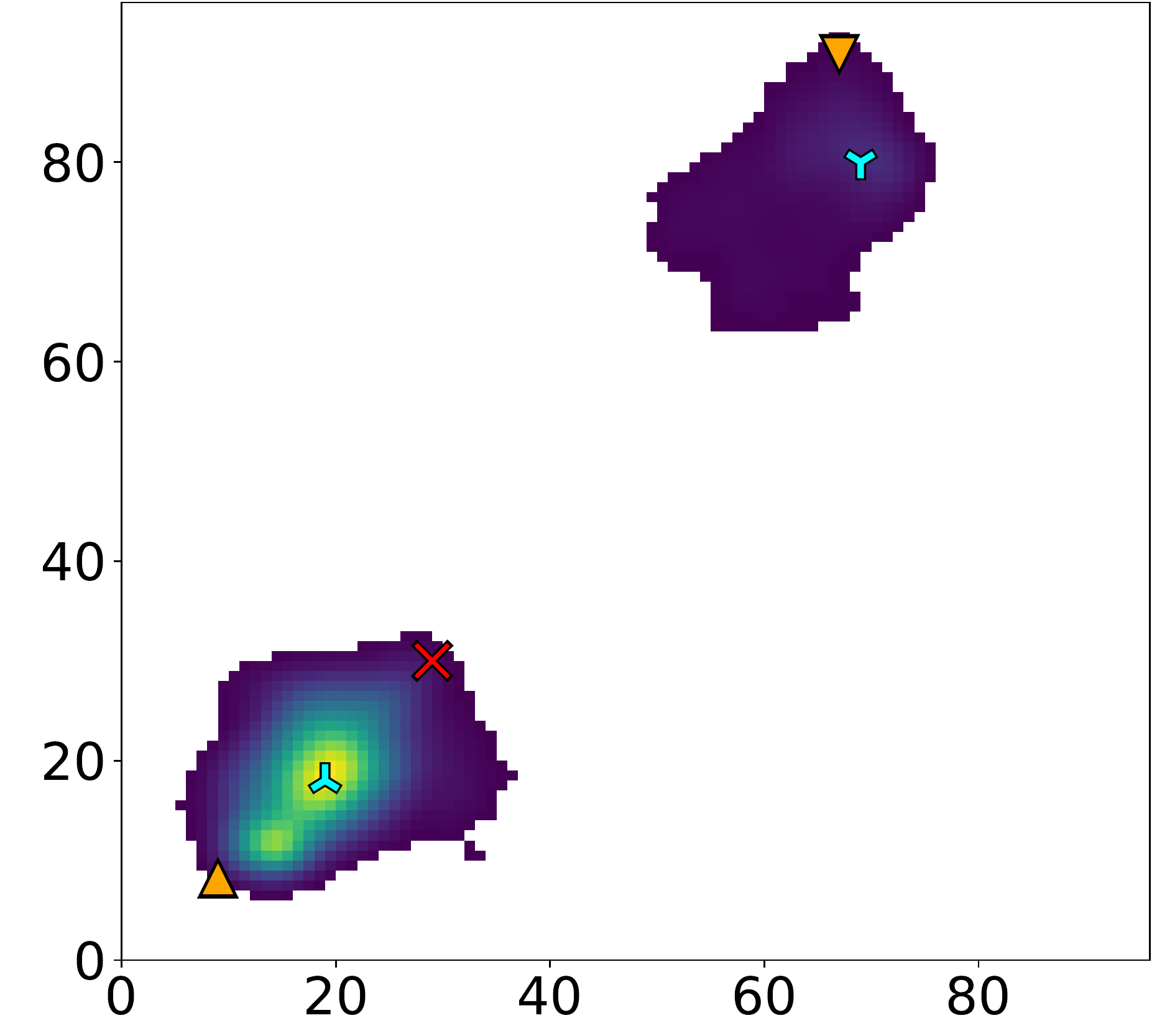}
	\includegraphics[width=0.2\textwidth, trim=1.4cm 1cm 6.8cm 2.4cm, clip=true]{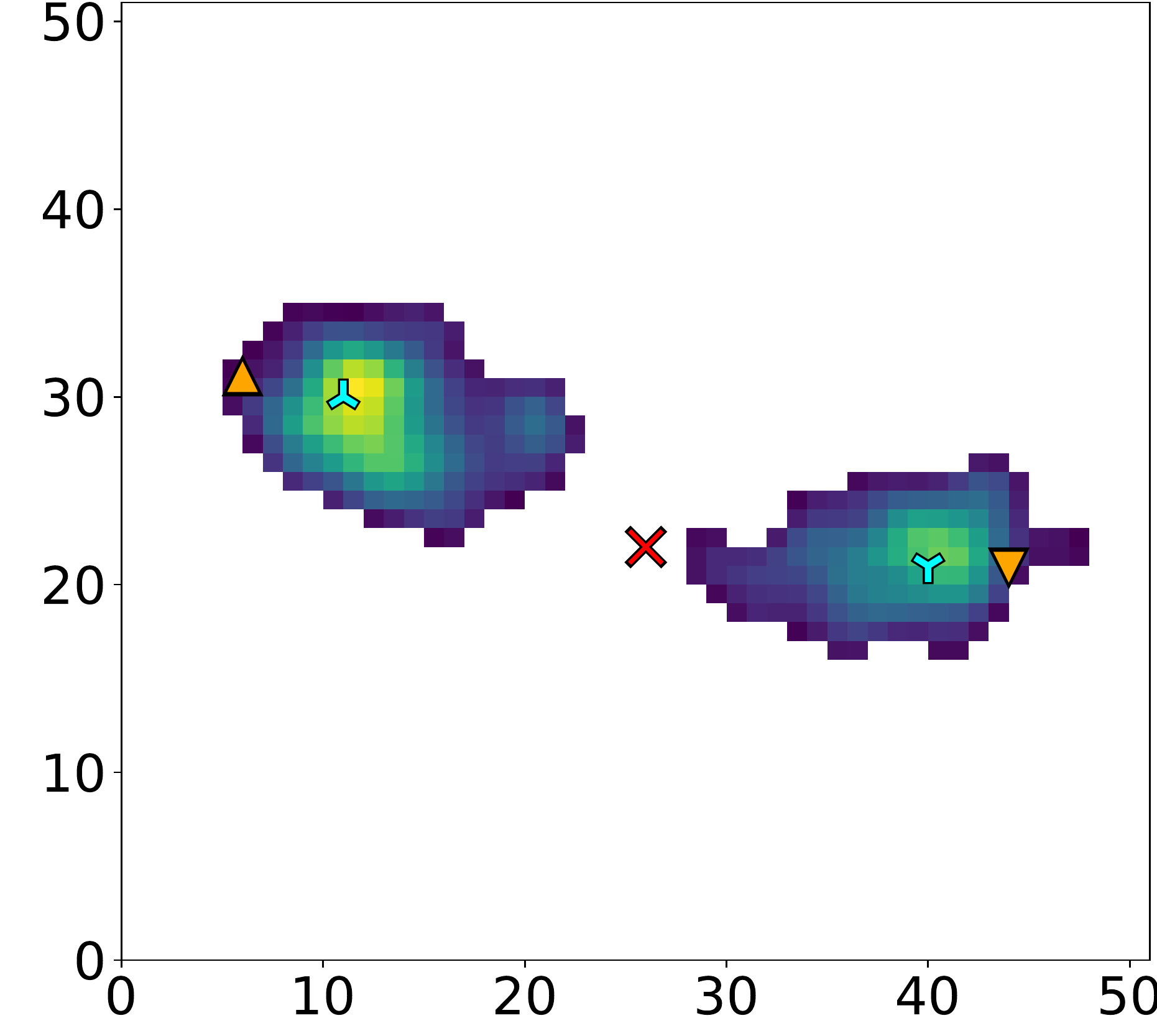} 
	\includegraphics[width=0.2\textwidth, trim=1.4cm 1cm 6.8cm 2.4cm, clip=true]{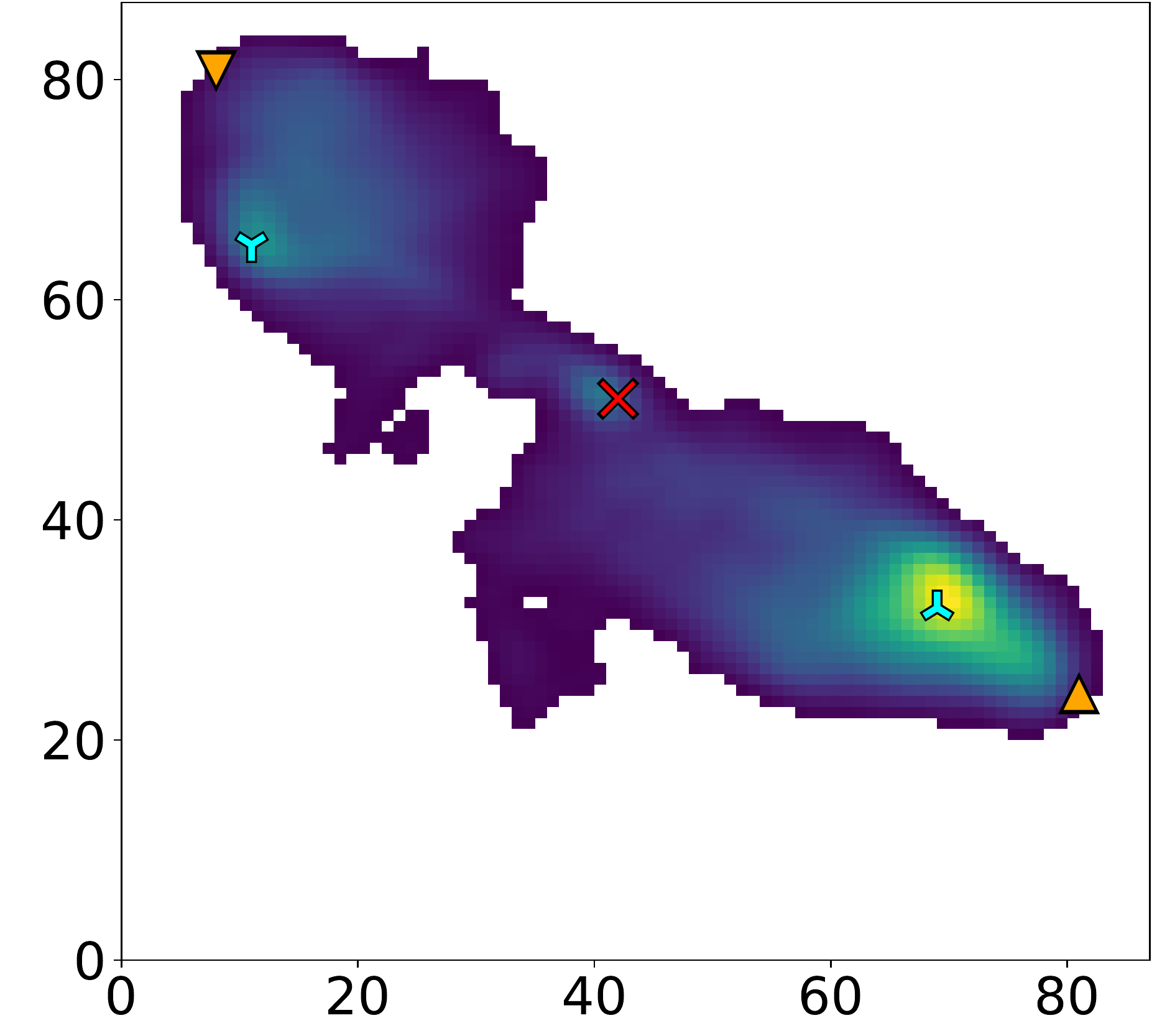} 
	\includegraphics[width=0.2\textwidth, trim=1.4cm 1cm 6.8cm 2.4cm, clip=true]{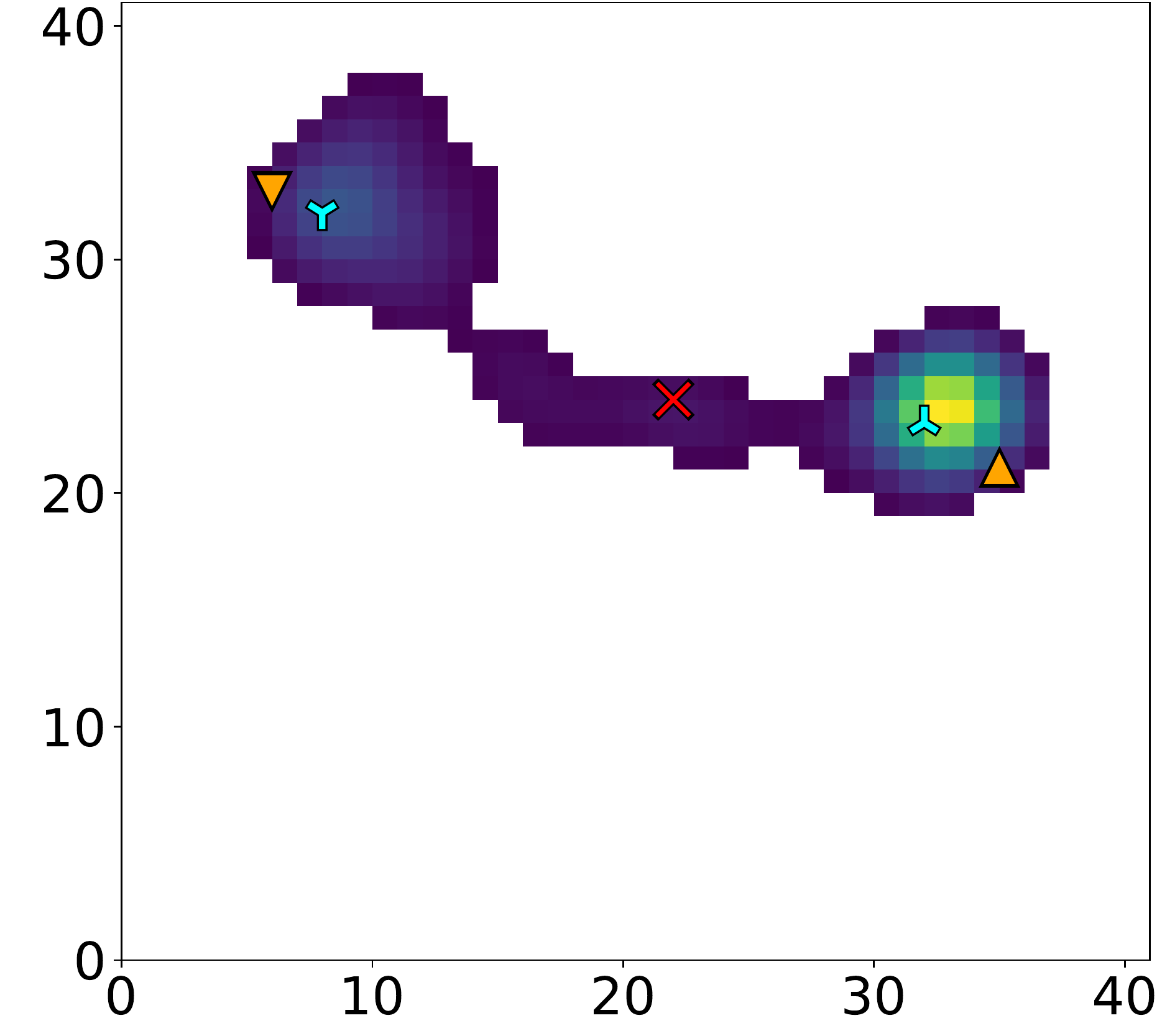} 
	\includegraphics[width=0.2\textwidth, trim=1.4cm 1cm 6.8cm 2.4cm, clip=true]{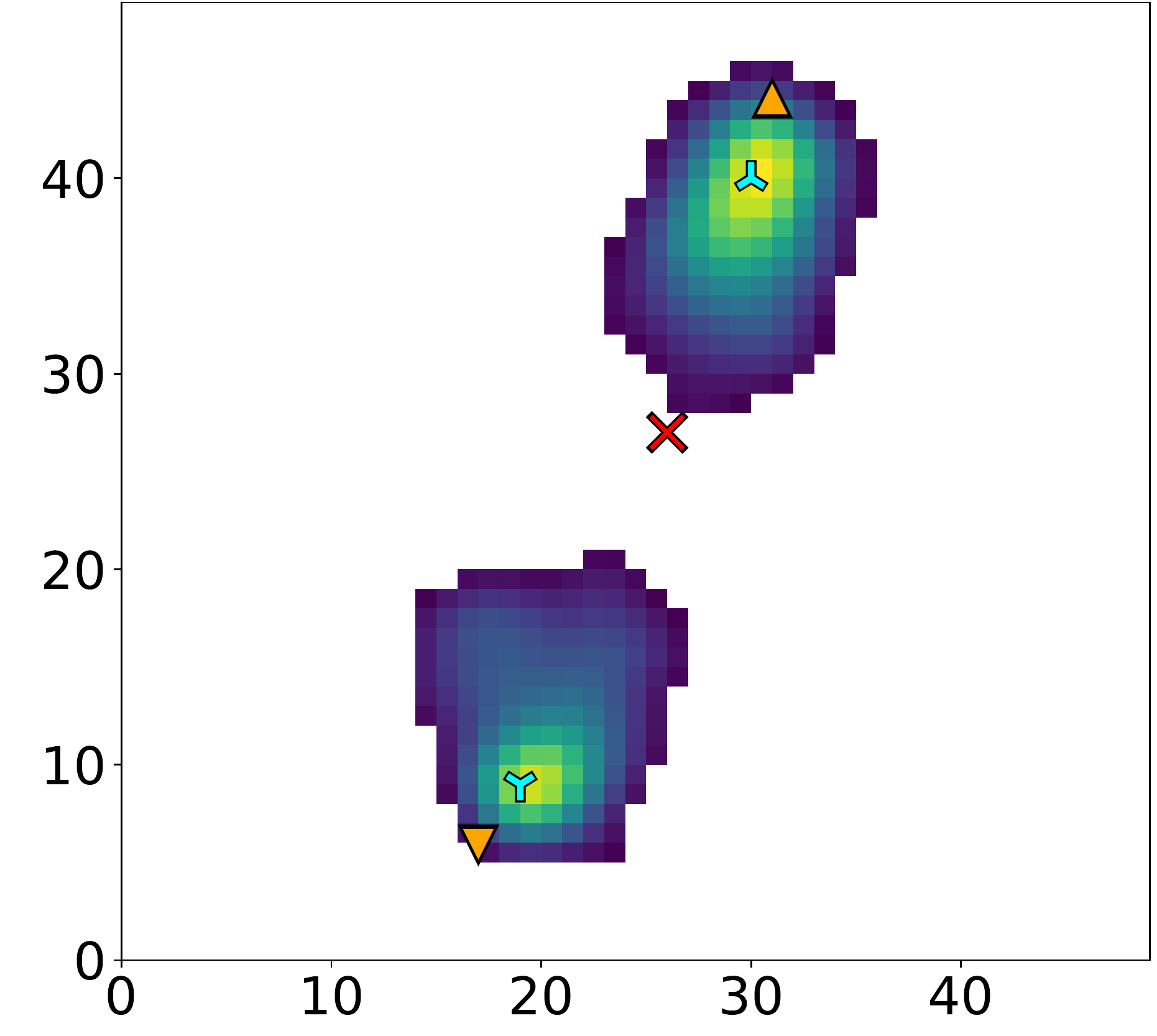}
	\includegraphics[width=0.2\textwidth, trim=1.4cm 1cm 6.8cm 2.4cm, clip=true]{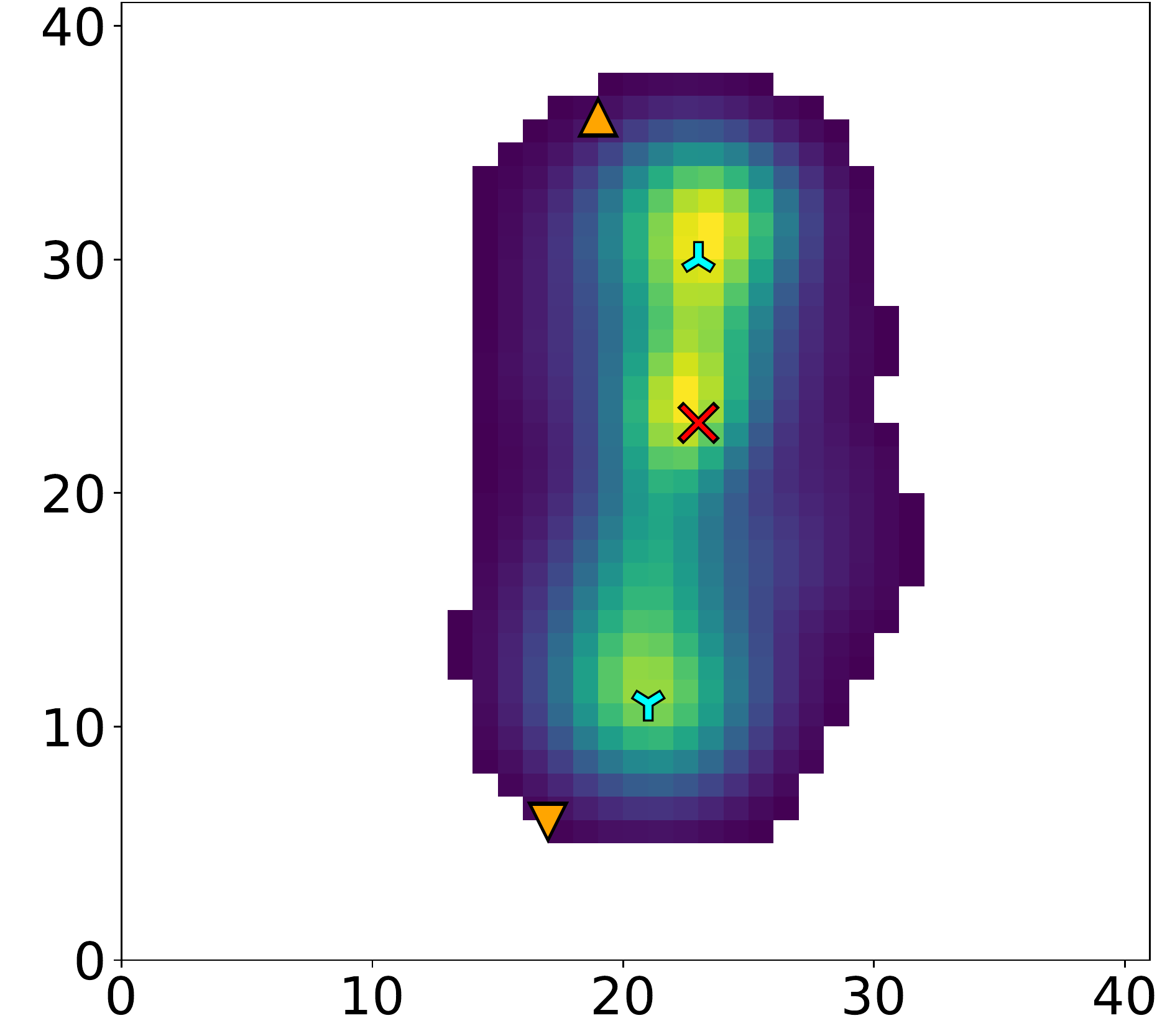} 
	\includegraphics[width=0.2\textwidth, trim=1.4cm 1cm 6.8cm 2.4cm, clip=true]{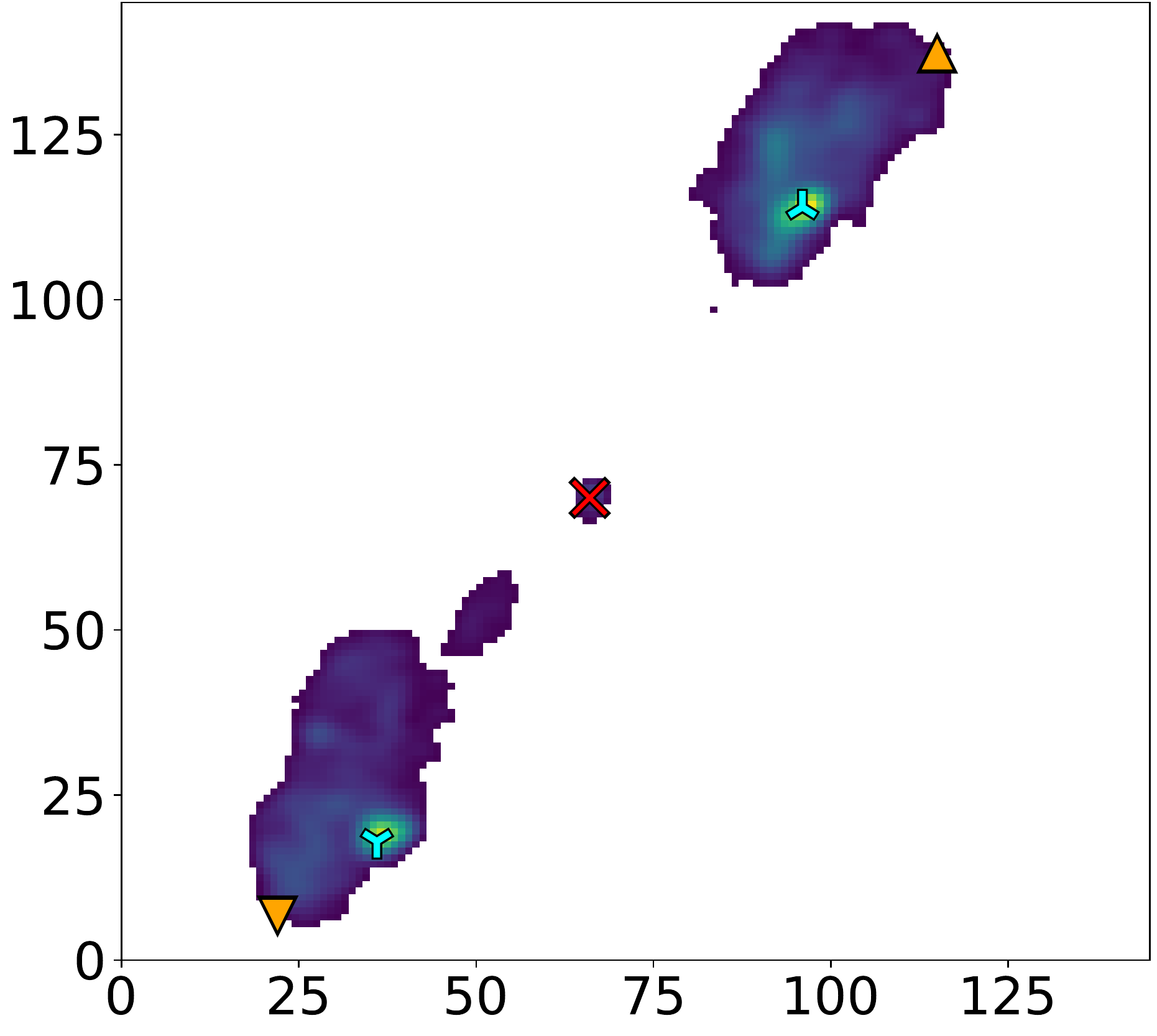} 
	\includegraphics[width=0.2\textwidth, trim=1.4cm 1cm 6.8cm 2.4cm, clip=true]{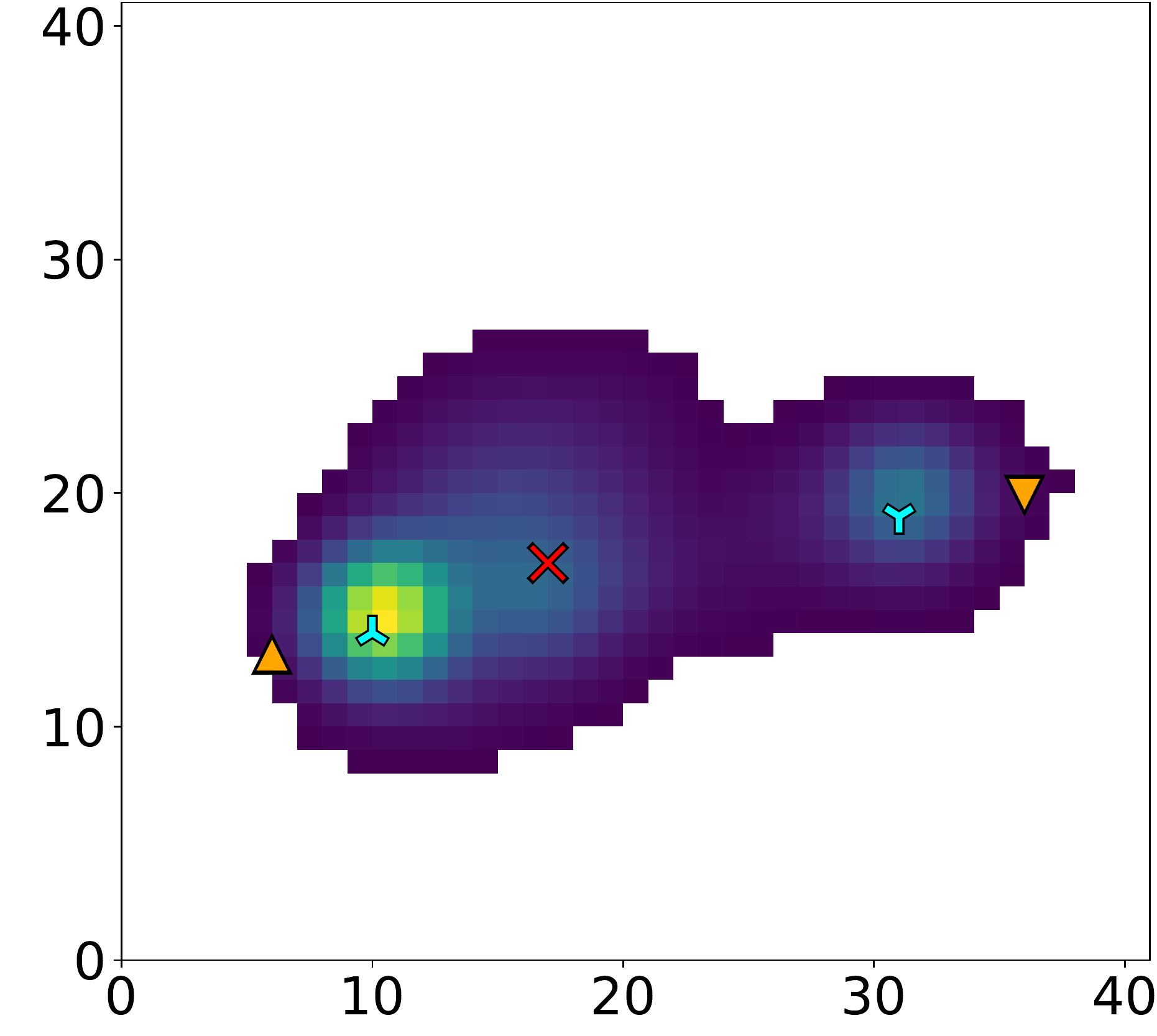} 
	\includegraphics[width=0.2\textwidth, trim=1.4cm 1cm 6.8cm 2.4cm, clip=true]{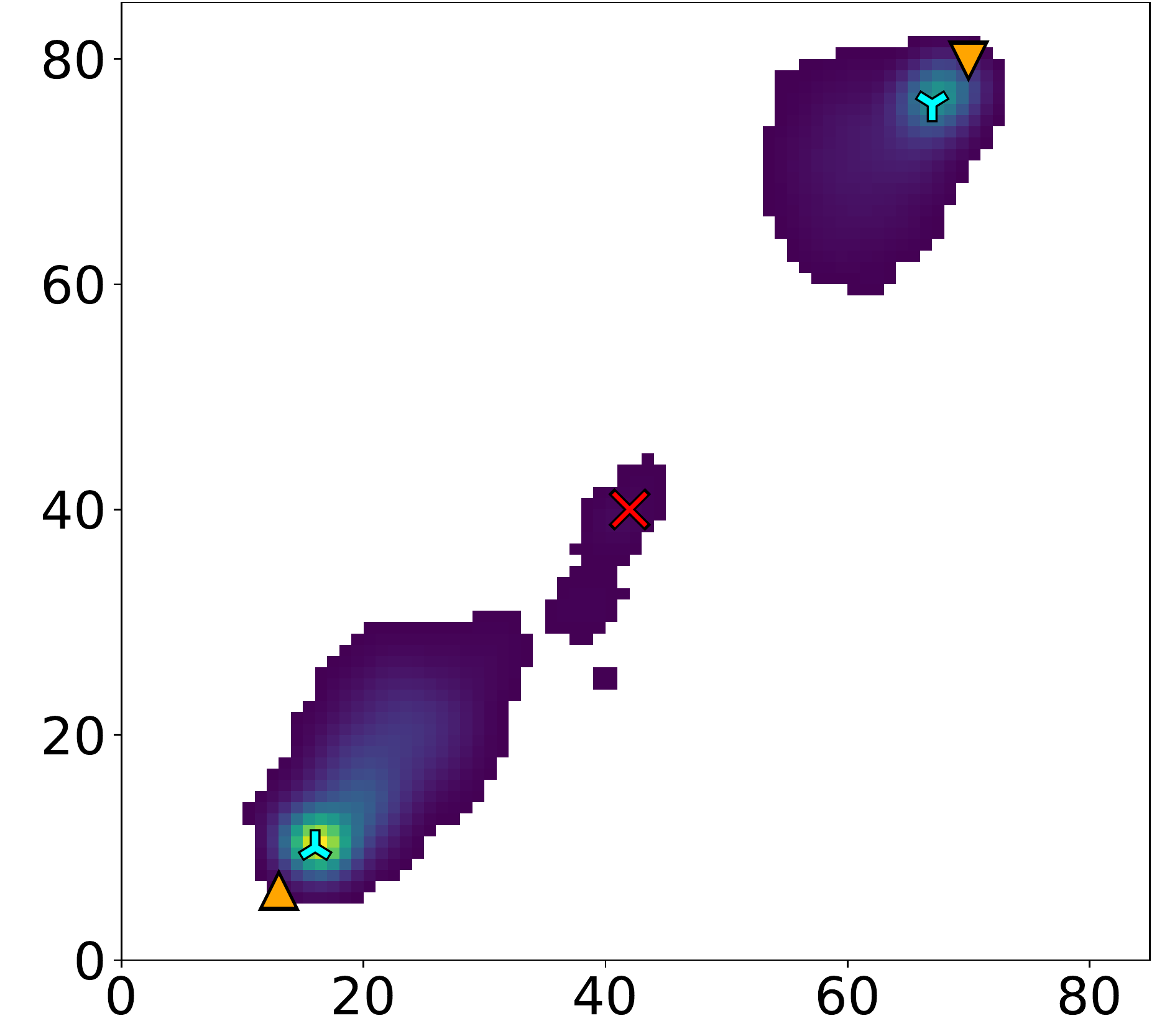}
	\includegraphics[width=0.2\textwidth, trim=1.4cm 1cm 6.8cm 2.4cm, clip=true]{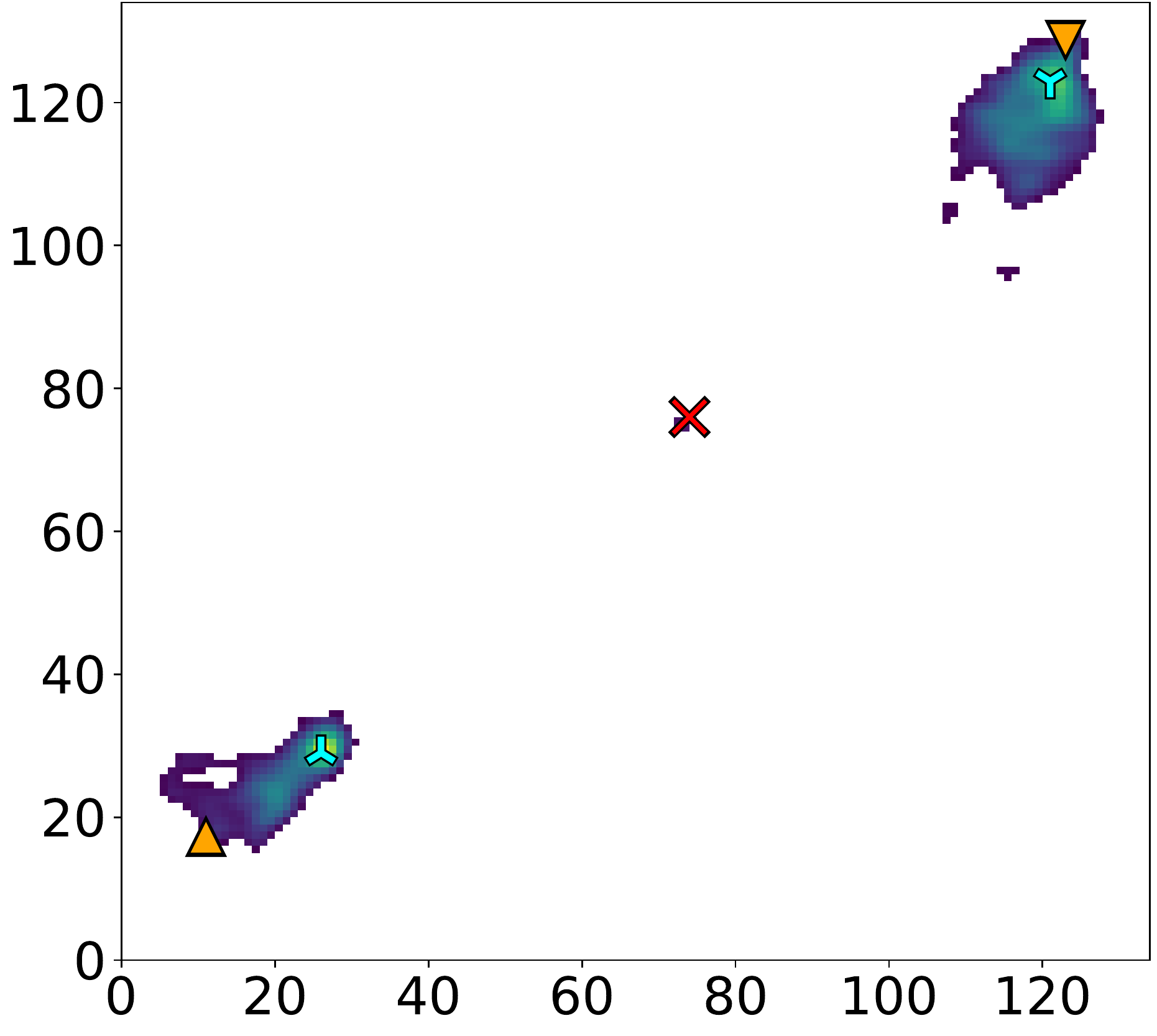} 
	\includegraphics[width=0.2\textwidth, trim=1.4cm 1cm 6.8cm 2.4cm, clip=true]{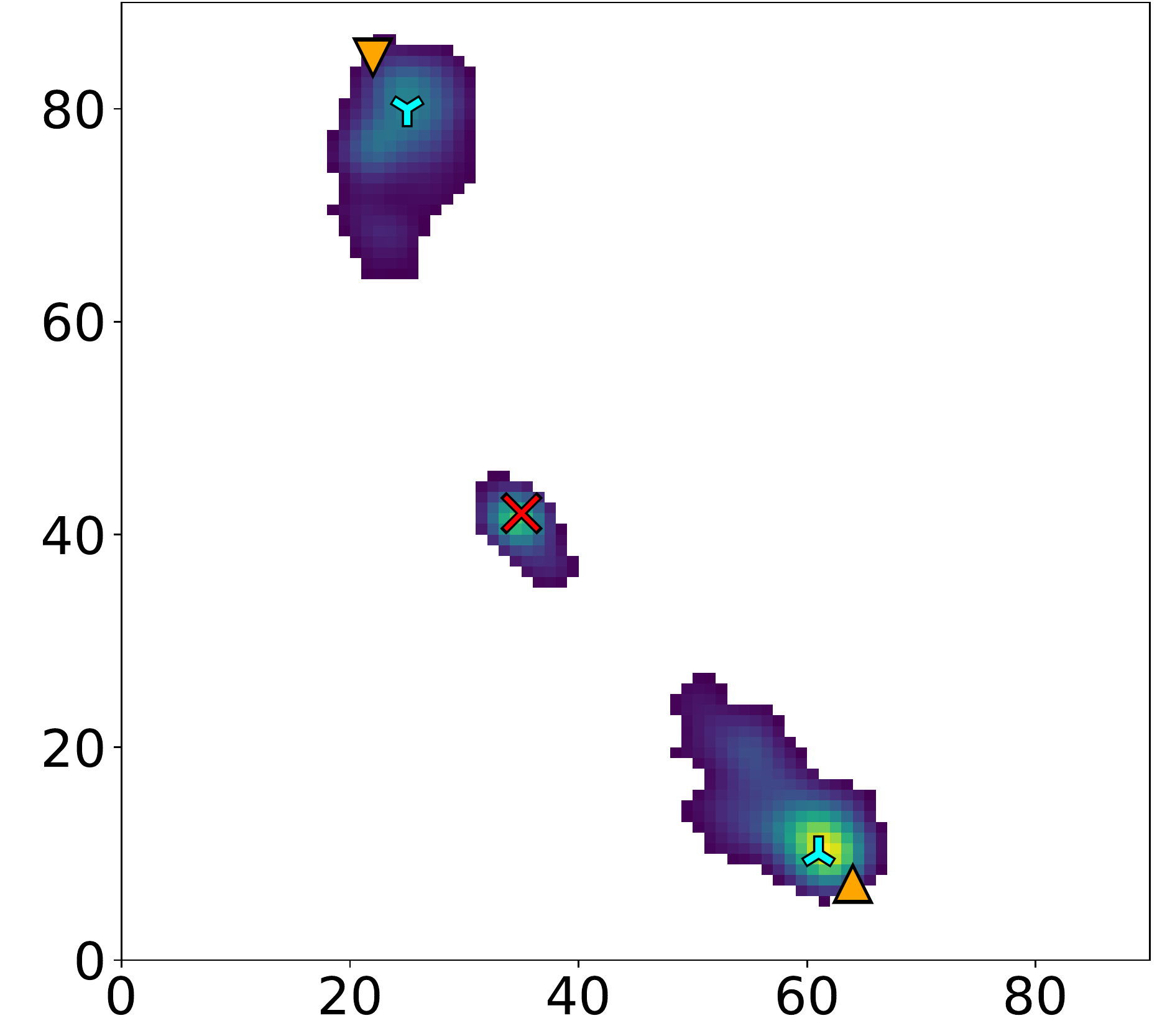} 
	\includegraphics[width=0.2\textwidth, trim=1.4cm 1cm 6.8cm 2.4cm, clip=true]{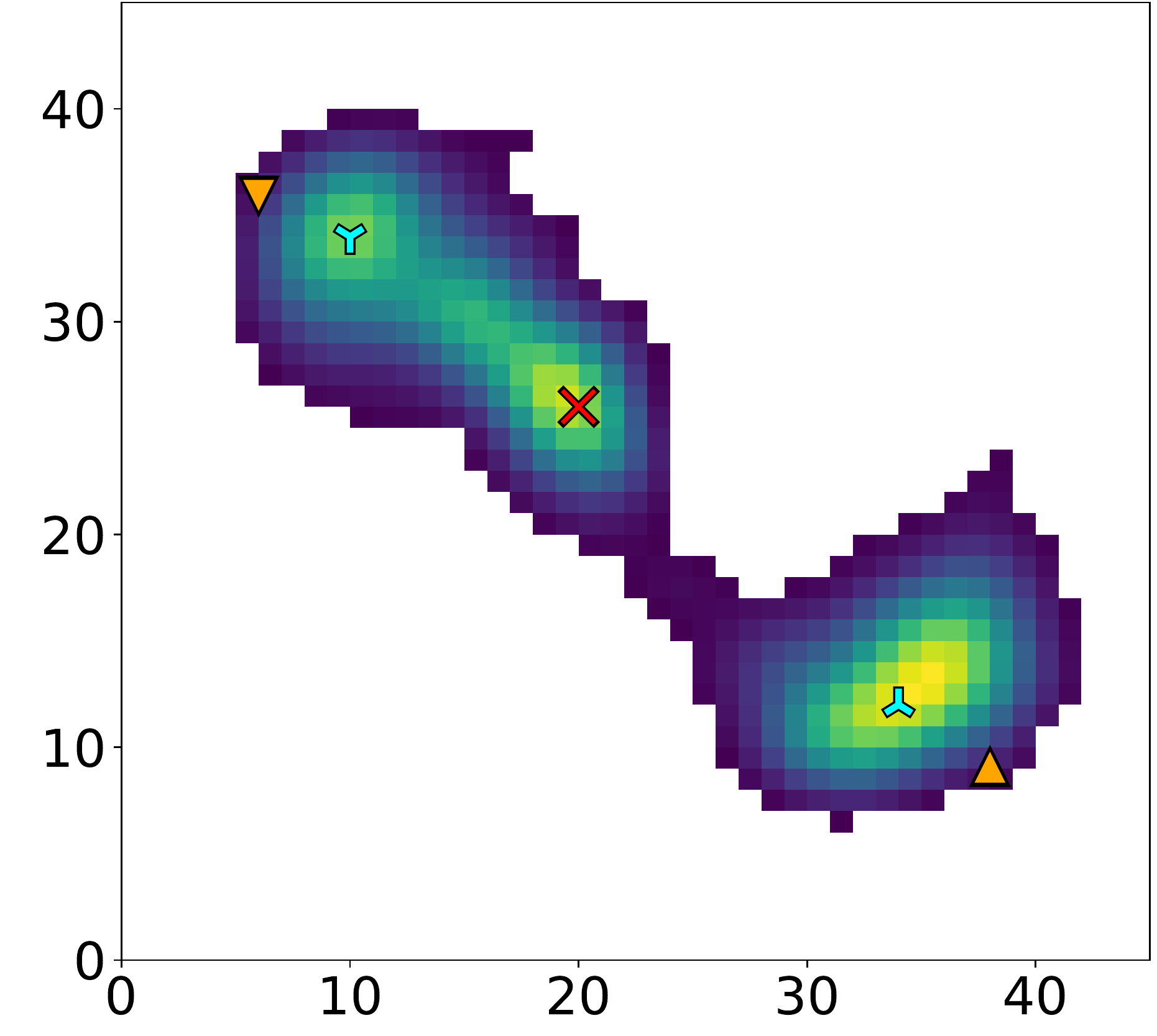} 
	\includegraphics[width=0.2\textwidth, trim=1.4cm 1cm 6.8cm 2.4cm, clip=true]{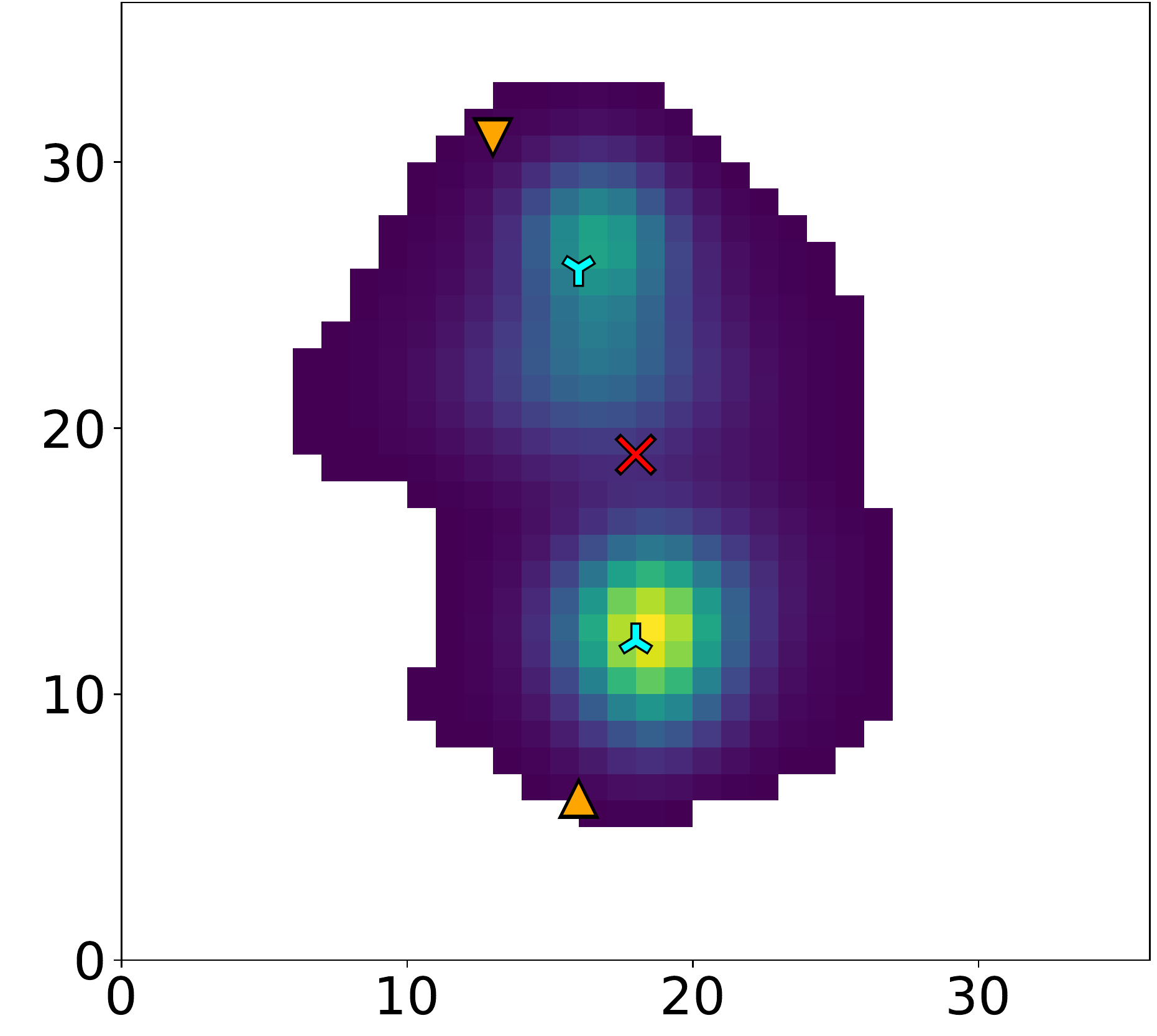}
	\includegraphics[width=0.2\textwidth, trim=1.4cm 1cm 6.8cm 2.4cm, clip=true]{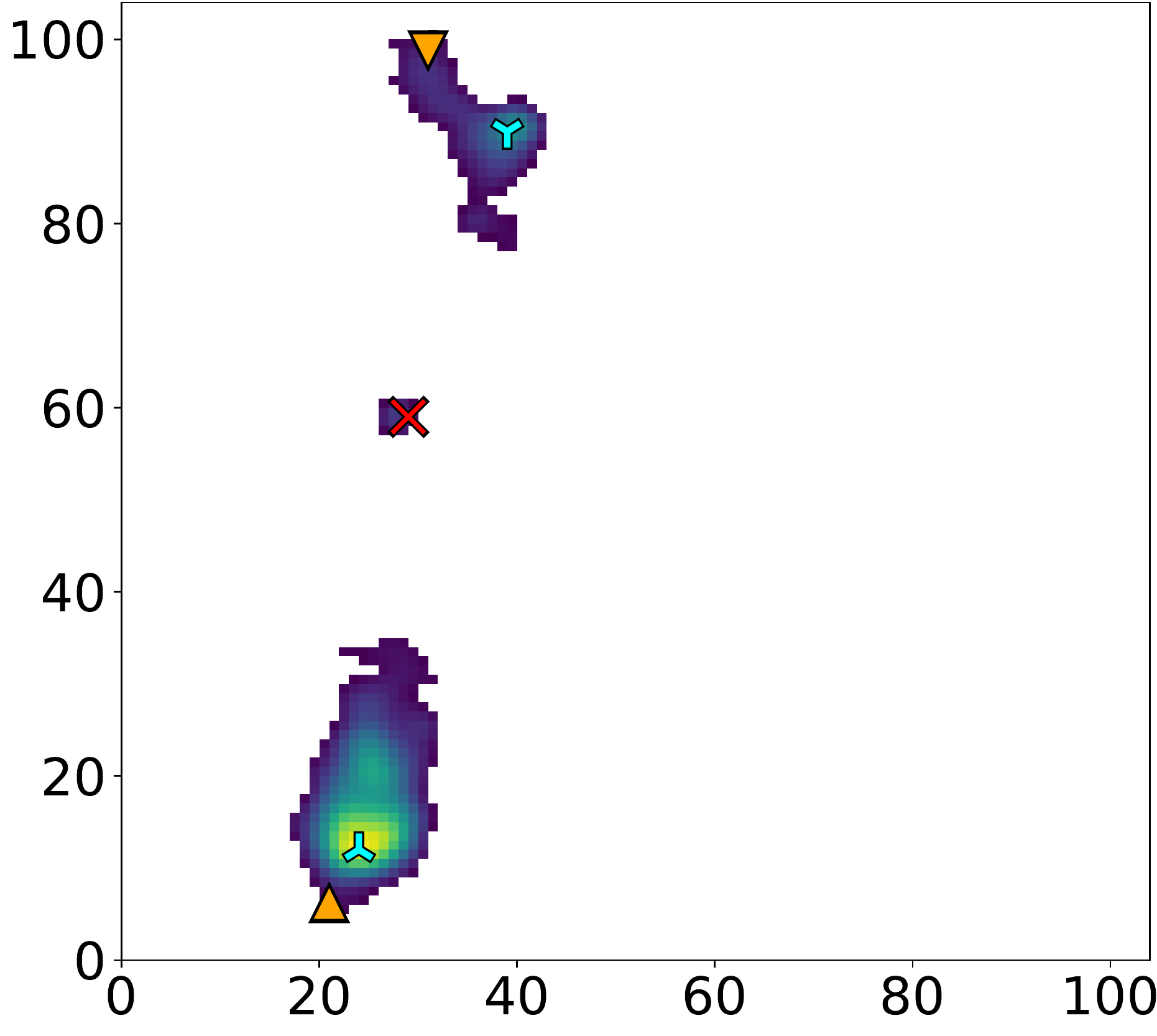} 
	\includegraphics[width=0.2\textwidth, trim=1.4cm 1cm 6.8cm 2.4cm, clip=true]{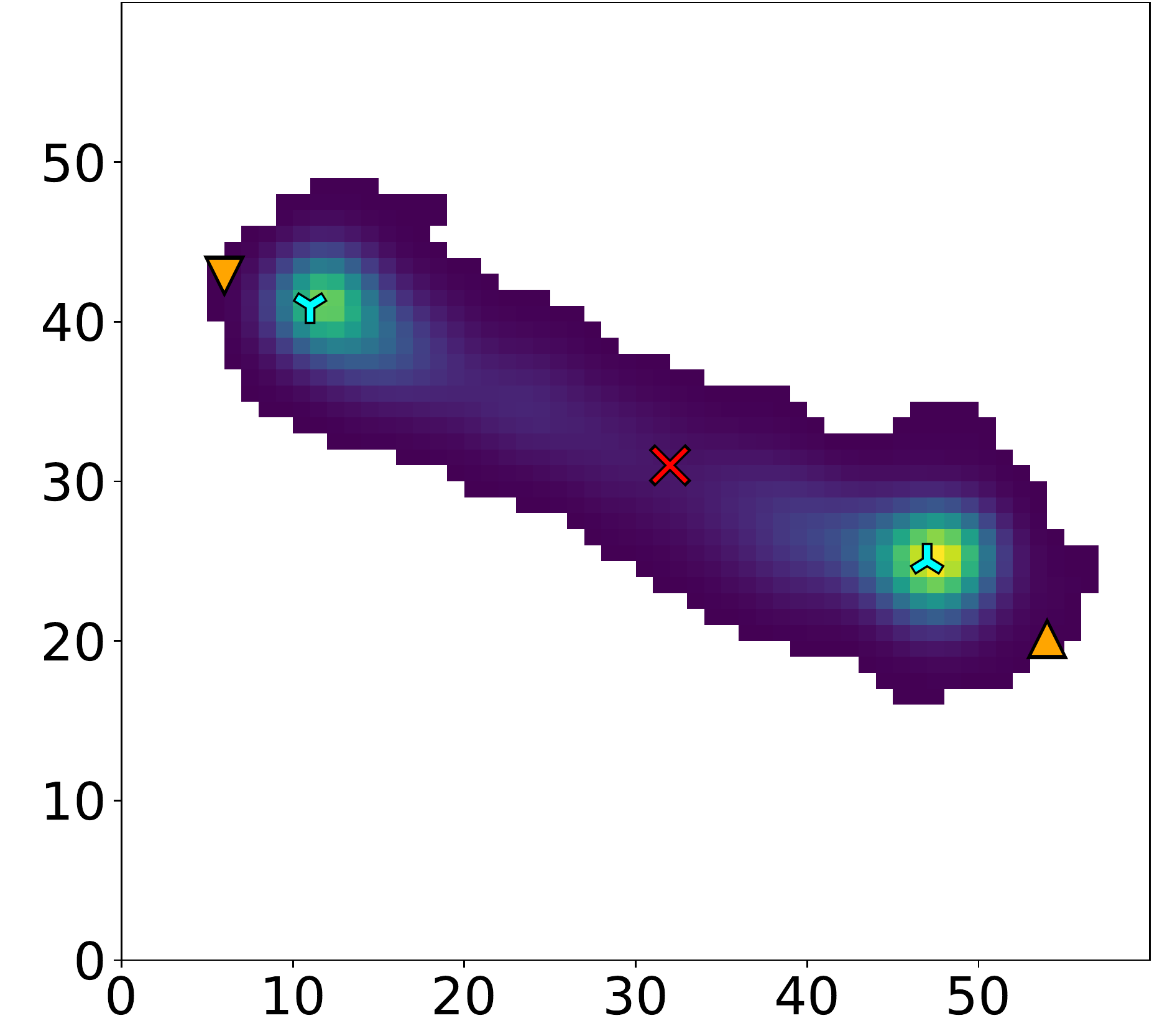} 
	\includegraphics[width=0.2\textwidth, trim=1.4cm 1cm 6.8cm 2.4cm, clip=true]{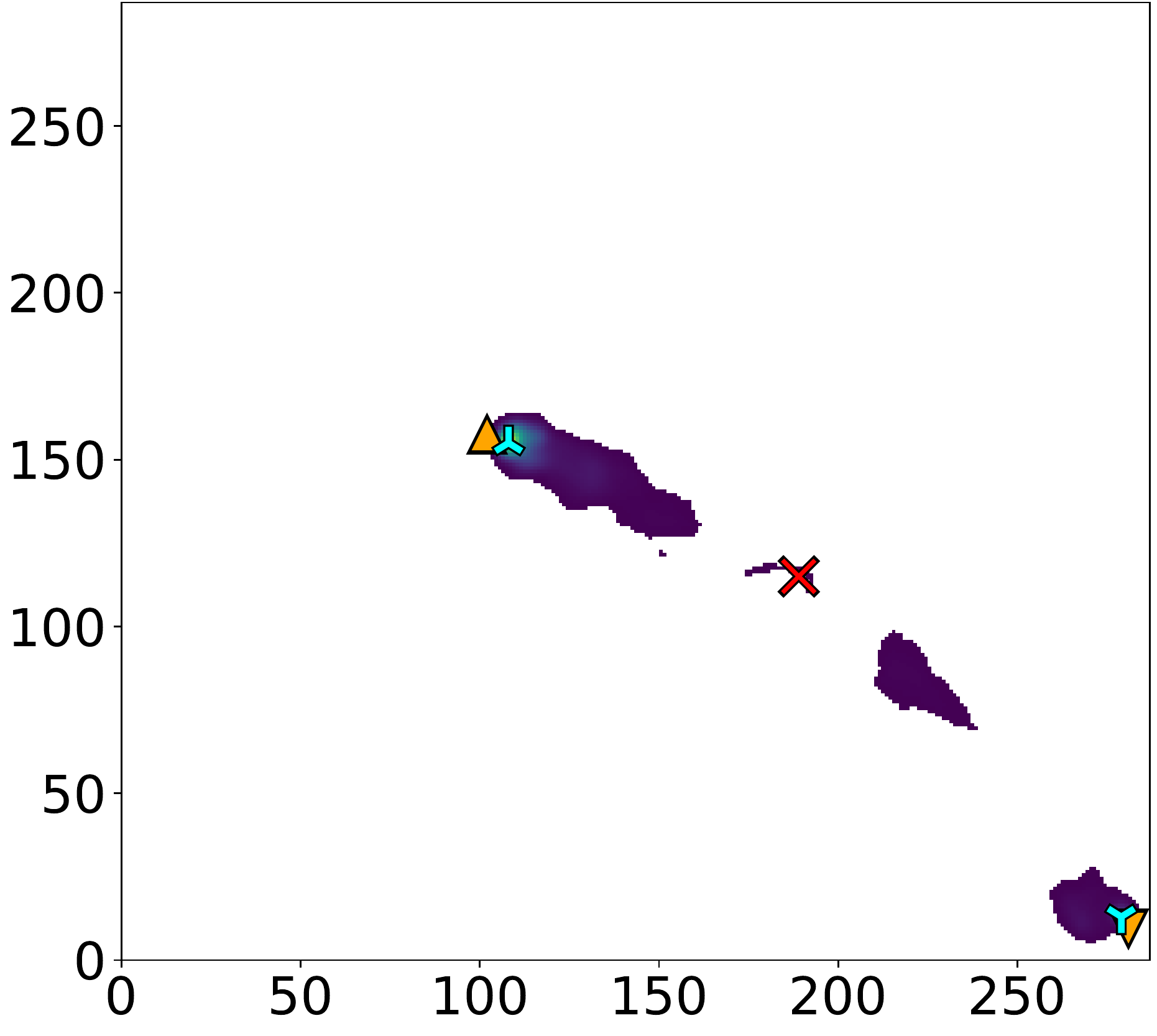} 
	\includegraphics[width=0.2\textwidth, trim=1.4cm 1cm 6.8cm 2.4cm, clip=true]{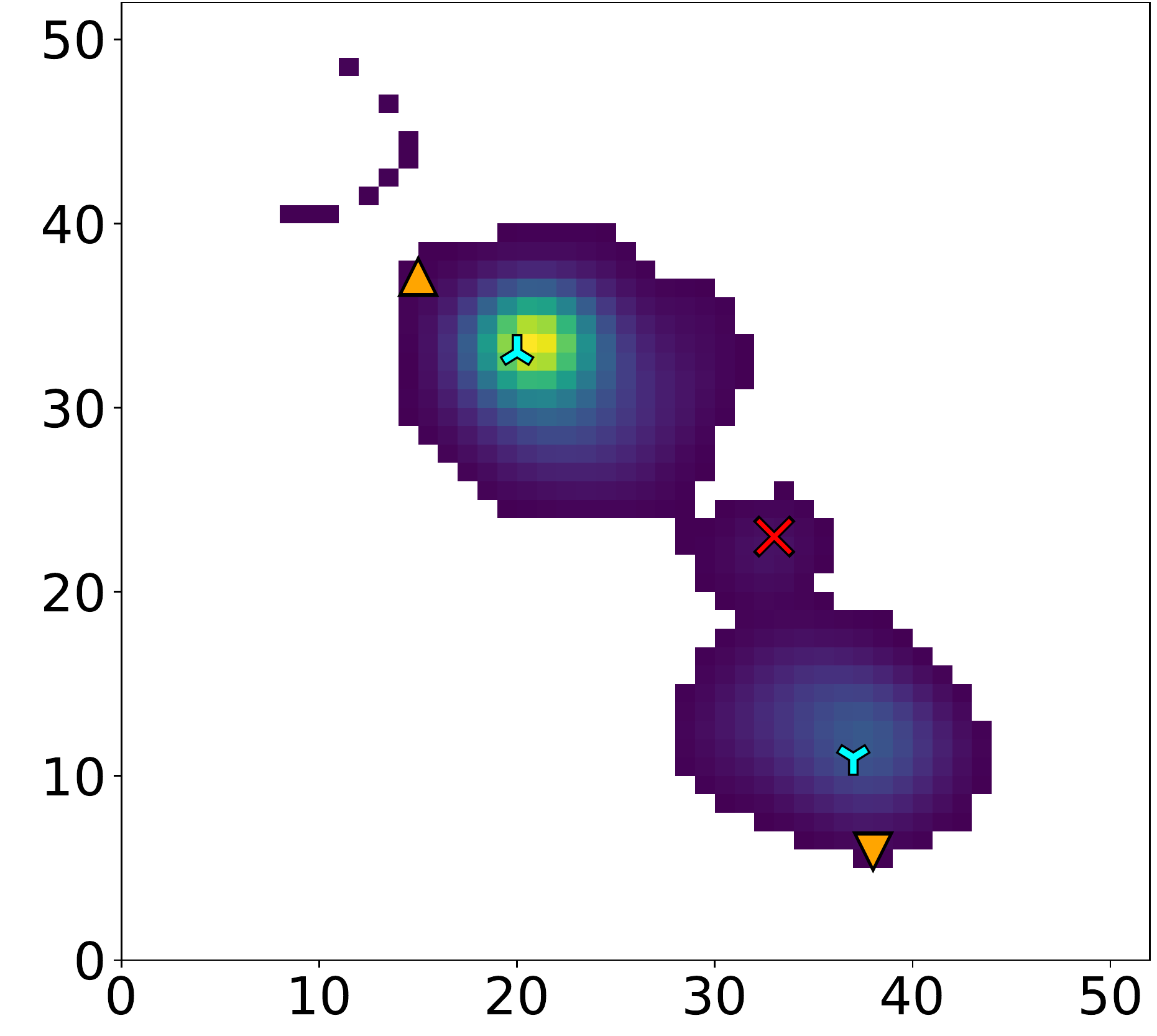}
	\includegraphics[width=0.2\textwidth, trim=1.4cm 1cm 6.8cm 2.4cm, clip=true]{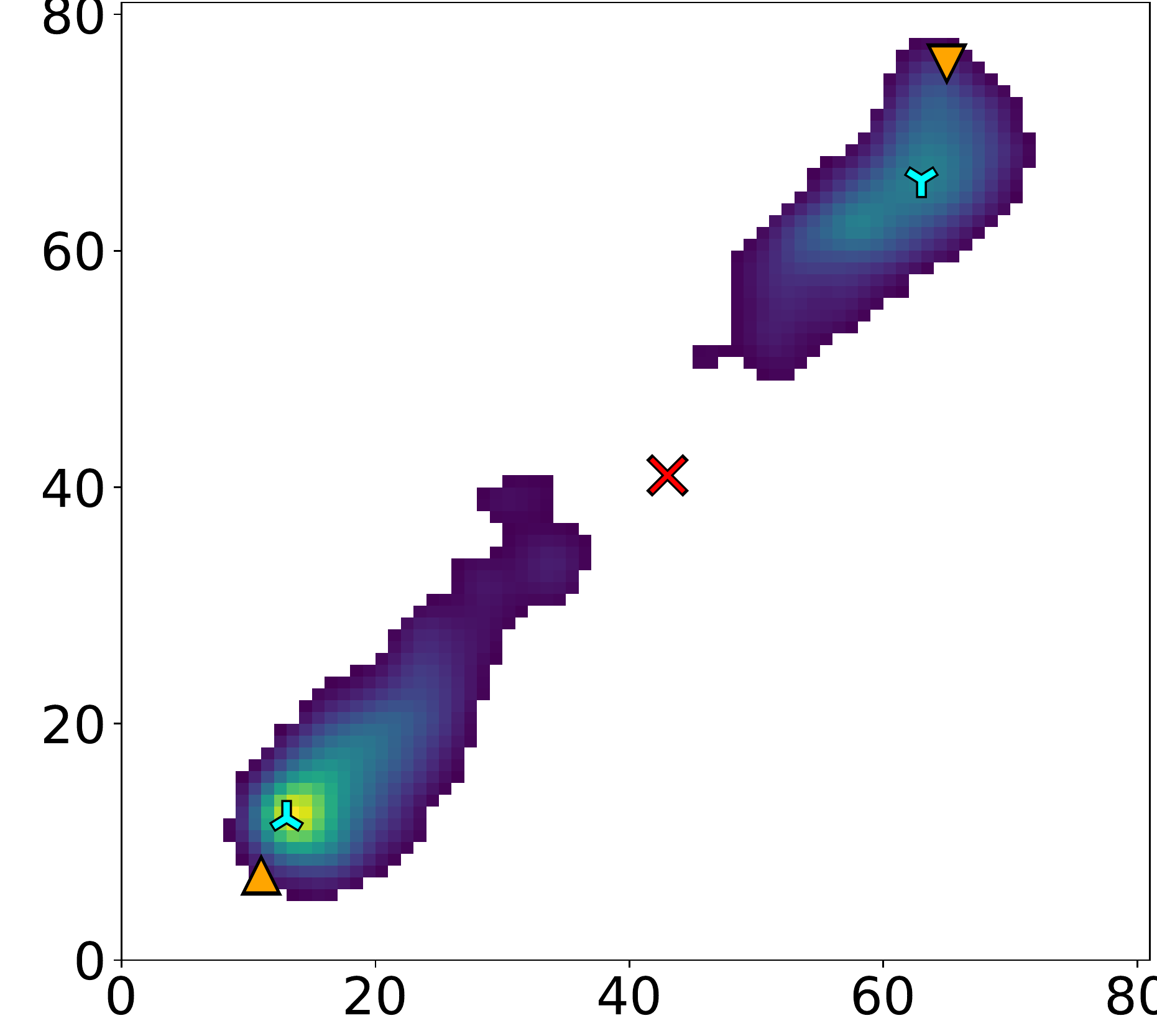}
	\includegraphics[width=0.2\textwidth, trim=1.4cm 1cm 6.8cm 2.4cm, clip=true]{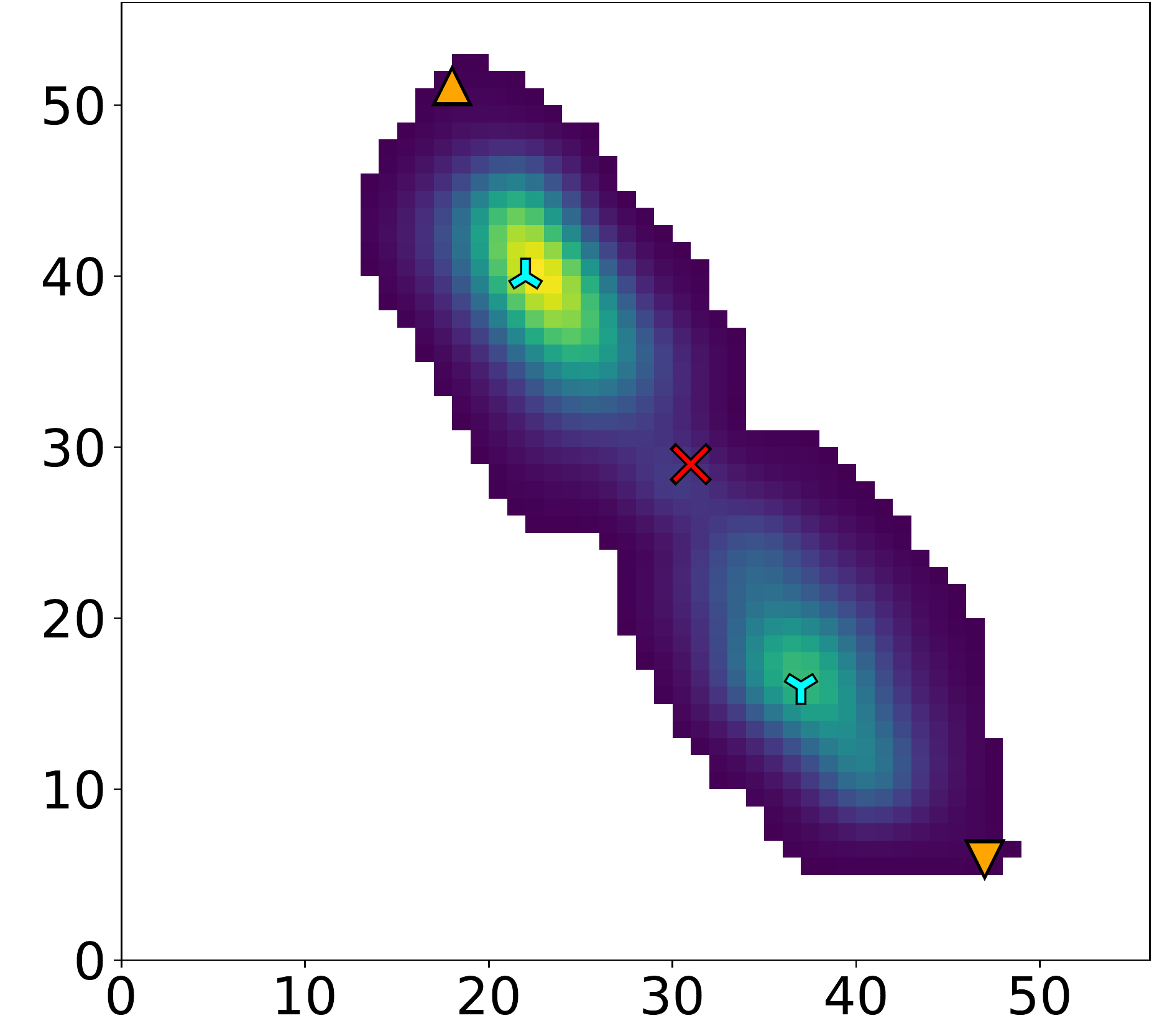} 
	\includegraphics[width=0.2\textwidth, trim=1.4cm 1cm 6.8cm 2.4cm, clip=true]{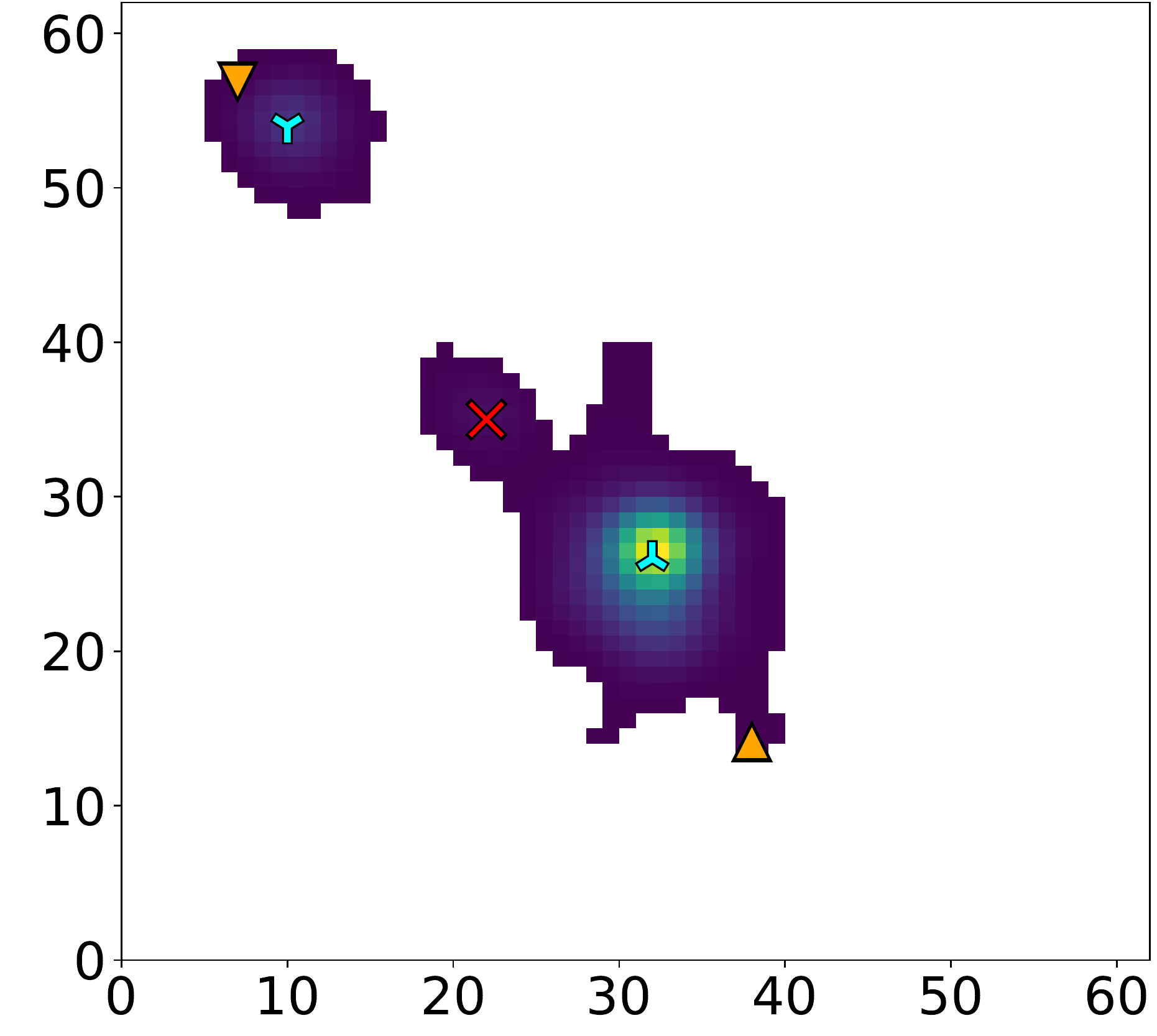} 
	\includegraphics[width=0.2\textwidth, trim=1.4cm 1cm 6.8cm 2.4cm, clip=true]{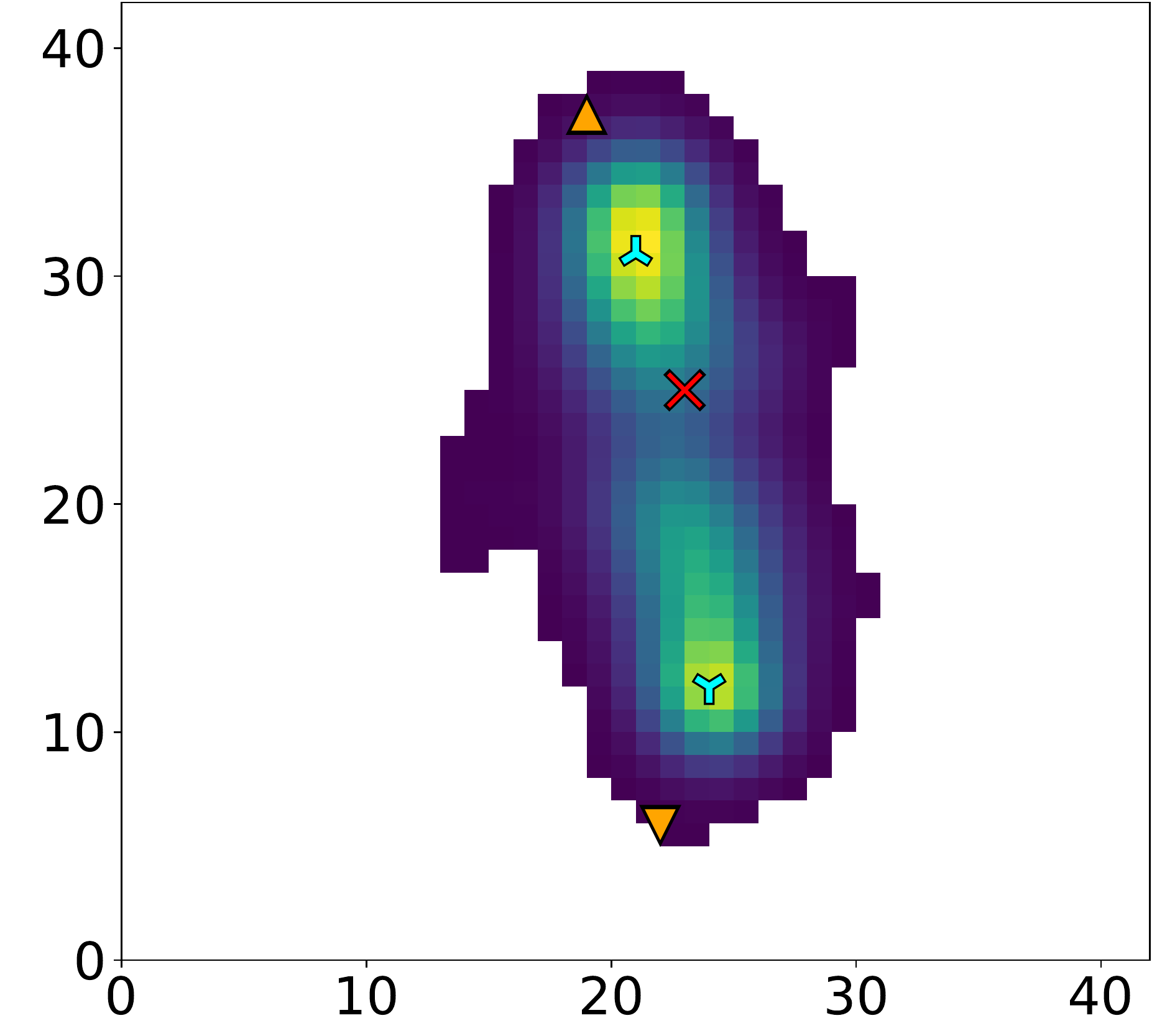}
	\includegraphics[width=0.2\textwidth, trim=1.4cm 1cm 6.8cm 2.4cm, clip=true]{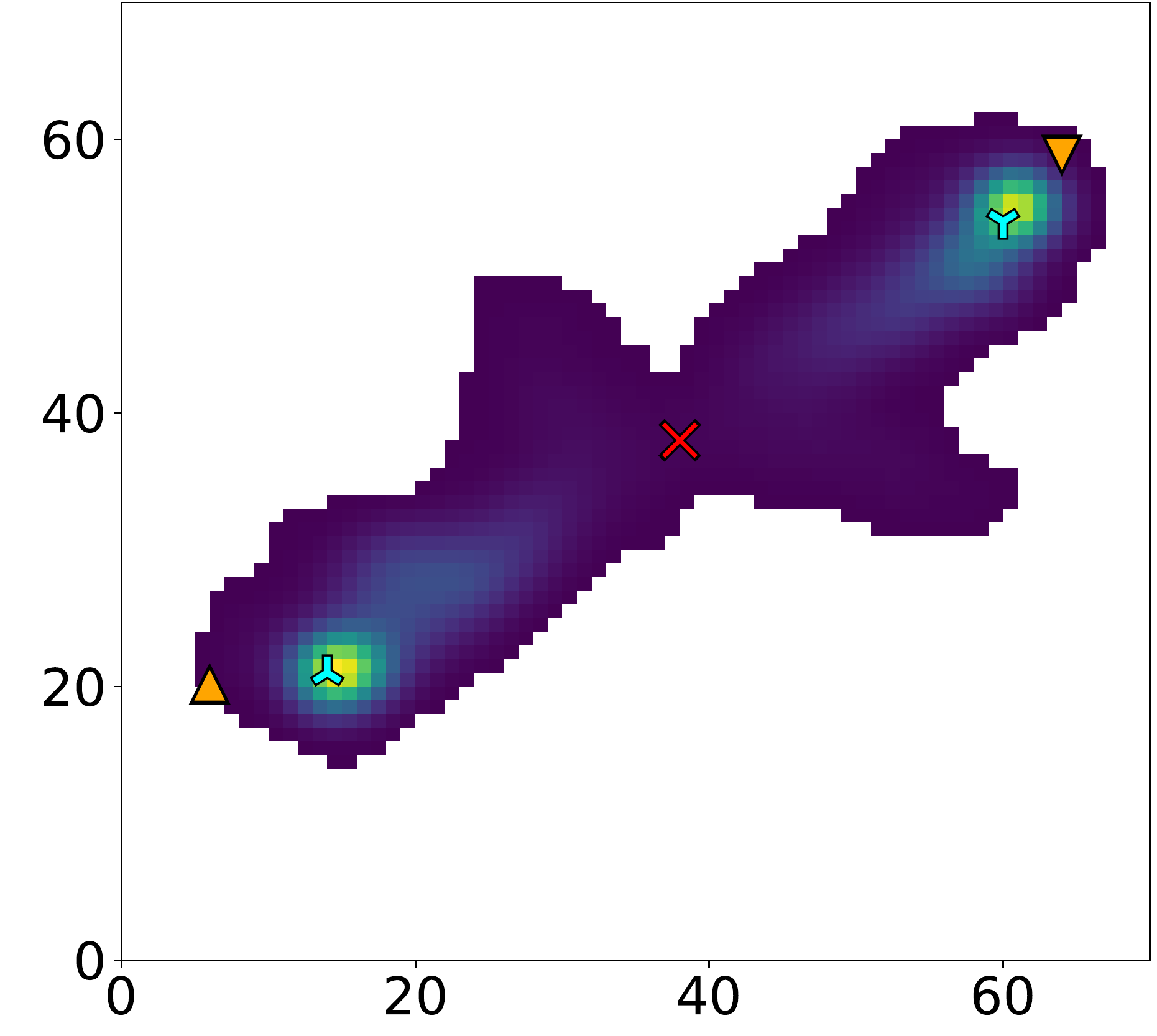} 
	\includegraphics[width=0.2\textwidth, trim=1.4cm 1cm 6.8cm 2.4cm, clip=true]{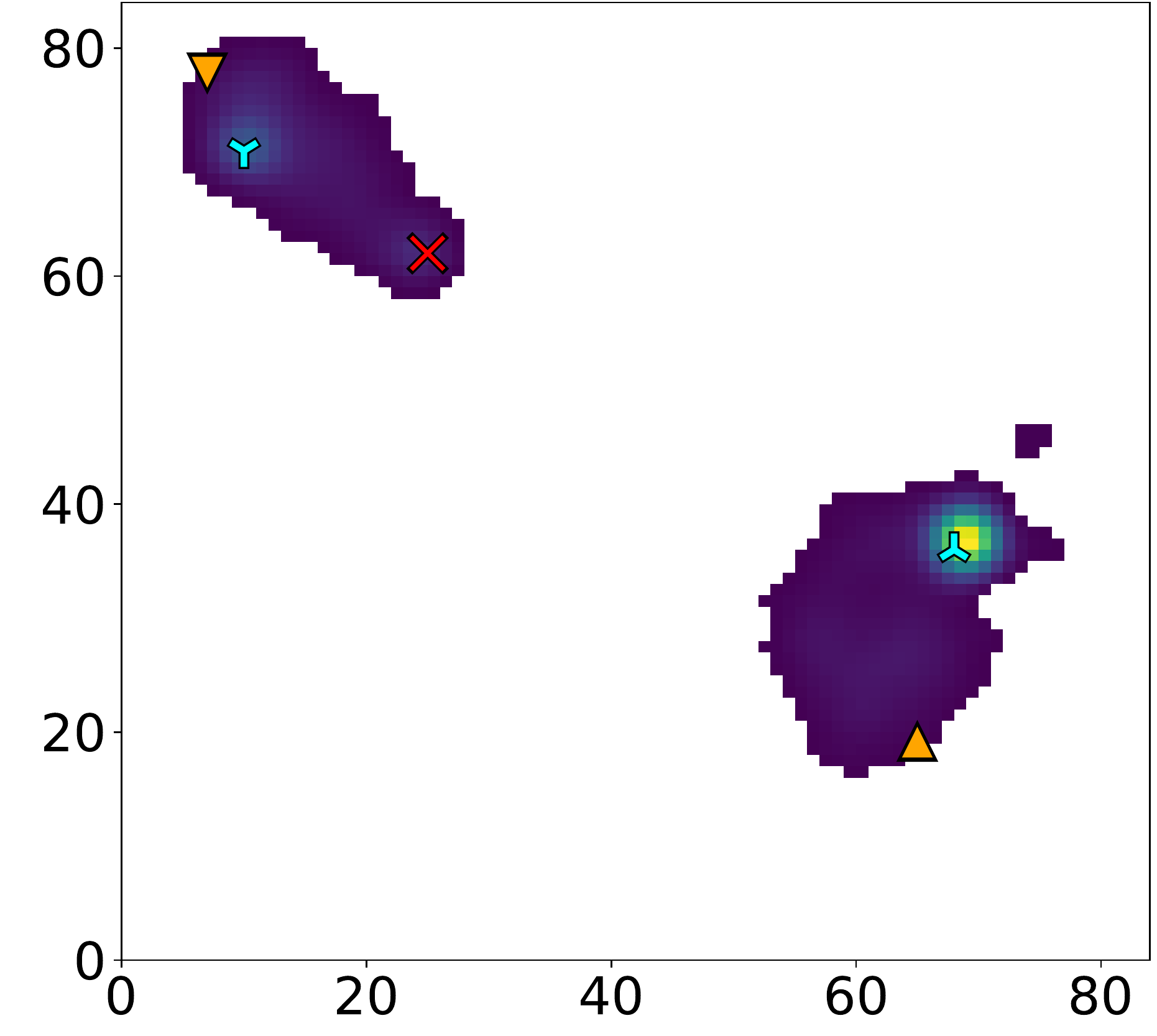} 
	\includegraphics[width=0.2\textwidth, trim=1.4cm 1cm 6.8cm 2.4cm, clip=true]{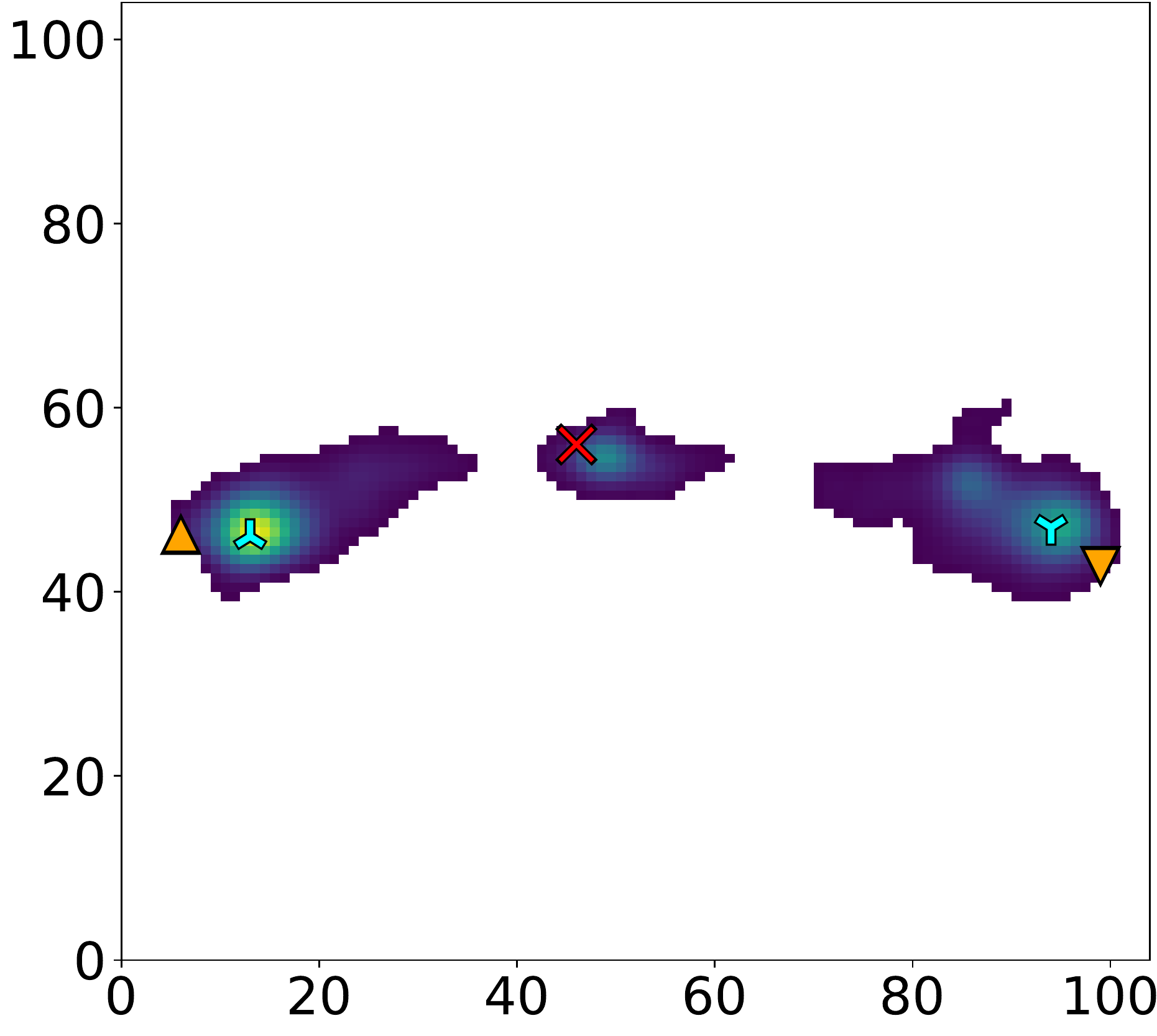} 
	\includegraphics[width=0.2\textwidth, trim=1.4cm 1cm 6.8cm 2.4cm, clip=true]{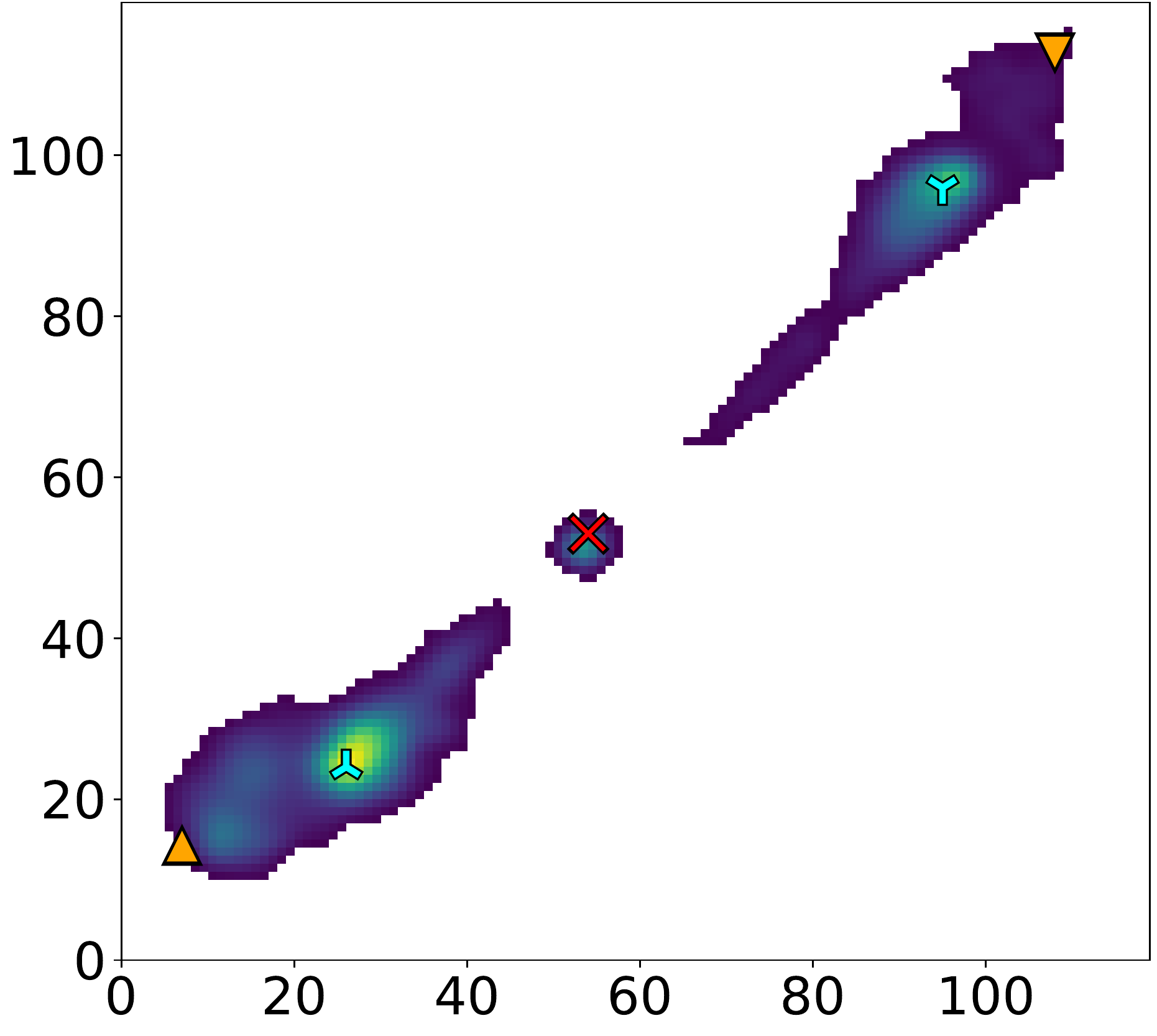}
	\includegraphics[width=0.2\textwidth, trim=1.4cm 1cm 6.8cm 2.4cm, clip=true]{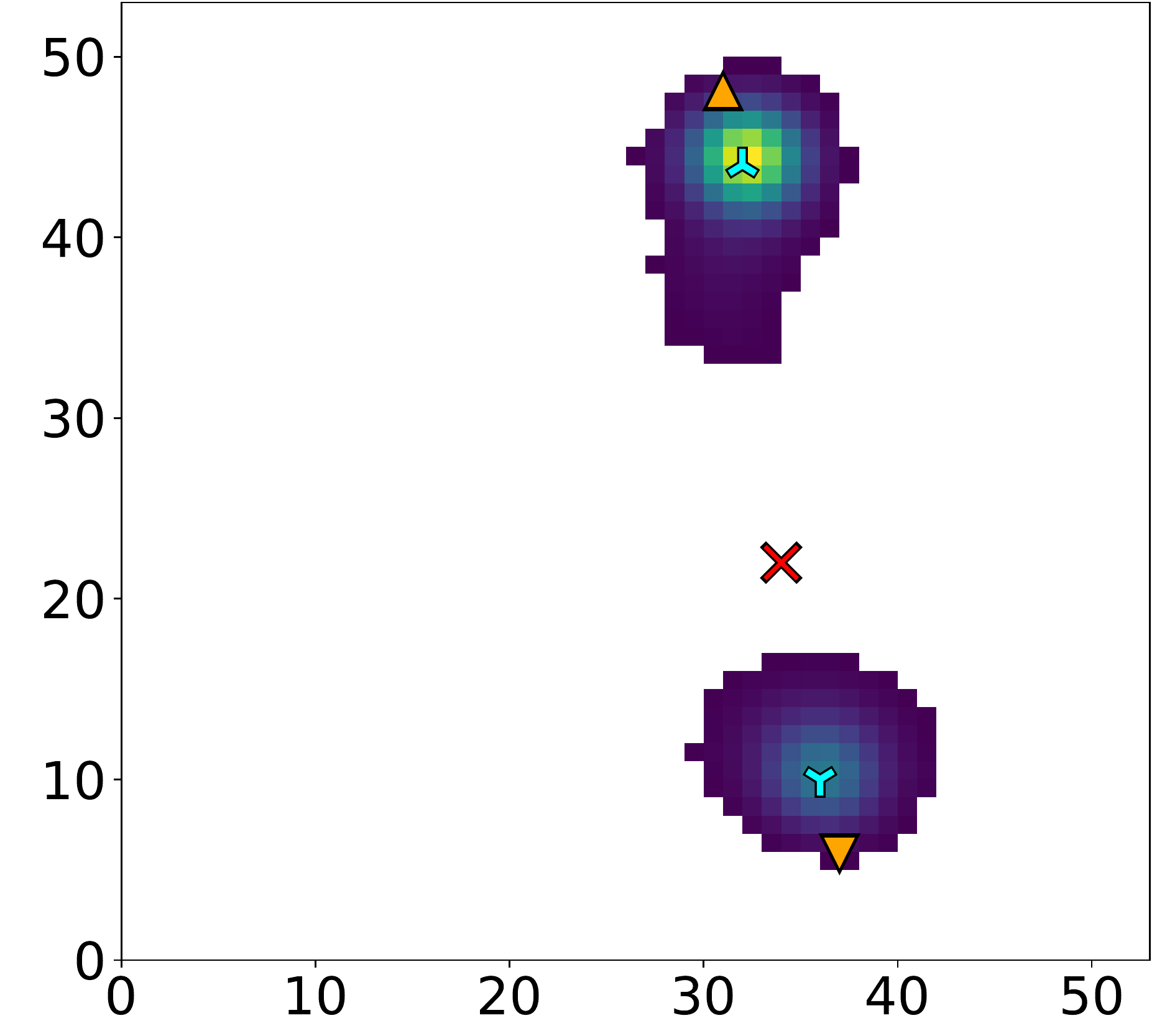}
	\caption{Gallery of randomly-selected (within each subset) FRII sources. Rows 1-2 show FRII-Low LERGs, row 3 FRII-Low HERGs, rows 4-5 FRII-High LERGs, and rows 6-7 FRII-High HERGs. Symbols and colours as in Fig. \ref{fig:FRI_LERG_gallery}. Scale in pixel coordinates, with 1.5 arcsec per pixel.} 
	\label{fig:FRII_gallery}
\end{figure*}

A key question this work was designed to address is whether morphology and accretion mode have any relation for the FRII population in particular, where both accretion modes have long been known. Fig. \ref{fig:FRII_gallery} demonstrates that it is not possible to visually identify whether an FRII drawn at random from our sample is a LERG or a HERG, and whether it is in the `High' or `Low' luminosity category. Morphologically, our FRIIs are quite uniform across the board. 

Conversely, our LoTSS-Deep results do show a difference in the accretion properties of FRII-Low and FRII-High, with the latter having a much higher HERG fraction than the former (see Table \ref{tab:accr_mode}). Table \ref{tab:AC} shows that the HERG/LERG statistics for both populations do not overlap accounting for uncertainties. Fig \ref{fig:L150_size_HERG_LERG} suggests there is no obvious size trend in HERG fraction, but Fig. \ref{fig:comb_LH_histo} does suggest a trend with luminosity, although it is unclear whether this continues to the highest luminosities. There appears to be a gradual transition in accretion properties between the FRII-Low and FRII-High, rather than a sharp dichotomy. It is thus not possible to unequivocally determine whether an FRII is an FRII-High or -Low based on either its accretion properties or its morphology.

In the following subsections we examine the morphology and accretion behaviour of the two FRII subclasses, including examining source morphology quantitatively using the source dynamic range (SDR) -- defined as the peak pixel flux in a source \citepalias[smoothed over a 4-pixel area, as described by][]{Mingo2019} divided by the mean pixel flux (after the 4 RMS cut) -- as a proxy for radio structure. Sources with high SDR show a large flux difference between the brightest and faintest structures, while those with low SDR have very uniform surface brightness.


\subsubsection{Low-luminosity FRII (FRII-Low)}\label{sec:FR2L}

While Fig. \ref{fig:FRII_gallery} shows that it is not possible to visually distinguish FRII-High and FRII-Low, in our sample the FRII-Low do have lower mean SDRs than the FRII-High at a level that is statistically significant ($3.9\pm0.6$ and $6.3\pm0.7$, respectively, with $2\sigma$ errors). This is not surprising as the population on average will have lower total flux, so that the ratio between the peak flux density and the survey RMS noise is likely to be systematically lower. A fraction of the low SDR FRII-Lows seem to have more relaxed (`fluffier') structures, with irregular lobe edges, not something we clearly observed for the FRII-Low in  LoTSS-DR1. This could indicate that with the lower flux density limit we might be picking up more fading sources in LoTSS-Deep. 

A spectral index analysis would be necessary to establish the fraction of fading sources in our sample and will be carried out in a future analysis. We argued in \citetalias{Mingo2019} based on spectral index and population demographics that the majority of FRII-Low are unlikely to be fading sources -- given that sources fade quickly it is implausible that such a large fraction of the total FRII population would be in this stage of evolution. The fact that nearly all FRII-Low are LERGs reinforces the hypothesis that these sources are likely to be more similar to their FRI counterparts in terms of accretion properties than to the powerful, high-accretion-rate, `traditional' FRIIs from the 2Jy and 3CRR surveys, as originally noted by \citet{Hardcastle2008} and recently highlighted by \citet{Grandi2021}. Our results also show that these radiatively-inefficient, low-luminosity FRIIs can be found well beyond the local Universe, with 44 per cent of our FRII-Low at $0.5<z\leq1$, and 18 per cent at $z>1$. This, as well as the presence of FRII-Low across a wide range of source sizes, seems to contradict the idea proposed by \citet{Grandi2021} that the FRII-Low are a late-stage evolution of luminous, high-accretion FRIIs and a phenomenon of the local Universe. 

As mentioned in Section \ref{sec:zLumSiz} we have found a slightly higher fraction of FRII-Low giant radio galaxies in our LoTSS-Deep sample compared to  LoTSS-DR1. With the low statistics it is impossible to tell if this indicates any redshift evolution, but further studies using future larger LoTSS datasets of the FRII-Low giants in particular will be key to understanding the impact, and evolution of the FRII-Low population.

It is clear that accretion mode or jet power on their own do not determine whether a low-power, low-accretion-rate source becomes an FRI or an FRII. We will address whether the host properties can help us find the answer to this question in Section \ref{sec:Mstar_SFR}. Given the lack of radiative signatures at other wavelengths and their low radio fluxes, meaning that before LOFAR the FRII-Low were only identified in small numbers, their contribution to cosmic evolution might have been underestimated. The importance of their role in galaxy and cluster evolution will be greatly dependent on the duration of their life cycles, which we will be able to constrain with lower frequency, LOFAR LBA observations in the near future (\citeauthor{DeGasperin2021} \citeyear{DeGasperin2021}; \citeauthor{Williams2021} \citeyear{Williams2021}).


\subsubsection{Luminous FRII (FRII-High)}\label{sec:FR2H}

\begin{figure}
\begin{subfigure}[b]{0.99\linewidth}
\centering
\includegraphics[width=.75\linewidth,trim={0cm 0cm 1.2cm 1.2cm},clip]{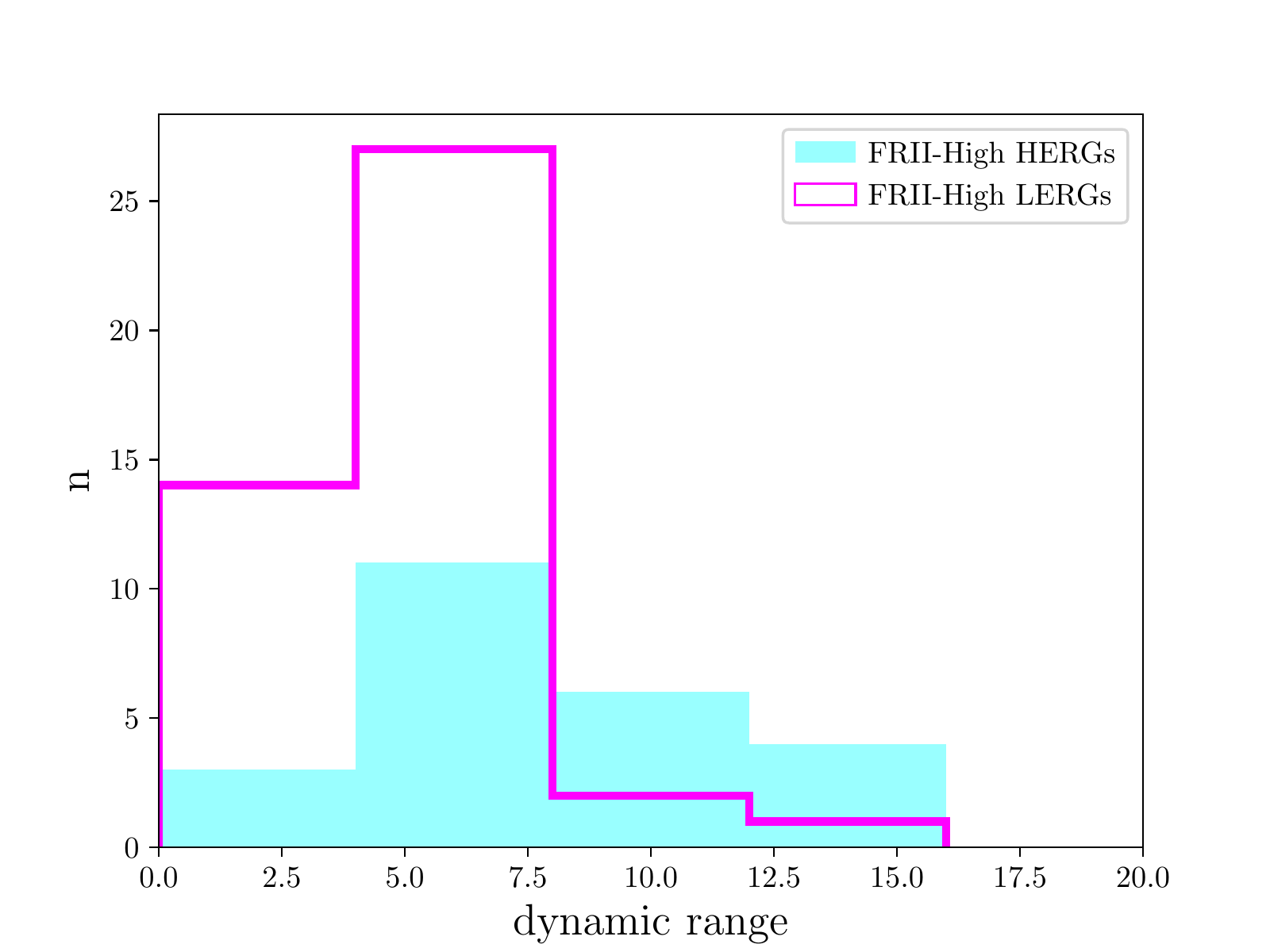}
\end{subfigure}
\begin{subfigure}[b]{0.99\linewidth}
\centering
\includegraphics[width=.75\linewidth,trim={0cm 0cm 1.2cm 1.2cm},clip]{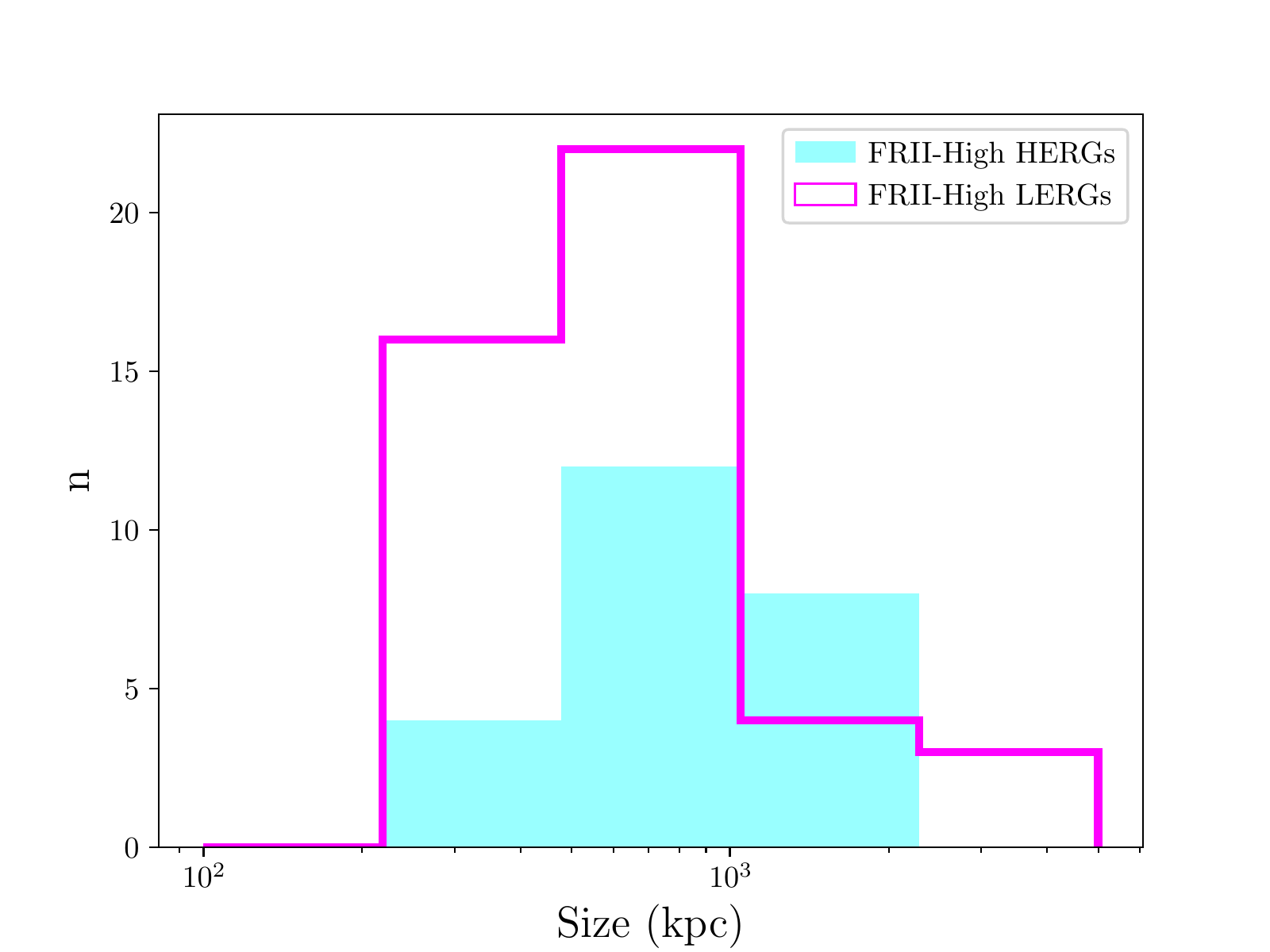}
\end{subfigure}
\caption{Source dynamic range and size comparison for the FRII-High HERG (cyan bars) and LERG (magenta outlines) subsets.}
\label{fig:FRIIH_LERG_HERG_comparison}
\end{figure}

\begin{table*}
	\caption{Median and mean values, and KS test statistics that the two samples are drawn from the same distribution, for the parameters shown in Fig. \ref{fig:FRIIH_LERG_HERG_comparison}.}\label{tab:KS}
	\centering
	\begin{tabular}{c|cc|cc|cc}
		Variable&HERG median &LERG median&HERG mean&LERG mean&KS stat&p value\\\hline
		SDR&7.5&5.1&7.6&5.6&0.367&0.021\\
		Size (kpc)&708&551&865&741&0.233&0.318\\
		Total flux (mJy)&151&61&1356&167&0.422&0.005\\
		Redshift&1.30&1.41&1.22&1.44&0.236&0.305\\
	\end{tabular}
\end{table*}

The FRII-High are particularly key to understanding the relationship between accretion mode and morphology. We found no visually identifiable difference between LERGs and HERGs among the luminous FRIIs (Fig. \ref{fig:FRII_gallery}); however, our data also allow us to examine this question quantitatively by comparing the source dynamic range (SDR) for the FRII-High HERGs and LERGs (Fig~\ref{fig:FRIIH_LERG_HERG_comparison}). Based on a Kolmogorov-Smirnov (KS) test, the hypothesis that the two SDR distributions are drawn from the same parent population cannot be ruled out at the 95 per cent confidence level (Table~\ref{tab:KS})-- this confirms our visual impression that across our large sample of 69 high-luminosity FRIIs, the accretion mode does not significantly affect the large-scale radio morphology.

The resolution of our LOFAR images is not sufficient to compare the inner jet morphology or the compactness of hotspots for the FRII-High HERGs and LERGs. We therefore cannot directly rule out differences in these structures. However, we note that travel time for fluid along a collimated highly relativistic FRII jet, even out to 500 kpc distances, is $\sim$ a few million years \citep[e.g.][]{MullinHardcastle2009} -- one to two orders of magnitude shorter than typical lobe evolution timescales (i.e. overall source ages) \citep[e.g.][]{Hardcastle2019}. It is therefore unlikely that systematic changes in the nucleus would not lead to identifiable differences in the radio brightness distribution of the two populations as a whole. We defer further investigation of FRII life cycles to a future paper. We already know from simulations and population statistics \citep[e.g.][]{Hardcastle2013,Hardcastle2014,Hardcastle2019,Shabala2020} and our spectral index results from \citetalias{Mingo2019} that the bulk of the FRII-High LERGs cannot be explained away as fading sources. 

Fig~\ref{fig:FRIIH_LERG_HERG_comparison} and Table~\ref{tab:KS} also show a lack of significant difference in the size distributions of HERGs and LERGs. This argues against a model in which there is a general tendency for sources to evolve from high accretion rate HERGs to low accretion rate LERGs over their lifetimes. We note that (as shown in the Table) there is evidence of a systematic difference in flux for the two subsamples, which cannot be fully explained by redshift effects. A systematic failure to identify the hosts of HERGs with low radio fluxes (but high enough radio luminosities to be FRII-High) could explain the observed flux difference between FRII-High HERGs and LERGs but this very unlikely, given the high host identification rate for LoTSS-Deep (97 per cent), and the fact that HERGs, by definition, are more luminous than LERGs beyond the radio. Our sample is also free of the issues with SED-template fitting for bright QSOs described by \citet{Duncan2019}, given the improved infrared coverage in LoTSS-Deep compared to  LoTSS-DR1 \citep{Duncan2021}.

\begin{figure}
  \resizebox{\hsize}{!}{\includegraphics[trim={0cm 0cm 1.2cm 1cm},clip]{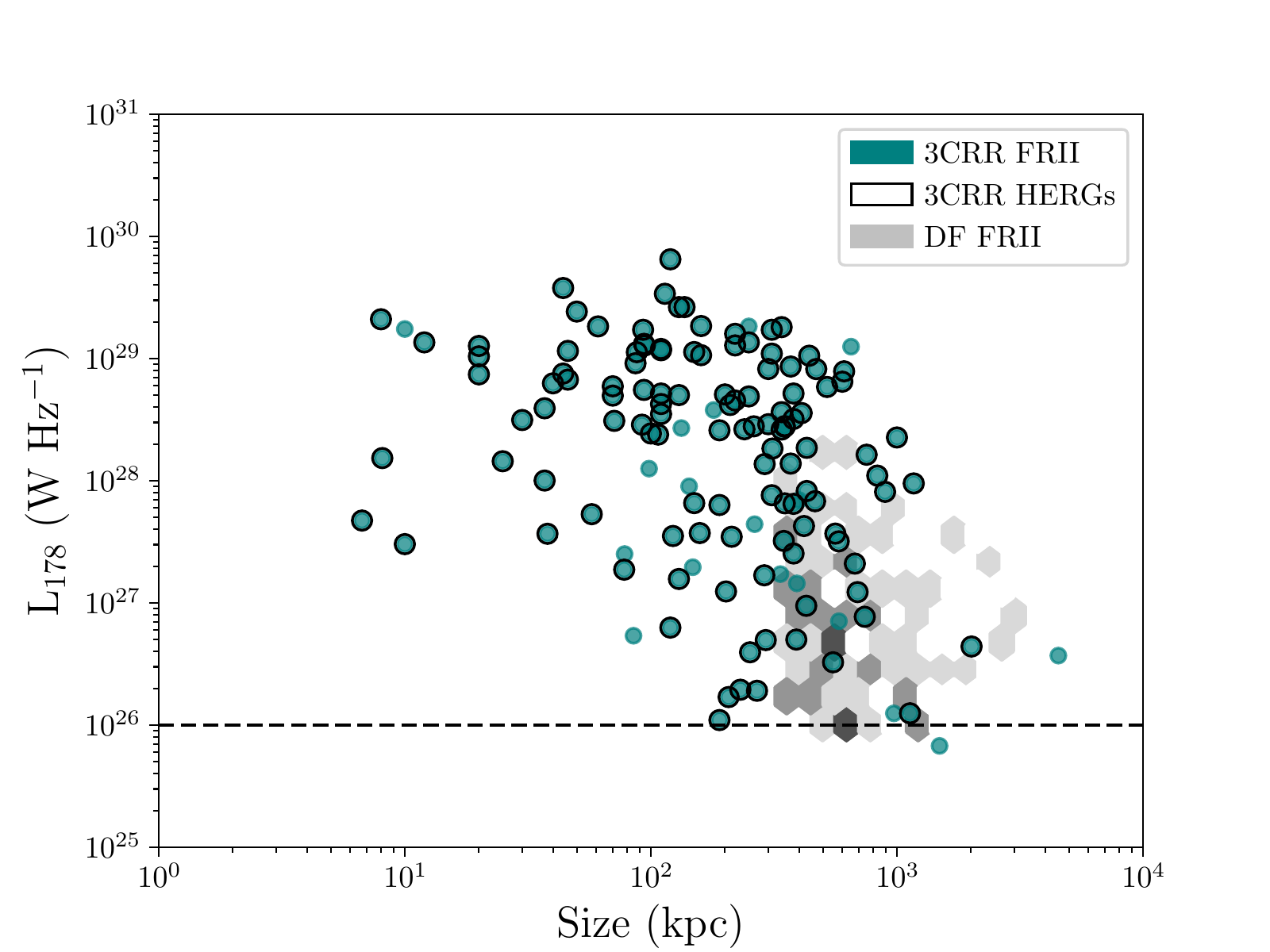}}
  \caption{Luminosity vs size distribution for all the FRII sources in the 3CRR survey. Empty black circles surround the HERGs. The plot also shows a density map of the DF FRIIs (regardless of accretion mode), with darker shades of gray denoting larger numbers of sources in a given hexagonal bin.}
  \label{fig:3CRR_L150_size_HERG_LERG}
\end{figure}

We now revisit the question of why the population seen in LoTSS-deep differs from older surveys of luminous radio galaxies, such as 3CRR, 2Jy, and others, whose FRII populations were dominated by luminous, radiatively-efficient FRIIs \citep[e.g.][]{Hardcastle2007,Hardcastle2009,Best2012,Gendre2013,Mingo2014,Ineson2015,Ineson2017,Tadhunter2016}. Fig. \ref{fig:3CRR_L150_size_HERG_LERG} shows all the FRIIs from the updated version of the 3CRR radio survey \citep{Laing1983}. The luminosities and sizes were taken from the 3CRR database\footnote{\url{https://3crr.extragalactic.info/}}, with sizes measured uniformly from the best available published low-frequency map (typically a multi-configuration Very Large Array map at 1.4 or 5 GHz). As above, 3C 343.1 was excluded from the sample. 79 per cent of all 3CRR sources are FRII, of which 87 per cent are HERGs. Comparing the 3CRR and DF FRII source distributions (the latter being shown in the density map and in Fig. \ref{fig:L150_size_HERG_LERG}), it is immediately obvious that the populations are different, even accounting for the different redshift ranges.

As we highlighted in \citetalias{Mingo2019}, surveys prior to LoTSS covered a very different part of the parameter space, which explains why the FRII-Low lay undiscovered as a population until recently. However, luminosity differences alone cannot explain the discrepancy in LERG/HERG fractions: Table \ref{tab:AC} shows that restricting the 3CRR FRIIs to the luminosity range comparable to our sample only increases the LERG fraction to 21 per cent. The difference in the LERG/HERG ratio for both samples is stark even when a dependence with luminosity is taken into account, as shown in the right-hand panel of Fig. \ref{fig:comb_LH_histo}.

One possible explanation relates to the significantly different size distributions of 3CRR and our sample: if HERGs are more common for small source sizes (perhaps because of the evolution of accretion over a source's lifetime) this could explain the discrepancy. We do not see a dependence of accretion mode on size within our FRII sample, but this consists primarily of physically large sources. We therefore looked at the accretion mode statistics at $L_{150}>1\times10^{26}$ W Hz$^{-1}$ (where sources have a high probability of being FRII) and $z<2.5$ for sources (i) in our small, unresolved, and faint categories (see Tables \ref{tab:auto_class}, \ref{tab:eye_class}, and \ref{tab:accr_mode}), and (ii) pre-filtered from the catalogue before the classification procedure (catalogued sizes below 8 arcseconds). These smaller source samples cover physical sizes of $\sim 50-400$ kpc and up to $\sim50$ kpc, respectively.

We found 176 sources in the first group of small, unresolved and faint sources (overwhelmingly in the small FRI, small FRII, and small Hybrid categories), of which 145 are LERGs ($\sim82$ per cent) and 31 HERGs ($\sim18$ per cent). We found 278 pre-filtered sources in the same redshift and luminosity range, of which 217 ($\sim78$ per cent) are LERGs, and 61 ($\sim22$ per cent) HERGs. Although we do not have reliable morphological classifications for these smaller sources we know from our earlier work and that of \citet{Gurkan2019} that even with LOFAR we are unlikely to observe a substantial enough number of FRIs in this luminosity regime to cause contamination concerns (see also Fig. \ref{fig:L150_size_HERG_LERG}). 

We therefore find no evidence that the HERG fraction increases to smaller sizes, and so the different size distributions cannot explain our discrepancy with 3CRR. This additional analysis of our smaller source categories also further supports the argument made earlier that the lack of significant difference in the size distribution of HERGs and LERGs in our sample contradicts a scenario in which all FRII sources start their lives as HERGs and then switch to a lower accretion rate once they have exhausted their initial supply of gas. This is also consistent with the small HERG fraction in the LOFAR galaxy-scale jet sample of \citet{Webster2021} and \citet{Webster2021b}. We note that of course this does not rule out a small fraction of the FRII LERGs being older/fading systems, but this scenario cannot explain the population as a whole.

A more viable explanation for the discrepancy in HERG fraction with 3CRR is the strong redshift dependence in 3CRR caused by its shallow flux density limit, which means that (beyond the local Universe) 3CRR sources will have the highest radio luminosities of any sources at their redshift, and therefore are likely to have among the highest jet powers. While accretion rate is not the only factor controlling jet power (black hole mass and spin are also important), it is nevertheless easier to produce a higher jet power at a high accretion rate \citep[e.g.]{Hardcastle2018}. It therefore seems plausible to expect HERGs to be particularly prevalent in a population that comprises the most powerful jets of their epoch; the fact that HERGs are less prevalent in a more representative sample spanning a wider range of luminosities at each epoch is perhaps then unsurprising.

Although there are still many open questions the more complete perspective on the radio-loud AGN population from our wide area deep survey enables several firm conclusions about the FRII population: 
\begin{itemize}
    \item It is not possible to tell whether an FRII is luminous (FRII-High) or less luminous (FRII-Low), or whether it is accreting at a low (LERG) or high (HERG) rate based solely on radio morphology inferred from LoTSS-Deep quality data (see Fig. \ref{fig:FRII_gallery}). While we cannot examine the jet structures of the two classes directly with our images, we consider it implausible that systematic differences in the inner jets and/or hotspot regions would not lead to a detectable systematic difference in the large-scale brightness distributions of the two classes.
    \item Luminous FRIIs are not predominantly HERGs, although conversely a majority ($\sim 2/3$) of the minority HERG population are luminous FRIIs. 
    \item There is no significant difference in the size distribution of FRII HERGs and LERGs, with many examples of both accretion modes in the giant radio galaxy regime ($>1$ Mpc), which argues against models in which there is a general tendency for individual sources to transition from the HERG to LERG regime as they age.
    \item While population studies based on luminosity functions include LERGs in general terms, the impact of \textit{FRII} LERGs (at all luminosities) on cosmic evolution and cosmic magnetism might need to be reassessed, given that they are more numerous than previously thought and their evolution and physical conditions are expected to be different to those of FRI LERGs \citep[e.g.][]{Croston2018}.
\end{itemize}


\section{The role of the hosts: stellar mass and star formation rate}\label{sec:Mstar_SFR}

From Section \ref{sec:FR_accretion_link}, we conclude that radio morphology is not primarily controlled by accretion mode, since we cannot identify systematic differences in the radio structures of FRII HERGs and LERGs. We also conclude from this work and from \citetalias{Mingo2019} that radio morphology is not solely controlled by jet power, because of the existence of the FRII-Low. Therefore, in this section we incorporate the high-quality host-galaxy information we have available for this sample to examine the role that host-galaxy properties play in determining both radio structures and accretion modes.

In the pc to kpc-scale jet disruption paradigm for the FR dichotomy \citep[e.g.][]{Bicknell1994,Laing2002,LaingBridle2014,Kaiser2007,Perucho2007}, jet deceleration to form FRI jets is primarily the result of interaction with the environment and entrainment of material from it. The primary source of material to be entrained is thought to be the central hot-gas environment, rather than winds from young stellar populations \citep[e.g.][]{LaingBridle2014,AnglesCastillo2021}, though stellar mass loading can play an important role for weaker jets \citep{Perucho2014}. For massive ellipticals, the typical hosts of radio-loud AGN, the hot-gas environment richness is expected to scale with galaxy mass \citep[e.g.][]{KimFabbiano2013}. Therefore if the jet disruption paradigm is correct, we would predict that stellar mass is the best easily measurable proxy for the environmental conditions that, along with jet power, determine the FR class of a source. We test this model in Section~\ref{sec:FR_class}.

Accretion mode, in contrast, is thought to be most closely linked to the availability of a supply of dense cold gas close to the nucleus, which can increase the accretion rate above that available only from fuelling linked to the hot gas environment \citep{Gaspari2013,Gaspari2015,HardcastleCroston2020}. In AGN the mass of the black hole also plays a more important role than in X-ray binaries in determining how much gas is needed to cross the threshold between RI and RE accretion \citep[e.g.][]{Hardcastle2018b}. We might expect that the specific star formation rate (sSFR, defined as the total star formation rate divided by the stellar mass), which is a better indicator of the fractional gas content of the galaxy than the global star formation rate \citep[e.g.][]{Abramson2014,Ilbert2015}, would be a good proxy for short-term cold gas availability, and hence accretion mode. This idea is investigated in Section~\ref{sec:Accr_mode}.


\subsection{What host galaxy properties control FR class?}\label{sec:FR_class}

\begin{figure}
  \resizebox{\hsize}{!}{\includegraphics[trim={0cm 0cm 1.1cm 1cm},clip]{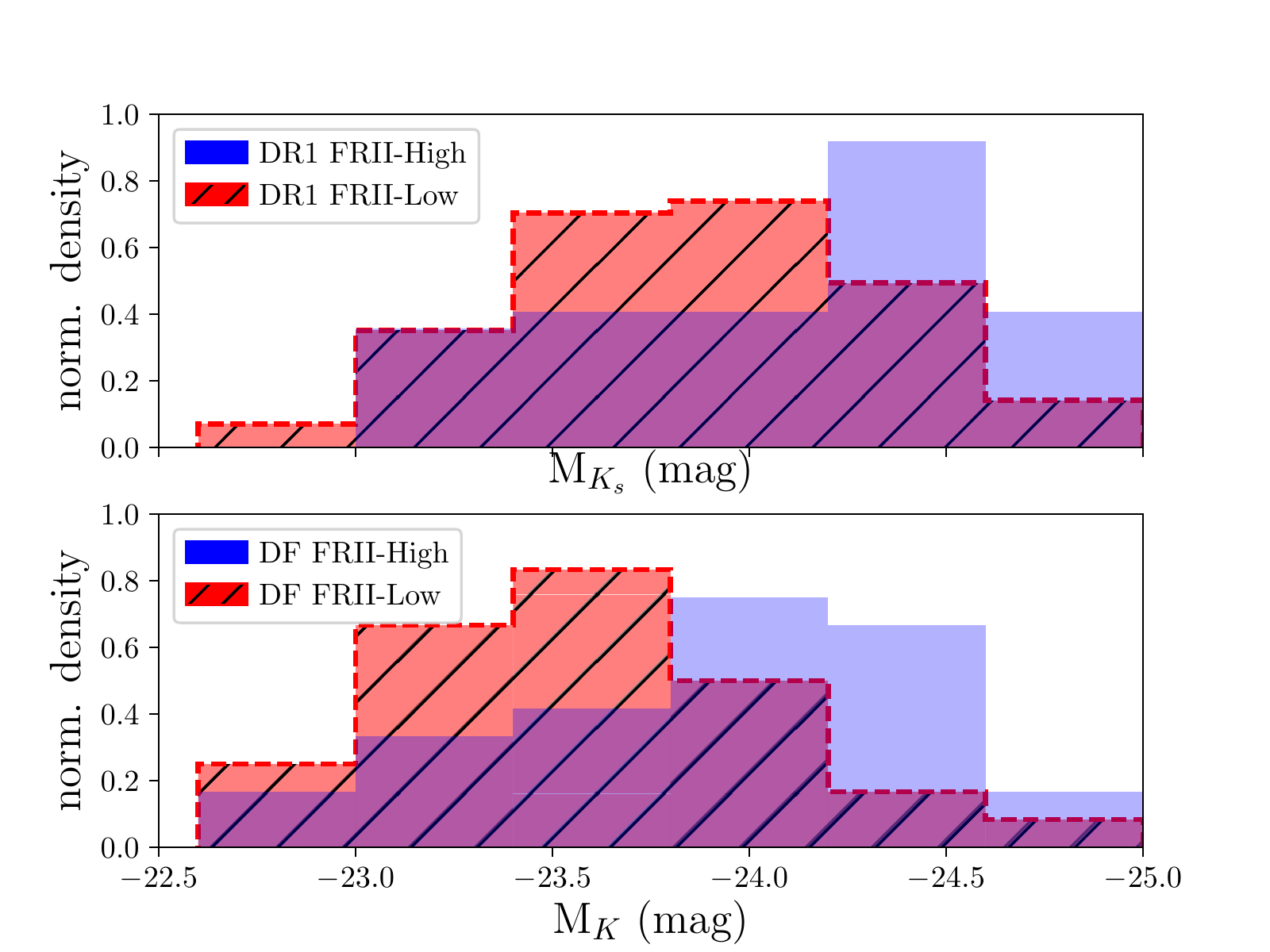}}
  \caption{Histogram of the near-IR K-band absolute magnitude distribution for the FRII-High (blue) vs FRII-Low (red, hatched). The top plot shows the rest-frame $M_{K_{s}}$  LoTSS-DR1 results from \citetalias{Mingo2019}, with slightly different binning. The bottom plot shows the rest-frame $M_{K}$ results for the three combined deep fields.}
  \label{fig:DF_Hetdex_K_histo}
\end{figure}

\begin{figure}
  \resizebox{\hsize}{!}{\includegraphics[trim={0cm 0cm 1.2cm 1cm},clip]{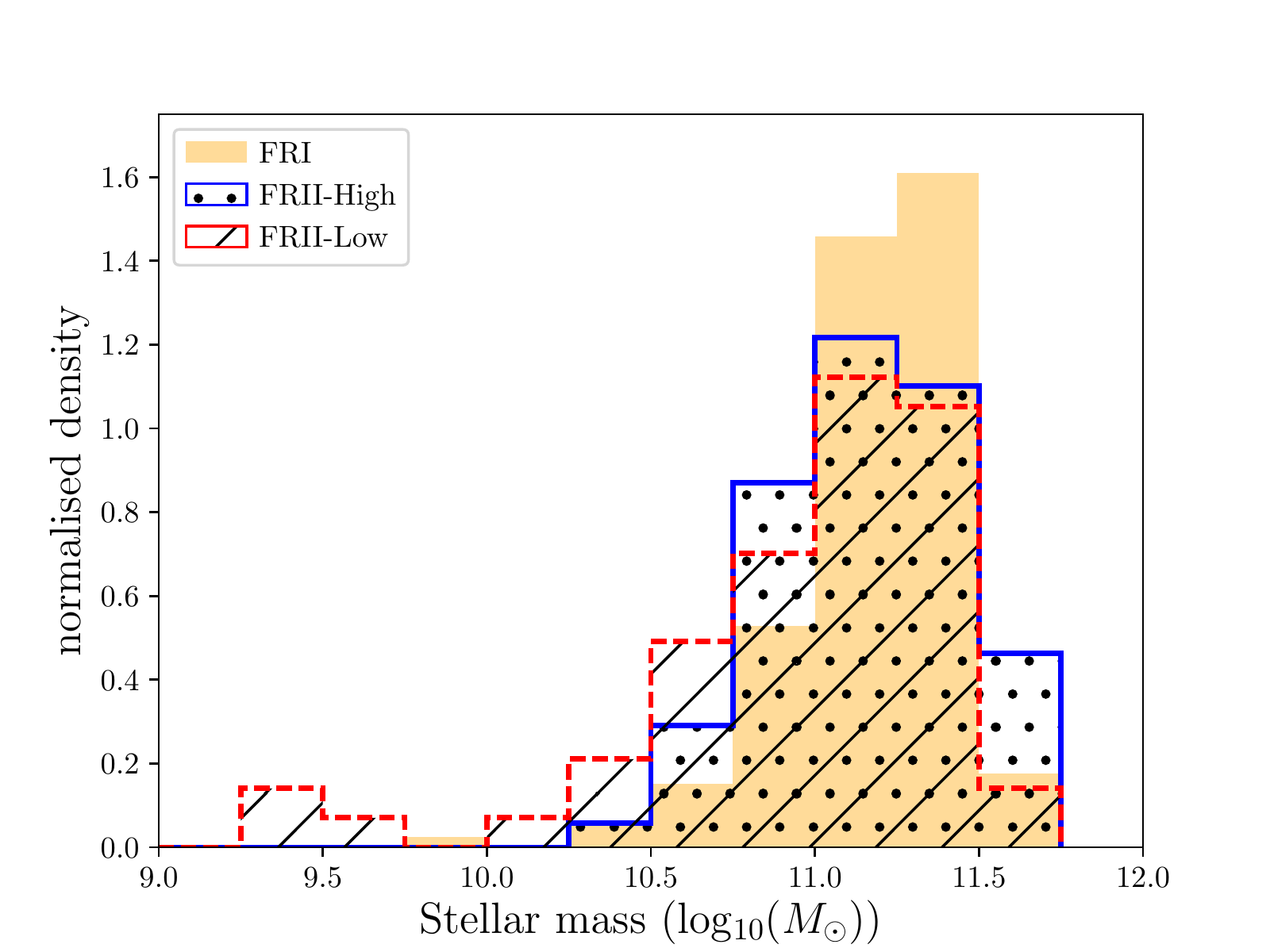}}
  \caption{Normalised histogram distribution of stellar masses for the LoTSS-Deep FRI (yellow, solid), FRII-High (blue solid outline, dotted) and FRII-Low (red dashed outline, hashed).}
  \label{fig:SMass}
\end{figure}

\begin{figure*}
\centering
\includegraphics[width=.5\textwidth,trim={0cm 0cm 0cm 0cm},clip]{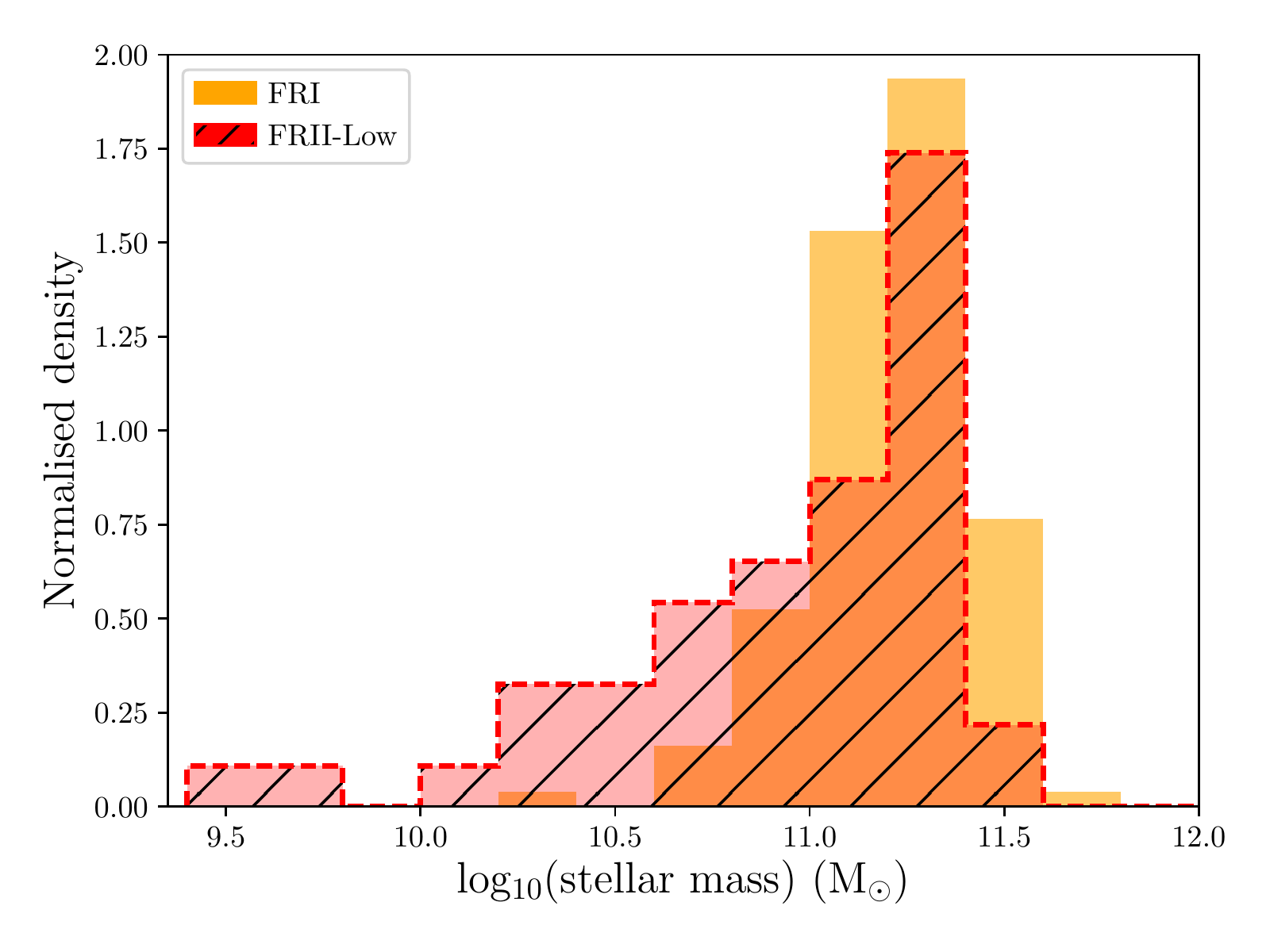}\hfill
\includegraphics[width=.5\textwidth,trim={0cm 0cm 0cm 0cm},clip]{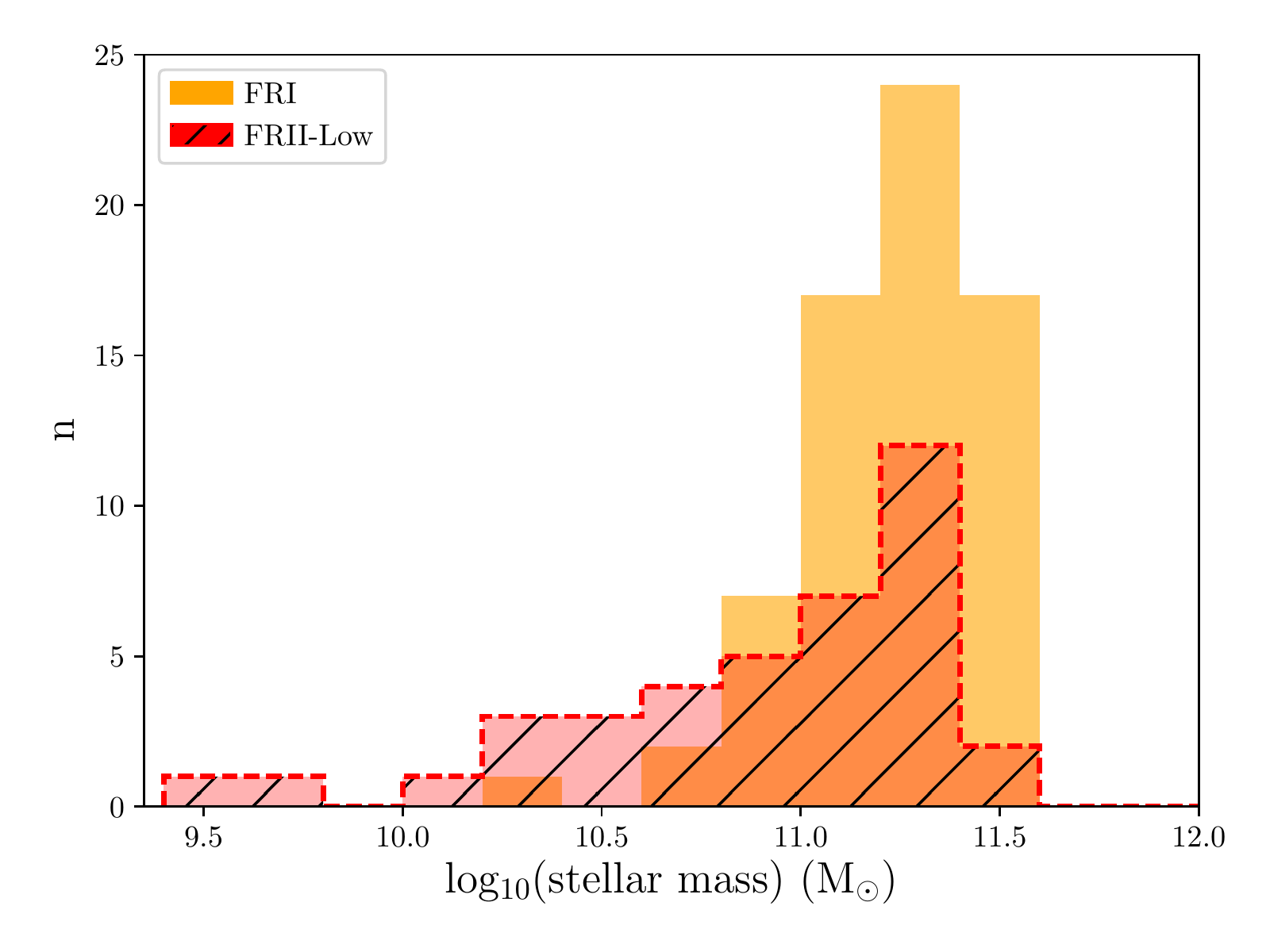}\hfill
\caption{Stellar mass distributions for a subset of matched FRI (yellow solid bars) and FRII-Low (red hatched bars, dashed outline). Left (normalised): samples matched in size and luminosity (200-1000 kpc, $24.3<\log_{10}(L_{150})<26$ W Hz$^{-1}$). Right: narrower luminosity range, focusing on the sources just below the traditional FRI/II boundary ($25<\log_{10}(L_{150})<26$ W Hz$^{-1}$).}
\label{fig:FRI_FRIIL_comparison_SM}
\end{figure*}

In our previous work with  LoTSS-DR1 we observed a significant difference between the host absolute magnitudes of FRII-High and FRII-Low. This holds true for our LoTSS-Deep sample as well, as shown in Fig. \ref{fig:DF_Hetdex_K_histo}, where we have plotted a comparison histogram of K$_{s}$ and K absolute magnitudes for both samples. Unlike in \citetalias{Mingo2019} we did not filter and match the populations in size and redshift for LoTSS-Deep, as the statistics would have been too low, which is also why we used wider bins for this comparison. We again filtered out any QSOs to avoid contamination. M$_{K_{S}}$ is a reliable proxy for stellar mass \citep[e.g.][]{Ziparo2016}, but with our SED fitting approach in LoTSS-Deep (covered in sections \ref{sec:DF_data} and \ref{sec:SED_fitting}) we have much better, more direct estimates of this parameter.

The histogram in Fig. \ref{fig:SMass} therefore shows the host mass comparison between FRI, FRII-High and FRII-Low with values derived through the SED fitting procedure. For all three populations most hosts have masses $10.5<\log_{10}(M_{*})<11.5$ M$_{\odot}$, but it is clear that the FRII-Low distribution has a tail of hosts with lower stellar masses. There are three FRII-Low sources with $\log_{10}(M_{*})<10$ M$_{\odot}$, two of which are at relatively high $z$ (0.9, 1.2) and have larger uncertainties on their stellar masses, as these sources are below the mass completeness limit of the sample \citep{Duncan2021}. The lower-mass tail of FRII-Lows is present even ignoring these outliers.

The trends in Fig. \ref{fig:SMass} are consistent with what the M$_{K_{S}}$ histograms in Fig. \ref{fig:DF_Hetdex_K_histo} show, and our conclusions in \citetalias{Mingo2019}. The greater overlap between the FRII-High and FRII-Low in stellar masses compared to the M$_{K}$ distribution could be due to some AGN contribution in this band for radiatively-efficient sources (which are overwhelmingly FRII-High).
 
It is possible to take this analysis further without this sample to consider only the luminosity ranges in which the FRII and FRI populations show most overlap. In Fig. \ref{fig:FRI_FRIIL_comparison_SM} we show a matched subset of FRII-Low and FRI in both luminosity and size, eliminating giant radio galaxies, which are more likely to be fading or restarting sources \citep{Dabhade2020}. The difference in host stellar masses between both populations becomes very clear. The left-hand panel shows the luminosity range where FRI and FRII coexist, highlighting that both distributions have significantly different dependences on stellar mass. The right-hand panel shows only the decade of luminosity around the canonical FR break, demonstrating that the absolute ratio of FRI/II depends on stellar mass. Using three bins in stellar mass, we used a $\chi^{2}$ fit to a line of slope zero and expected y value corresponding to the weighted mean percentage of FRII-Lows to confirm that the null hypothesis that the percentage of FRII-Low is independent of stellar mass can be ruled out at the 95 per cent confidence level. The binned FRII-Low percentages were $86\pm28$, $45\pm13$, and $19\pm12$ for the three mass ranges $\log_{10}(M_{*})<10.5$, $10.5<\log_{10}(M_{*})<11.25$, $\log_{10}(M_{*})>11.25$ M$_{\odot}$, showing a decreasing fraction of FRII-Low with increasing stellar mass. In other words, for intermediate jet powers, the probability of a galaxy producing an FRI or FRII is strongly dependent on stellar mass; for masses $\geq10^{11}$ M$_{\odot}$ the probability of forming an FRI becomes higher than the probability of forming an FRII-Low. This confirms the long-standing theory that morphology for jets of similar power is determined by interaction with the host galaxy, likely on pc to kpc scales, consistent with the conclusions of \citetalias{Mingo2019}.

There is nevertheless a non-negligible number of FRII-Low in high-mass galaxies. It is possible that this minority of FRII-Low are the fading remains of former FRII-High, or that different gas density and pressure distributions in the inner few kpc of these host galaxies could enable the jets to remain collimated at low powers. 


\subsection{What host galaxy properties control accretion mode?}\label{sec:Accr_mode}

\begin{figure*}
\centering
\includegraphics[width=.5\textwidth,trim={0cm 0cm 1.2cm 1cm},clip]{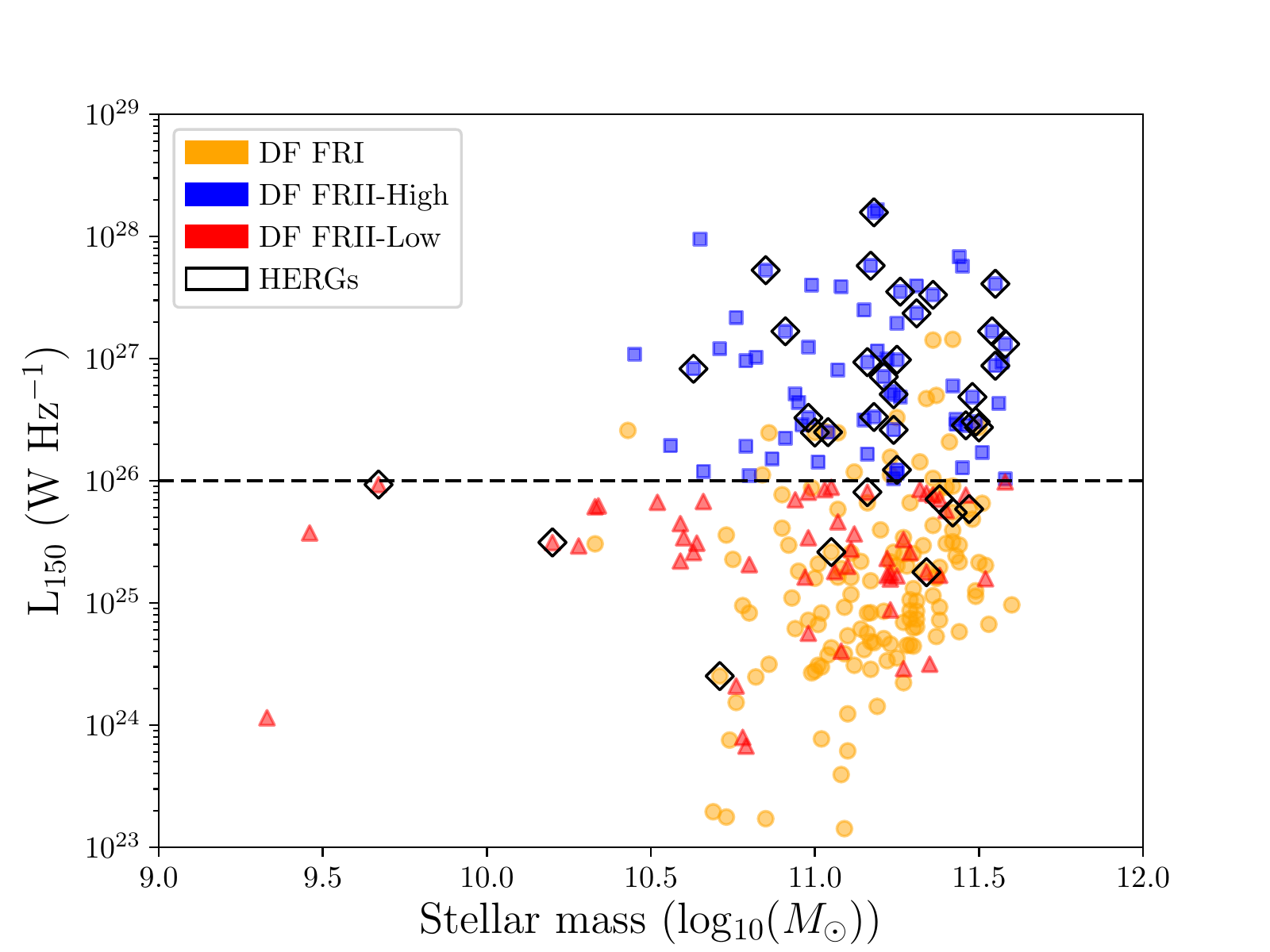}\hfill
\includegraphics[width=.5\textwidth,trim={0cm 0cm 1.2cm 1cm},clip]{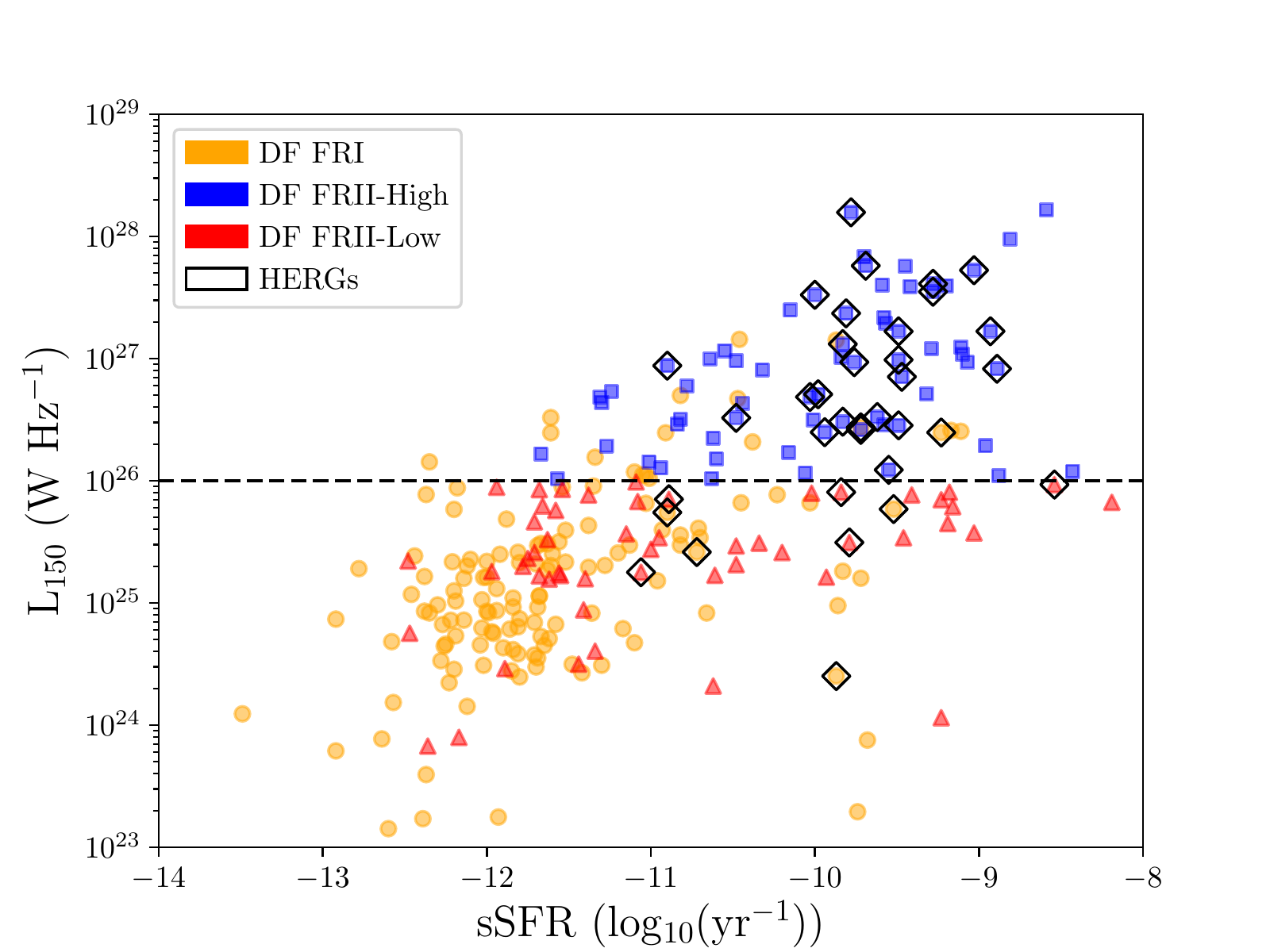}\hfill
\caption{\textbf{Left:} luminosity vs stellar mass. \textbf{Right:} luminosity vs specific star formation rate. Symbols and colours as in fig. \ref{fig:IRAC}.}
\label{fig:L150_SM_sSFR}
\end{figure*}

As introduced earlier, we consider the (specific) star formation rate to be a good proxy for gas availability near the black hole. Stellar mass may also be relevant for accretion mode due to its tight scaling with black hole mass. The left-hand panel of Fig. \ref{fig:L150_SM_sSFR} shows 150-MHz luminosities against stellar masses for our FRI, FRII-High, and FRII-Low. Most FRIs are found towards the bottom-right of the plot (lower luminosities, higher stellar masses), as expected. For comparison, star-forming galaxies are known to span a wide range of sSFR \citep{Gurkan2018a} with significant redshift dependence -- values around $10^{-10}$ to $10^{-9}$ yr$^{-1}$ are typical of star-forming galaxies from the nearby Universe to the median redshift of our sample \citep[e.g.][]{Damen09}.

We found no statistically significant difference between the stellar mass distributions of all LERG and HERG hosts ($p$ value of 0.203), nor between those of FRII-High LERG and HERG hosts ($p$ value of 0.141). Our results appear to contradict the idea that HERGs, overall, favour systems with lower host masses than LERGs \citep[e.g.][]{Tasse2008,Smolcic2009,Best2012,Hardcastle2013b}, but selection effects and measurement methods are crucial to consider \citep[see e.g. the discussions by][]{Fernandes2015,Gurkan2018a}. Unlike in previous surveys, our LERG population encompasses not just FRIs, but a large number of FRIIs (across the entire luminosity range, $10^{23}<$L$_{150}<10^{29}$ W Hz$^{-1}$) in lower-mass hosts, while our HERGs are likely more dominated by QSOs than Seyfert galaxies. As the survey is not statistically complete up to $z=2.5$, at the high-$z$ end our sample is dominated by bright sources. 

\begin{figure*}
\centering
\includegraphics[width=.5\textwidth,trim={0cm 0cm 1.2cm 1cm},clip]{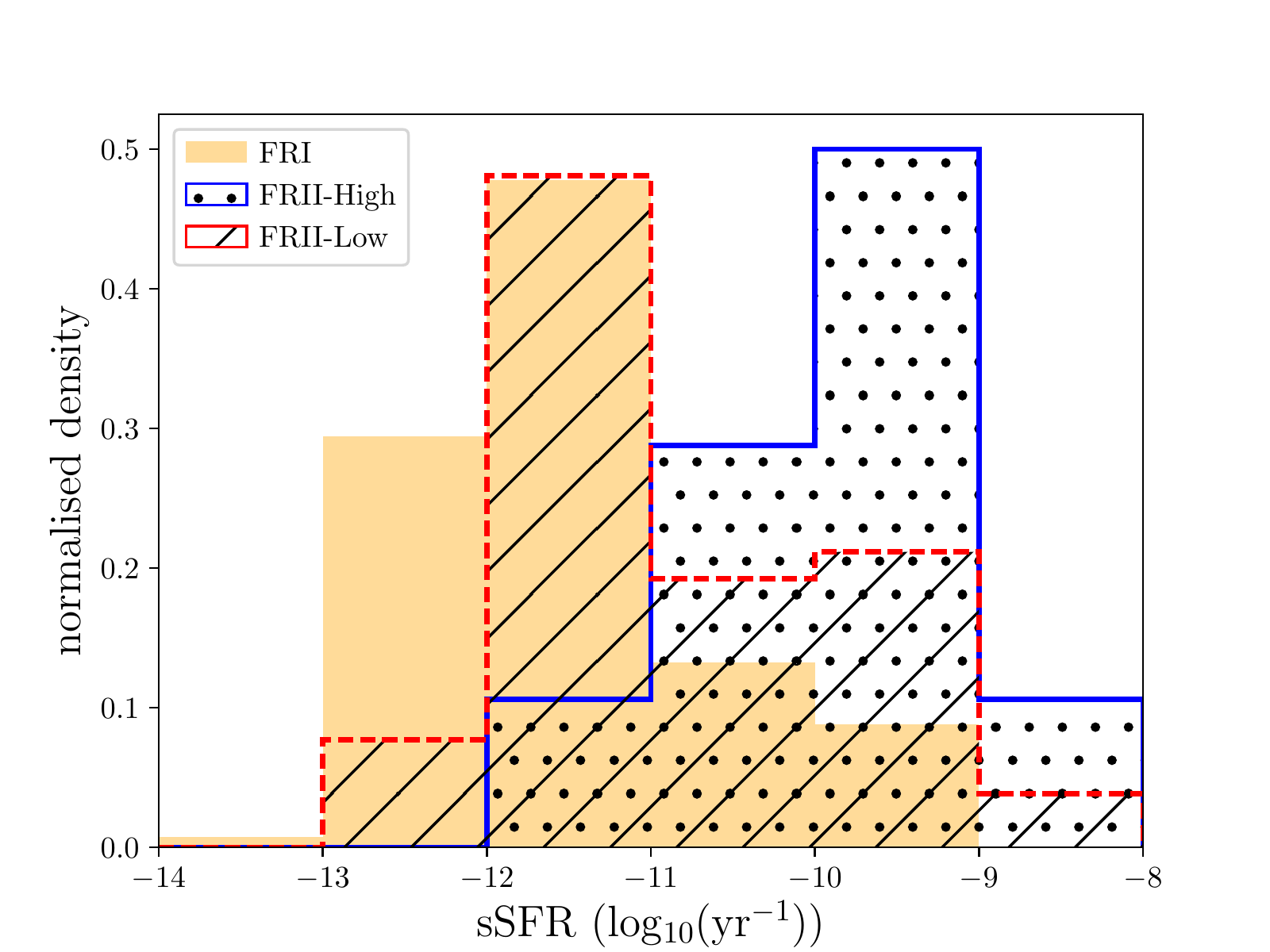}\hfill
\includegraphics[width=.5\textwidth,trim={0cm 0cm 1.2cm 1cm},clip]{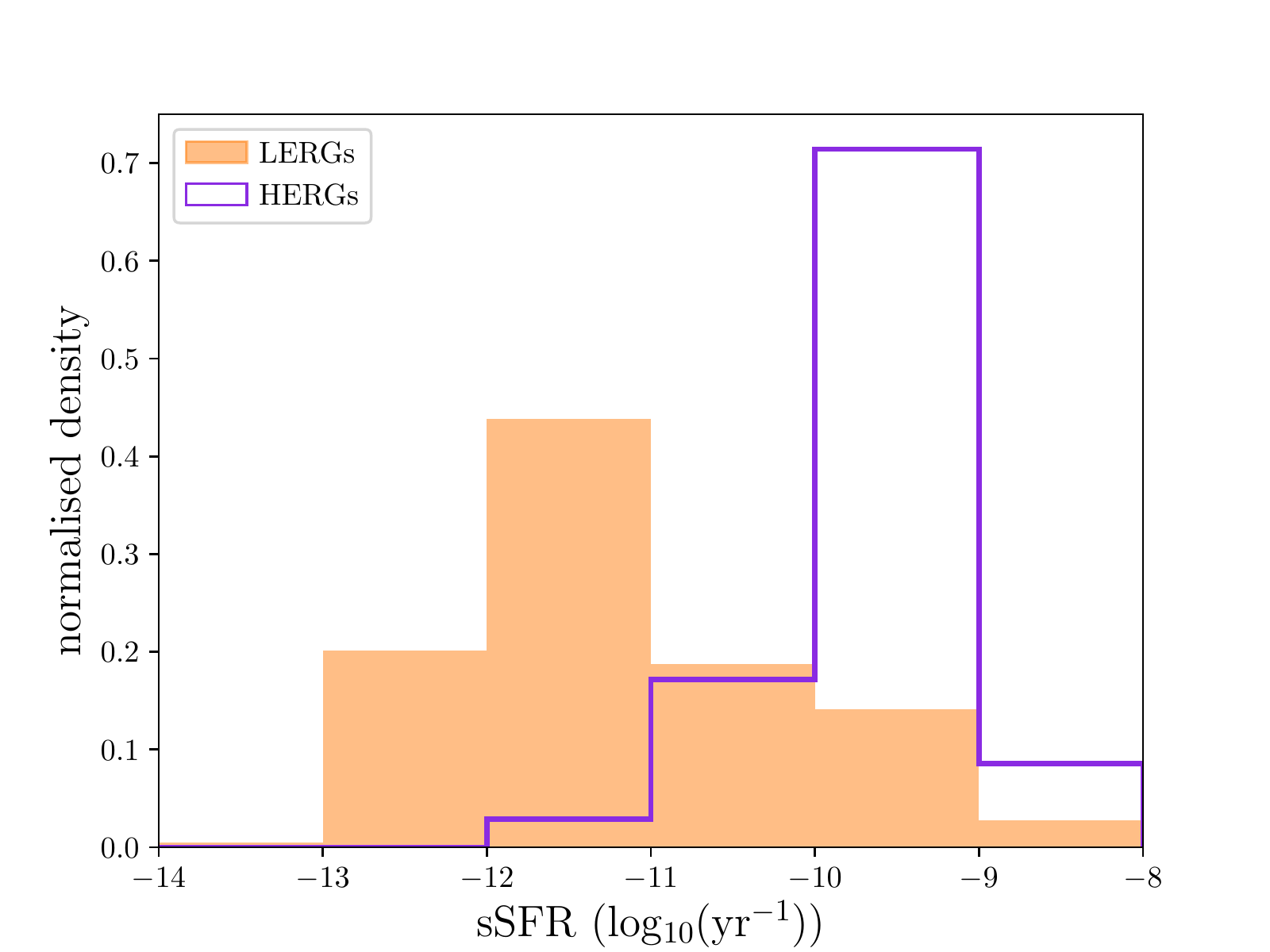}\hfill
\caption{\textbf{Left:} normalised histogram distribution of specific star formation rates for the LoTSS-Deep FRI, FRII-Low, and FRII-High. Colours and symbols as in Fig. \ref{fig:SMass}. The median sSFRs are -11.7, -11.4, and -9.8, respectively. \textbf{Right:} same histogram for all LERGs (orange, solid bars) and HERGs (purple outline). The median sSFRs are -11.6 and -9.8, respectively. Note that LERGs outnumber HERGs by a factor of 7; see Table \ref{tab:AC} for detailed statistics.}
\label{fig:sSFR}
\end{figure*}

The right-hand panel of Fig. \ref{fig:L150_SM_sSFR} shows the distribution of our sources in terms of luminosity versus specific star formation rate. Unlike with the stellar mass, it is very clear from the plot that there is a very strong preference for HERGs to be in star-forming systems (sSFR$>-10$ yr$^{-1}$), regardless of their radio morphology (see also the right-hand panel of Fig. \ref{fig:sSFR}). This is consistent with what has been found in the past \citep[e.g.][]{Baldi2008,Janssen2012,Hardcastle2013b,Gurkan2014,Karouzos2014,Ineson2015,Ineson2017,Mingo2016,Miraghaei2017,Weigel2017,Williams2018a}, and despite some caveats in terms of the timescales required to fuel star formation versus powering the AGN \citep[see e.g.][]{Wild2010} it seems very likely that both mechanisms are fuelled from the same gas. 

In terms of FR populations, the FRII-High show a very broad distribution of specific star formation rates, but they are more star-forming than the FRIs, as can be seen in the left-hand panel of Fig. \ref{fig:sSFR}. The histogram also shows that, as a whole, the FRII-Low tend to have specific star formation rates intermediate between those of the FRIs and the FRII-High, but they show a very large scatter. Most of the star-forming FRII-Low are quite close to the traditional FRI/II boundary (Fig. \ref{fig:L150_SM_sSFR}), hinting that they might belong to a similar population to the FRII-High above the line, just fainter. A large fraction (40--50 per cent) of the FRII-Low occupy the same L$_{150}$/sSFR parameter space as the FRIs even when the environmental dependence of the luminosity is considered \citep{Hardcastle2013,Croston2018}. These FRIIs therefore appear to have been produced in different host-galaxy conditions to their more luminous counterparts.

The right-hand panel of Fig. \ref{fig:sSFR} shows a fairly bimodal sSFR distribution between LERGs and HERGs. There is, however, a noticeable tail of LERGs in star forming systems, as also noted by Kondapally et al. (subm.). This seems to be driven mostly by a subgroup of FRII sources, both above and below (but close to) the traditional luminosity boundary (see the right-hand panel of Fig. \ref{fig:L150_SM_sSFR}).

Another conclusion we can draw from the right-hand panel of Fig. \ref{fig:sSFR} is that AGN host galaxies can have high sSFR without having a high accretion rate (tail of high sSFR LERGs), but not the other way around: the HERG distribution drops very sharply as sSFR decreases. It is thus clear that sSFR is the host-galaxy property most closely linked to accretion mode; in other words: a gas supply (traced by ongoing star formation) is necessary to produce a HERG, but crucially, gas availability somewhere in the host does not guarantee that a HERG will be present (tail of LERGs). We note that this result is compatible with star formation and RE AGN activity occurring on different physical and temporal scales, as the HERG phase is shorter than most star formation episodes \citep[e.g.][]{Wild2010}.

If the integrated star formation rate is a proxy for the black hole accretion rate, and the stellar mass is a proxy for the black hole mass \citep{Magorrian1998}, then we would expect that the sSFR is a proxy for the key accretion switch quantity $\dot{M}/\dot{M}_{\text{Edd}}$. We can substitute in some representative values to test this. The Eddington accretion rate is defined as $\dot{M}_{\text{Edd}}=(4 \pi G M_{\text{BH}} m_{p})/(\epsilon c \sigma_{T})$, where M$_{\text{BH}}$ is the black hole mass, m$_{p}$ the mass of the proton, c the speed of light, $\epsilon$ the radiative efficiency (typically assumed to be $\sim0.1$), and $\sigma_{T}$ the Thomson scattering cross-section for the electron. Substituting in all values except the black hole mass gives an Eddington accretion rate of $2.3\times10^{-9}(M_{\text{BH}}/M_{\odot})$ M$_{\odot}$ yr$^{-1}$. If we considered a critical accretion rate for the transition between the RI and RE regimes of 0.03 $\dot{M}_{\text{Edd}}$ \citep[as observed for e.g. the 2Jy sample, see][]{Mingo2014}, we would obtain a specific accretion rate of $\dot{M}_{\text{crit}}/M_{\text{BH}}=6.9\times10^{-11}$ yr$^{-1}$. Assuming a ratio between the black hole mass and the galaxy stellar mass of 0.025 per cent \citep[e.g.][]{Reines2015}, we would get $\dot{M}_{\text{crit}}/M_{\text{galaxy}}=1.7\times10^{-14}$ yr$^{-1}$. And assuming that the AGN accretion rate is roughly a factor of $10^{-3}$ smaller than the star formation rate \citep[e.g.][]{Mullaney2012}, we would obtain a critical sSFR value for the transition between the RI and RE regimes of $1.7\times10^{-11}$ yr$^{-1}$. While this is a very rough calculation, and there is a broad range of possible values for most of the parameters involved, it is very interesting that the result is in the same order of magnitude as the transition seen in Fig. \ref{fig:sSFR}. 

Despite these interesting insights into the nature of an accretion mode `switch', there are nevertheless some LERGs in star-forming systems ($\sim17$ per cent have sSFR$>-10$ yr$^{-1}$), so that sources cannot be classified by accretion mode based only on sSFR. A significant fraction of LERGs in star-forming systems have lower-mass hosts: 53 per cent of LERGs with sSFR$>-11$ yr$^{-1}$ have $\log_{10}(M_{*})<11.0$ M$_{\odot}$, compared with just 17 per cent of LERGs with sSFR$<-11$ yr$^{-1}$. However, it is not possible to determine how much of this difference is driven by selection effects. Both host mass and star formation rate are also linked to the larger-scale environment, which we know also plays a role in directly fuelling not just LERGs \citep[see e.g.][]{Ineson2015,Miraghaei2017,Morganti2018,Massaro2019,Croston2019} but also less powerful HERGs, through mergers and interactions \citep{Pierce2019}. 

In summary, we have found a clear link between accretion mode and sSFR in our sample, which provides insight into the nature of the accretion mode `switch' in radio-loud AGN. There remain gaps in our understanding of the relationship between host properties and accretion behaviour, however, which await larger samples and better understanding of activity timescales for different source populations as a function of epoch.

\subsection{The overall picture}
\label{sec:Host_Summary}

In Sections~\ref{sec:FR_accretion_link},~\ref{sec:FR_class} and~\ref{sec:Accr_mode} we were able to draw a number of firm conclusions about the relationships between radio morphology (FR class), accretion mode and host galaxy properties. Here we briefly summarise how these results fit together.

We have shown that the observed radio morphology (FR class) of a radio-loud AGN is not a direct consequence of its accretion mode, because edge-brightened, FRII lobe structures are found across all radio luminosities and both accretion modes. This led us to take a detailed look at the influence of host galaxy properties on both radio morphology and accretion mode in the two preceding subsections. We have found several important relationships that help us to understand what controls radio-loud AGN behaviour. 

Firstly, we have used our large sample numbers in the radio luminosity range where both FRI and FRII morphologies are observed to demonstrate that for systems that -- broadly -- are expected to have similar jet powers, the probability of a jet forming FRI or FRII structures on large scales is strongly linked to stellar mass (Fig.~\ref{fig:FRI_FRIIL_comparison_SM}). As discussed above, this provides the best evidence to date that host-galaxy environment (likely inner hot-gas density) controls whether jets of similar power are disrupted. In contrast, the previous section shows that accretion mode is not primarily determined by the same parameter, but instead is closely linked with specific star-formation rate (Fig.~\ref{fig:sSFR}), which we interpret as tracing situations of high fuel availability enabling high accretion rates. 

By separating out the influence of different host galaxy parameters, our analysis has demonstrated for the first time that accretion mode and radio morphology are controlled in different ways by their host-galaxy environment, consistent with the observational landscape in which accretion mode and radio morphology have patterns of connection but not a simple one-to-one link.

\subsection{Sample completeness considerations}

Sample completeness is always a concern, but particularly so when working at the intersection of flux-limited samples and complex selection techniques. In this section we briefly summarise the key considerations we have addressed in this work.

In terms of radio flux densities, LoTSS-Deep is highly complete down to values well below those that we consider in our sample (see Section \ref{sec:Analysis}). The main survey papers cover this in more detail \citep{Tasse2021,Sabater2021}. As mentioned in Section \ref{sec:DF_data}, extended sample completeness (association of components via Galaxy Zoo) is also very high in our sample, as any sources that were flagged up either by conflicting associations from the volunteers, classified as blends, or identified as non-standard were checked multiple times \citep{Kondapally2021}. This process also flagged up any issues with host identification in extended sources, and those identified via the maximum likelihood process were also checked repeatedly \citep{Duncan2021}. The host identification completeness is greater than 97 per cent for LoTSS-Deep.

Identification of faint radio AGN, particularly in heavily star-forming systems, is an important concern. Our selection requires a radio excess for any sources to be identified as radio AGN (see Section \ref{sec:SED_fitting}). However, for our morphological classification we require the radio emission to be extended on large angular scales, and for sources to be included in our sample of FRIs and FRIIs they must present a morphology that is not consistent with that of radio emission in star forming galaxies. Proof of our success in selecting clean samples is presented in Section \ref{sec:AGN_SF}. Our sample is missing radio AGN with small angular scales (as discussed in detail in section \ref{sec:FR2H}), including those with high star formation rates, and radio-quiet AGN. These systems, and sample completeness at low radio luminosities, are covered in more detail by Best et al. (subm.) and Kondapally et al. (subm.).

We have also discussed completeness in terms of HERG/LERG classifications in Section \ref{sec:FR_accretion_link}, concluding that we are unlikely to have missed faint HERGs among luminous FRIIs because of the broad correlation between kinetic and radiative output in radio AGN \citep{Mingo2016}. We might have missed faint HERGs at lower radio luminosities, but given the consensus approach at the SED fitting stage the fraction is likely to be very low in our sample (see Section \ref{sec:SED_fitting} and the discussions by Best et al., subm. and Kondapally et al., subm.). Hybrid AGN/star-forming systems with small jets are a population where this selection bias might be of key importance, however.

There are also considerations regarding cosmic variance and small sample sizes, which we have mostly covered in Section \ref{sec:FR2H}. LoTSS-Deep is a pencil-beam survey, and as such it does not contain many examples of rare populations, such as the extremely bright FRIIs found in e.g. the 3CRR and 2Jy surveys. These issues will be addressed as the LOFAR samples grow in depth and coverage, and eventually by the advent of the SKA surveys. As stated in Section \ref{sec:DF_data}, we have observed no systematic or statistically significant differences between our three fields.

\smallskip


\section{Conclusions}\label{sec:Conclusions}

In this work we have carried out an investigation into the relationship between black hole accretion mode and large-scale radio morphology, using new data from the LOFAR Deep Fields. The deep LOFAR data and exquisite multiwavelength coverage have enabled us to analyse the properties of sources up to z=2.5, much farther than we were able to achieve for  LoTSS-DR1  \citepalias{Mingo2019}, and to use broad-band SED fitting to constrain the host and black hole accretion properties for all our sources. While our samples are not complete, they are representative of the general radio-galaxy population.

Our results have led to the following conclusions:
\begin{itemize}
    \item In contrast with what has been found in most previous surveys, the majority (65 per cent) of the luminous FRIIs (FRII-High, $L_{150}>10^{26}$ W Hz$^{-1}$) in our sample are LERGs. We attribute this difference to the impact of higher flux density limits on selection of older samples. 
    \item Using our large samples of luminous, well-resolved FRII HERGs and LERGs we identify no significant morphological differences between 100-kpc-scale FRIIs of different accretion modes, demonstrating that the FR class is not primarily controlled by the central engine. While the quality of our data does not allow us to rule out kpc- and/or pc- scale differences in the jet structures of FRII LERGs and HERGs, we consider it unlikely that such differences would lead to no systematic difference in the large-scale brightness distributions. 
    \item As in our earlier work, a significant population of low-luminosity FRIIs (FRII-Low, $L_{150}\leq 10^{26}$ W Hz$^{-1}$) is present. Even accounting for the large scatter in relating luminosity to jet power, this demonstrates that FR class is also not controlled solely by jet power.
    \item The overwhelming majority of low-luminosity FRIIs ($>91$ per cent) and FRIs ($>95$ per cent) are LERGs, demonstrating that radiatively efficient accretion in low-power sources is very rare, regardless of their FR class.
    \item By examining FRIs and low-luminosity FRIIs in the luminosity range where the populations most overlap, we demonstrate conclusively that the probability of a low-power jet becoming an FRI or an FRII depends strongly on the host galaxy stellar mass, consistent with the kpc-scale jet disruption model for FRIs.
    \item Accretion mode is not closely linked to stellar mass in our sample, but across all morphologies and luminosities HERGs are found at high specific star formation rates, demonstrating a close link between fuel availability and accretion behaviour.
\end{itemize}

FRIIs play a larger role in the low-power and low-accretion rate AGN population than previously anticipated. To determine whether we need to adjust our AGN feedback recipes for large-scale cosmological simulations to account for this we will need to consider the luminosity functions of low-power and low-accretion rate AGN (e.g. Kondapally et al., subm.), energetic and particle content constraints on different FR populations \citep[e.g.][]{Hardcastle2013,Croston2005,Croston2018}, and the duty cycles of these sources \citep[e.g][]{Brienza2017,Sabater2019,Jurlin2020,Shabala2020}. We are now also moving into a regime in which we have more than just morphological (FR) and accretion (LERG/HERG) information to allow us to investigate the underlying physics of these sources, thanks to the increased depth and resolution of polarisation surveys \citep[e.g.][]{OSullivan2017,Mahatma2021}.

Given that LoTSS-Deep \citep[as well as the MeerKAT surveys, see e.g.][]{MIGHTEE2016,Fanaroff2021} is representative of what future all-sky radio surveys with the SKA will accomplish in terms of depth, achieving good classifications for SKA data with our simple approach would be challenging, as visually inspecting millions of sources is not desirable. It is likely that machine learning approaches will suffer similarly, but training using both the visually-filtered results from LoTSS and LoTSS-Deep as truth sets might mitigate this. Incorporating intensity profile methods such as that described by \citet{Barkus2022} could produce a valuable initial filter to minimise the number of sources that need to be visually inspected, and provide some direction to machine learning methods.


\section*{Acknowledgements}\label{sec:Acknowledgements}

We thank the anonymous referee for their comments, which have helped improve the paper. BM and JHC acknowledge support from the Science and Technology Facilities Council (STFC) under grants ST/R00109X/1, ST/R000794/1, and ST/T000295/1. PNB and JS are grateful for support from the UK STFC via grant ST/R000972/1. PNB is also grateful for support via grant ST/V000594/1. KJD acknowledges funding from the European Union’s Horizon 2020 research and innovation programme under the Marie Sk\l{}odowska-Curie grant agreement No. 892117 (HIZRAD). MJH and JP acknowledge support from STFC [ST/V000624/1]. RK acknowledges support from the Science and Technology Facilities Council (STFC) through an STFC studentship via grant ST/R504737/1. IP and MB acknowledge support from INAF under the SKA/CTA PRIN `FORECaST' and the PRIN MAIN STREAM `SAuROS' projects. WLW acknowledges support from the CAS-NWO programme for radio astronomy with project number 629.001.024, which is financed by the Netherlands Organisation for Scientific Research (NWO). MB also acknowledges support from the Ministero degli Affari Esteri e della Cooperazione Internazionale - Direzione Generale per la Promozione del Sistema Paese Progetto di Grande Rilevanza ZA18GR02. 

This paper is based (in part) on data obtained with the International LOFAR Telescope (ILT) under project codes LC0 015, LC2 024, LC2 038, LC3 008, LC4 008, LC4 034, and LT10 01. LOFAR \citep{vanHaarlem2013} is the LOw Frequency ARray designed and constructed by ASTRON. It has observing, data processing, and data storage facilities in several countries, which are owned by various parties (each with their own funding sources), and are collectively operated by the International LOFAR Telescope (ILT) foundation under a joint scientific policy. The ILT resources have benefited from the following recent major funding sources: CNRS-INSU, Observatoire de Paris and Universit{\'e} d'Orl{\'e}ans, France; BMBF, MIWF-NRW, MPG, Germany; Science Foundation Ireland (SFI), Department of Business, Enterprise and Innovation (DBEI), Ireland; NWO, The Netherlands; the Science and Technology Facilities Council, UK; Ministry of Science and Higher Education, Poland. 
Part of this work was carried out on the Dutch national e-infrastructure with the support of the SURF Cooperative through grant e-infra 160022 \& 160152. The LOFAR software and dedicated reduction packages on \url{https://github.com/apmechev/GRID_LRT} were deployed on the e-infrastructure by the LOFAR einfragroup, consisting of J. B. R. Oonk (ASTRON \& Leiden Observatory), A. P. Mechev (Leiden Observatory) and T. Shimwell (ASTRON) with support from N. Danezi (SURFsara) and C. Schrijvers (SURFsara). 

Part of this research made use of the Dutch national e-infrastructure with support of the SURF Cooperative (e-infra 180169) and the LOFAR e-infra group. The J{\"u}lich LOFAR Long Term Archive and the German LOFAR network are both coordinated and operated by the J{\"u}lich Supercomputing Centre (JSC), and computing resources on the Supercomputer JUWELS at JSC were provided by the Gauss Centre for Supercomputing e.V. (grant CHTB00) through the John von Neumann Institute for Computing (NIC). This research has made use of the University of Hertfordshire high-performance computing facility (\url{http://uhhpc.herts.ac.uk/}) and the LOFAR-UK computing facility located at the University of Hertfordshire and supported by STFC [ST/P000096/1]. 

This research made use of \textsc{Astropy}, a community-developed core \textsc{Python} package for astronomy \citep{astropy2} hosted at \url{http://www.astropy.org/}, of \textsc{Matplotlib} \citep{Matplotlib} hosted at \url{https://matplotlib.org/}, \textsc{NumPy} \citep{numpyArray,NumPy} hosted at \url{https://numpy.org/}, \textsc{IPython} \citep{IPython} hosted at \url{https://ipython.org/}, and of \textsc{TOPCAT} \citep{Taylor2005} hosted at \url{http://www.star.bris.ac.uk/~mbt/topcat/}. 

This publication made use of data products from the Widefield Infrared Survey Explorer, which is a joint project of the University of California, Los Angeles, and the Jet Propulsion Laboratory/California Institute of Technology, and NEOWISE, which is a project of the Jet Propulsion Laboratory/California Institute of Technology. \textit{WISE} and NEOWISE are funded by the National Aeronautics and Space Administration. 

The Pan-STARRS1 Surveys (PS1) have been made possible through contributions by the Institute for Astronomy, the University of Hawaii, the Pan-STARRS Project Office, the Max-Planck Society and its participating institutes, the Max Planck Institute for Astronomy, Heidelberg and the Max Planck Institute for Extraterrestrial Physics, Garching, The Johns Hopkins University, Durham University, the University of Edinburgh, the Queen's University Belfast, the Harvard-Smithsonian Center for Astrophysics, the Las Cumbres Observatory Global Telescope Network Incorporated, the National Central University of Taiwan, the Space Telescope Science Institute, and the National Aeronautics and Space Administration under Grant No. NNX08AR22G issued through the Planetary Science Division of the NASA Science Mission Directorate, the National Science Foundation Grant No. AST-1238877, the University of Maryland, Eotvos Lorand University (ELTE), the Los Alamos National Laboratory, and the Gordon and Betty Moore Foundation.

This work has made use of data from the European Space Agency (ESA) mission Gaia (\url{https://www.cosmos.esa.int/gaia}), processed by the Gaia Data Processing and Analysis Consortium (DPAC, \url{https://www.cosmos.esa.int/web/gaia/dpac/consortium}). Funding for the DPAC has been provided by national institutions, in particular the institutions participating in the Gaia Multilateral Agreement. 

This work is based on observations obtained with MegaPrime/MegaCam, a joint project of CFHT and CEA/DAPNIA, at the Canada-France-Hawaii Telescope (CFHT) which is operated by the National Research Council (NRC) of Canada, the Institut National des Sciences de l’Univers of the Centre National de la Recherche Scientifique (CNRS) of France, and the University of Hawaii. This research used the facilities of the Canadian Astronomy Data Centre operated by the National Research Council of Canada with the support of the Canadian Space Agency. RCSLenS data processing was made possible thanks to significant computing support from the NSERC Research Tools and Instruments grant program. This work is based in part on observations made with the \textit{Spitzer} Space Telescope, which was operated by the Jet Propulsion Laboratory, California Institute of Technology under a contract with NASA.

\textit{Herschel} is an ESA space observatory with science instruments provided by European-led Principal Investigator consortia and with important participation from NASA. This research has made use of data from the HerMES project. HerMES is a \textit{Herschel} Key Programme utilising Guaranteed Time from the SPIRE instrument team, ESAC scientists and a mission scientist. The HerMES data was accessed through the \textit{Herschel} Database in Marseille (HeDaM -- \url{http://hedam.lam.fr}) operated by CeSAM and hosted by the Laboratoire d’Astrophysique de Marseille. This work is based in part on observations made with the Galaxy Evolution Explorer (\textit{GALEX}). \textit{GALEX} is a NASA Small Explorer,
launched in 2003 April. We gratefully acknowledge NASA’s support for construction, operation, and science analysis for the \textit{GALEX} mission, developed in cooperation with the Centre National d’Etudes Spatiales of France and the Korean Ministry of Science and Technology.

\section*{Data availability}\label{sec:Data_sources}

The datasets used for the analysis in this paper, and a catalogue listing the morphological and accretion mode classes, are publicly available from the LOFAR surveys website at \url{https://lofar-surveys.org/deepfields.html} and can be freely accessed. The radio morphology classification code, \textsc{LoMorph}, is publicly available on Github, at \url{https://github.com/bmingo/LoMorph/}.



\bibliographystyle{mnras}
\bibliography{bmingo}

\begin{thebibliography}{}
\makeatletter
\relax
\def\mn@urlcharsother{\let\do\@makeother \do\$\do\&\do\#\do\^\do\_\do\%\do\~}
\def\mn@doi{\begingroup\mn@urlcharsother \@ifnextchar [ {\mn@doi@}
  {\mn@doi@[]}}
\def\mn@doi@[#1]#2{\def\@tempa{#1}\ifx\@tempa\@empty \href
  {http://dx.doi.org/#2} {doi:#2}\else \href {http://dx.doi.org/#2} {#1}\fi
  \endgroup}
\def\mn@eprint#1#2{\mn@eprint@#1:#2::\@nil}
\def\mn@eprint@arXiv#1{\href {http://arxiv.org/abs/#1} {{\tt arXiv:#1}}}
\def\mn@eprint@dblp#1{\href {http://dblp.uni-trier.de/rec/bibtex/#1.xml}
  {dblp:#1}}
\def\mn@eprint@#1:#2:#3:#4\@nil{\def\@tempa {#1}\def\@tempb {#2}\def\@tempc
  {#3}\ifx \@tempc \@empty \let \@tempc \@tempb \let \@tempb \@tempa \fi \ifx
  \@tempb \@empty \def\@tempb {arXiv}\fi \@ifundefined
  {mn@eprint@\@tempb}{\@tempb:\@tempc}{\expandafter \expandafter \csname
  mn@eprint@\@tempb\endcsname \expandafter{\@tempc}}}

\bibitem[\protect\citeauthoryear{{Abramson}, {Kelson}, {Dressler}, {Poggianti},
  {Gladders}, {Oemler}  \& {Vulcani}}{{Abramson} et~al.}{2014}]{Abramson2014}
{Abramson} L.~E.,  {Kelson} D.~D.,  {Dressler} A.,  {Poggianti} B.,  {Gladders}
  M.~D.,  {Oemler} Augustus J.,   {Vulcani} B.,  2014, \mn@doi [\apjl]
  {10.1088/2041-8205/785/2/L36}, \href
  {https://ui.adsabs.harvard.edu/abs/2014ApJ...785L..36A} {785, L36}

\bibitem[\protect\citeauthoryear{{Ahumada} et~al.,}{{Ahumada}
  et~al.}{2020}]{Ahumada2020}
{Ahumada} R.,  et~al., 2020, \mn@doi [\apjs] {10.3847/1538-4365/ab929e}, \href
  {https://ui.adsabs.harvard.edu/abs/2020ApJS..249....3A} {249, 3}

\bibitem[\protect\citeauthoryear{{Aihara} et~al.,}{{Aihara}
  et~al.}{2018}]{Aihara2018}
{Aihara} H.,  et~al., 2018, \mn@doi [\pasj] {10.1093/pasj/psx066}, \href
  {https://ui.adsabs.harvard.edu/abs/2018PASJ...70S...4A} {70, S4}

\bibitem[\protect\citeauthoryear{{Angl{\'e}s-Castillo}, {Perucho}, {Mart{\'\i}}
   \& {Laing}}{{Angl{\'e}s-Castillo} et~al.}{2021}]{AnglesCastillo2021}
{Angl{\'e}s-Castillo} A.,  {Perucho} M.,  {Mart{\'\i}} J.~M.,   {Laing} R.~A.,
  2021, \mn@doi [\mnras] {10.1093/mnras/staa3291}, \href
  {https://ui.adsabs.harvard.edu/abs/2021MNRAS.500.1512A} {500, 1512}

\bibitem[\protect\citeauthoryear{{Arp}, {Burbidge}  \& {Burbidge}}{{Arp}
  et~al.}{2004}]{Arp2004}
{Arp} H.,  {Burbidge} E.~M.,   {Burbidge} G.,  2004, \mn@doi [\aap]
  {10.1051/0004-6361:20031745}, \href
  {https://ui.adsabs.harvard.edu/abs/2004A&A...414L..37A} {414, L37}

\bibitem[\protect\citeauthoryear{{Assef} et~al.,}{{Assef}
  et~al.}{2013}]{Assef2013}
{Assef} R.~J.,  et~al., 2013, \mn@doi [\apj] {10.1088/0004-637X/772/1/26},
  \href {http://cdsads.u-strasbg.fr/abs/2013ApJ...772...26A} {772, 26}

\bibitem[\protect\citeauthoryear{{Astropy Collaboration} et~al.,}{{Astropy
  Collaboration} et~al.}{2018}]{astropy2}
{Astropy Collaboration} et~al., 2018, \mn@doi [\aj] {10.3847/1538-3881/aabc4f},
  \href {https://ui.adsabs.harvard.edu/\#abs/2018AJ....156..123A} {156, 123}

\bibitem[\protect\citeauthoryear{{Baldi} \& {Capetti}}{{Baldi} \&
  {Capetti}}{2008}]{Baldi2008}
{Baldi} R.~D.,  {Capetti} A.,  2008, \mn@doi [\aap]
  {10.1051/0004-6361:20078745}, \href
  {https://ui.adsabs.harvard.edu/abs/2008A&A...489..989B} {489, 989}

\bibitem[\protect\citeauthoryear{{Baldi} et~al.,}{{Baldi}
  et~al.}{2010}]{Baldi2010}
{Baldi} R.~D.,  et~al., 2010, \mn@doi [\apj] {10.1088/0004-637X/725/2/2426},
  \href {https://ui.adsabs.harvard.edu/abs/2010ApJ...725.2426B} {725, 2426}

\bibitem[\protect\citeauthoryear{{Baldi}, {Capetti}  \& {Giovannini}}{{Baldi}
  et~al.}{2015}]{Baldi2015}
{Baldi} R.~D.,  {Capetti} A.,   {Giovannini} G.,  2015, \mn@doi [\aap]
  {10.1051/0004-6361/201425426}, \href
  {https://ui.adsabs.harvard.edu/abs/2015A&A...576A..38B} {576, A38}

\bibitem[\protect\citeauthoryear{{Barkus} et~al.,}{{Barkus}
  et~al.}{2022}]{Barkus2022}
{Barkus} B.,  et~al., 2022, \mn@doi [\mnras] {10.1093/mnras/stab2952}, \href
  {https://ui.adsabs.harvard.edu/abs/2022MNRAS.509....1B} {509, 1}

\bibitem[\protect\citeauthoryear{{Becker}, {White}  \& {Helfand}}{{Becker}
  et~al.}{1995}]{FIRST1995}
{Becker} R.~H.,  {White} R.~L.,   {Helfand} D.~J.,  1995, \mn@doi [\apj]
  {10.1086/176166}, \href {http://adsabs.harvard.edu/abs/1995ApJ...450..559B}
  {450, 559}

\bibitem[\protect\citeauthoryear{{Best} \& {Heckman}}{{Best} \&
  {Heckman}}{2012}]{Best2012}
{Best} P.~N.,  {Heckman} T.~M.,  2012, \mn@doi [\mnras]
  {10.1111/j.1365-2966.2012.20414.x}, \href
  {http://cdsads.u-strasbg.fr/abs/2012MNRAS.421.1569B} {421, 1569}

\bibitem[\protect\citeauthoryear{{Best}, {Ker}, {Simpson}, {Rigby}  \&
  {Sabater}}{{Best} et~al.}{2014}]{Best2014}
{Best} P.~N.,  {Ker} L.~M.,  {Simpson} C.,  {Rigby} E.~E.,   {Sabater} J.,
  2014, \mn@doi [\mnras] {10.1093/mnras/stu1776}, \href
  {http://cdsads.u-strasbg.fr/abs/2014MNRAS.445..955B} {445, 955}

\bibitem[\protect\citeauthoryear{{Bicknell}}{{Bicknell}}{1994}]{Bicknell1994}
{Bicknell} G.~V.,  1994, \mn@doi [\apj] {10.1086/173748}, \href
  {http://adsabs.harvard.edu/abs/1994ApJ...422..542B} {422, 542}

\bibitem[\protect\citeauthoryear{{Blandford}, {Meier}  \&
  {Readhead}}{{Blandford} et~al.}{2019}]{Blandford2019}
{Blandford} R.,  {Meier} D.,   {Readhead} A.,  2019, \mn@doi [\araa]
  {10.1146/annurev-astro-081817-051948}, \href
  {https://ui.adsabs.harvard.edu/abs/2019ARA&A..57..467B} {57, 467}

\bibitem[\protect\citeauthoryear{{Blanton} et~al.,}{{Blanton}
  et~al.}{2017}]{Blanton2017}
{Blanton} M.~R.,  et~al., 2017, \mn@doi [\aj] {10.3847/1538-3881/aa7567}, \href
  {https://ui.adsabs.harvard.edu/abs/2017AJ....154...28B} {154, 28}

\bibitem[\protect\citeauthoryear{{Brienza} et~al.,}{{Brienza}
  et~al.}{2017}]{Brienza2017}
{Brienza} M.,  et~al., 2017, \mn@doi [\aap] {10.1051/0004-6361/201730932},
  \href {http://adsabs.harvard.edu/abs/2017A%26A...606A..98B} {606, A98}

\bibitem[\protect\citeauthoryear{{Burgarella}, {Buat}  \&
  {Iglesias-P{\'a}ramo}}{{Burgarella} et~al.}{2005}]{cigale2005}
{Burgarella} D.,  {Buat} V.,   {Iglesias-P{\'a}ramo} J.,  2005, \mn@doi
  [\mnras] {10.1111/j.1365-2966.2005.09131.x}, \href
  {https://ui.adsabs.harvard.edu/abs/2005MNRAS.360.1413B} {360, 1413}

\bibitem[\protect\citeauthoryear{{Calistro Rivera}, {Lusso}, {Hennawi}  \&
  {Hogg}}{{Calistro Rivera} et~al.}{2016}]{agnfitter2016}
{Calistro Rivera} G.,  {Lusso} E.,  {Hennawi} J.~F.,   {Hogg} D.~W.,  2016,
  \mn@doi [\apj] {10.3847/1538-4357/833/1/98}, \href
  {https://ui.adsabs.harvard.edu/abs/2016ApJ...833...98C} {833, 98}

\bibitem[\protect\citeauthoryear{{Capetti}, {Massaro}  \& {Baldi}}{{Capetti}
  et~al.}{2017a}]{Capetti2017FRI}
{Capetti} A.,  {Massaro} F.,   {Baldi} R.~D.,  2017a, \mn@doi [\aap]
  {10.1051/0004-6361/201629287}, \href
  {https://ui.adsabs.harvard.edu/abs/2017A&A...598A..49C} {598, A49}

\bibitem[\protect\citeauthoryear{{Capetti}, {Massaro}  \& {Baldi}}{{Capetti}
  et~al.}{2017b}]{Capetti2017FRII}
{Capetti} A.,  {Massaro} F.,   {Baldi} R.~D.,  2017b, \mn@doi [\aap]
  {10.1051/0004-6361/201630247}, \href
  {https://ui.adsabs.harvard.edu/abs/2017A%26A...601A..81C} {601, A81}

\bibitem[\protect\citeauthoryear{{Carnall}, {McLure}, {Dunlop}  \&
  {Dav{\'e}}}{{Carnall} et~al.}{2018}]{bagpipes2018}
{Carnall} A.~C.,  {McLure} R.~J.,  {Dunlop} J.~S.,   {Dav{\'e}} R.,  2018,
  \mn@doi [\mnras] {10.1093/mnras/sty2169}, \href
  {https://ui.adsabs.harvard.edu/abs/2018MNRAS.480.4379C} {480, 4379}

\bibitem[\protect\citeauthoryear{{Carnall} et~al.,}{{Carnall}
  et~al.}{2019}]{bagpipes2019}
{Carnall} A.~C.,  et~al., 2019, \mn@doi [\mnras] {10.1093/mnras/stz2544}, \href
  {https://ui.adsabs.harvard.edu/abs/2019MNRAS.490..417C} {490, 417}

\bibitem[\protect\citeauthoryear{{Chhetri}, {Kimball}, {Ekers}, {Mahony},
  {Sadler}  \& {Jarrett}}{{Chhetri} et~al.}{2020}]{Chhetri2020}
{Chhetri} R.,  {Kimball} A.,  {Ekers} R.~D.,  {Mahony} E.~K.,  {Sadler} E.~M.,
   {Jarrett} T.,  2020, \mn@doi [\mnras] {10.1093/mnras/staa513}, \href
  {https://ui.adsabs.harvard.edu/abs/2020MNRAS.494..923C} {494, 923}

\bibitem[\protect\citeauthoryear{{Chiaberge}, {Capetti}  \&
  {Celotti}}{{Chiaberge} et~al.}{2000}]{Chiaberge2000}
{Chiaberge} M.,  {Capetti} A.,   {Celotti} A.,  2000, \aap, \href
  {https://ui.adsabs.harvard.edu/abs/2000A&A...355..873C} {355, 873}

\bibitem[\protect\citeauthoryear{{Croston}, {Hardcastle}, {Harris}, {Belsole},
  {Birkinshaw}  \& {Worrall}}{{Croston} et~al.}{2005}]{Croston2005}
{Croston} J.~H.,  {Hardcastle} M.~J.,  {Harris} D.~E.,  {Belsole} E.,
  {Birkinshaw} M.,   {Worrall} D.~M.,  2005, \mn@doi [\apj] {10.1086/430170},
  626, 733

\bibitem[\protect\citeauthoryear{{Croston}, {Ineson}  \&
  {Hardcastle}}{{Croston} et~al.}{2018}]{Croston2018}
{Croston} J.~H.,  {Ineson} J.,   {Hardcastle} M.~J.,  2018, \mn@doi [\mnras]
  {10.1093/mnras/sty274}, \href
  {http://adsabs.harvard.edu/abs/2018MNRAS.476.1614C} {476, 1614}

\bibitem[\protect\citeauthoryear{{Croston} et~al.,}{{Croston}
  et~al.}{2019}]{Croston2019}
{Croston} J.~H.,  et~al., 2019, \mn@doi [\aap] {10.1051/0004-6361/201834019},
  \href {http://adsabs.harvard.edu/abs/2019A%26A...622A..10C} {622, A10}

\bibitem[\protect\citeauthoryear{{Cutri}}{{Cutri}}{2014}]{AllWISE2014}
{Cutri} R.~M. e.~a.,  2014, VizieR Online Data Catalog, \href
  {https://ui.adsabs.harvard.edu/abs/2014yCat.2328....0C} {p. II/328}

\bibitem[\protect\citeauthoryear{{Dabhade}, {Gaikwad}, {Bagchi},
  {Pandey-Pommier}, {Sankhyayan}  \& {Raychaudhury}}{{Dabhade}
  et~al.}{2017}]{Dabhade2017}
{Dabhade} P.,  {Gaikwad} M.,  {Bagchi} J.,  {Pandey-Pommier} M.,  {Sankhyayan}
  S.,   {Raychaudhury} S.,  2017, \mn@doi [\mnras] {10.1093/mnras/stx860},
  \href {https://ui.adsabs.harvard.edu/abs/2017MNRAS.469.2886D} {469, 2886}

\bibitem[\protect\citeauthoryear{{Dabhade} et~al.,}{{Dabhade}
  et~al.}{2020a}]{Dabhade2020}
{Dabhade} P.,  et~al., 2020a, \mn@doi [\aap] {10.1051/0004-6361/201935589},
  \href {https://ui.adsabs.harvard.edu/abs/2020A&A...635A...5D} {635, A5}

\bibitem[\protect\citeauthoryear{{Dabhade} et~al.,}{{Dabhade}
  et~al.}{2020b}]{Dabhade2020b}
{Dabhade} P.,  et~al., 2020b, \mn@doi [\aap] {10.1051/0004-6361/202038344},
  \href {https://ui.adsabs.harvard.edu/abs/2020A&A...642A.153D} {642, A153}

\bibitem[\protect\citeauthoryear{{Damen}, {Labb{\'e}}, {Franx}, {van Dokkum},
  {Taylor}  \& {Gawiser}}{{Damen} et~al.}{2009}]{Damen09}
{Damen} M.,  {Labb{\'e}} I.,  {Franx} M.,  {van Dokkum} P.~G.,  {Taylor} E.~N.,
    {Gawiser} E.~J.,  2009, \mn@doi [\apj] {10.1088/0004-637X/690/1/937}, \href
  {https://ui.adsabs.harvard.edu/abs/2009ApJ...690..937D} {690, 937}

\bibitem[\protect\citeauthoryear{{Dewdney}, {Hall}, {Schilizzi}  \&
  {Lazio}}{{Dewdney} et~al.}{2009}]{SKA2009}
{Dewdney} P.~E.,  {Hall} P.~J.,  {Schilizzi} R.~T.,   {Lazio} T.~J.~L.~W.,
  2009, \mn@doi [IEEE Proceedings] {10.1109/JPROC.2009.2021005}, \href
  {https://ui.adsabs.harvard.edu/abs/2009IEEEP..97.1482D} {97, 1482}

\bibitem[\protect\citeauthoryear{{Donley} et~al.,}{{Donley}
  et~al.}{2012}]{Donley2012}
{Donley} J.~L.,  et~al., 2012, \mn@doi [\apj] {10.1088/0004-637X/748/2/142},
  \href {https://ui.adsabs.harvard.edu/abs/2012ApJ...748..142D} {748, 142}

\bibitem[\protect\citeauthoryear{{Donoso}, {Best}  \& {Kauffmann}}{{Donoso}
  et~al.}{2009}]{Donoso2009}
{Donoso} E.,  {Best} P.~N.,   {Kauffmann} G.,  2009, \mn@doi [\mnras]
  {10.1111/j.1365-2966.2008.14068.x}, \href
  {http://cdsads.u-strasbg.fr/abs/2009MNRAS.392..617D} {392, 617}

\bibitem[\protect\citeauthoryear{{Drouart} et~al.,}{{Drouart}
  et~al.}{2014}]{Drouart2014}
{Drouart} G.,  et~al., 2014, \mn@doi [\aap] {10.1051/0004-6361/201323310},
  \href {https://ui.adsabs.harvard.edu/abs/2014A&A...566A..53D} {566, A53}

\bibitem[\protect\citeauthoryear{{Duncan} et~al.,}{{Duncan}
  et~al.}{2019}]{Duncan2019}
{Duncan} K.~J.,  et~al., 2019, \mn@doi [\aap] {10.1051/0004-6361/201833562},
  \href {http://adsabs.harvard.edu/abs/2019A%26A...622A...3D} {622, A3}

\bibitem[\protect\citeauthoryear{{Duncan} et~al.,}{{Duncan}
  et~al.}{2021}]{Duncan2021}
{Duncan} K.~J.,  et~al., 2021, \mn@doi [\aap] {10.1051/0004-6361/202038809},
  \href {https://ui.adsabs.harvard.edu/abs/2021A&A...648A...4D} {648, A4}

\bibitem[\protect\citeauthoryear{{Falkendal} et~al.,}{{Falkendal}
  et~al.}{2019}]{Falkendal2019}
{Falkendal} T.,  et~al., 2019, \mn@doi [\aap] {10.1051/0004-6361/201732485},
  \href {https://ui.adsabs.harvard.edu/abs/2019A&A...621A..27F} {621, A27}

\bibitem[\protect\citeauthoryear{{Fanaroff} \& {Riley}}{{Fanaroff} \&
  {Riley}}{1974}]{FR1974}
{Fanaroff} B.~L.,  {Riley} J.~M.,  1974, \mnras, \href
  {http://adsabs.harvard.edu/abs/1974MNRAS.167P..31F} {167, 31P}

\bibitem[\protect\citeauthoryear{{Fanaroff} et~al.,}{{Fanaroff}
  et~al.}{2021}]{Fanaroff2021}
{Fanaroff} B.,  et~al., 2021, \mn@doi [\mnras] {10.1093/mnras/stab1540}, \href
  {https://ui.adsabs.harvard.edu/abs/2021MNRAS.505.6003F} {505, 6003}

\bibitem[\protect\citeauthoryear{{Fawcett}, {Alexander}, {Rosario}, {Klindt},
  {Fotopoulou}, {Lusso}, {Morabito}  \& {Calistro Rivera}}{{Fawcett}
  et~al.}{2020}]{Fawcett2020}
{Fawcett} V.~A.,  {Alexander} D.~M.,  {Rosario} D.~J.,  {Klindt} L.,
  {Fotopoulou} S.,  {Lusso} E.,  {Morabito} L.~K.,   {Calistro Rivera} G.,
  2020, \mn@doi [\mnras] {10.1093/mnras/staa954}, \href
  {https://ui.adsabs.harvard.edu/abs/2020MNRAS.494.4802F} {494, 4802}

\bibitem[\protect\citeauthoryear{{Fazio} et~al.,}{{Fazio}
  et~al.}{2004}]{SpitzerIRAC2004}
{Fazio} G.~G.,  et~al., 2004, \mn@doi [\apjs] {10.1086/422843}, \href
  {https://ui.adsabs.harvard.edu/abs/2004ApJS..154...10F} {154, 10}

\bibitem[\protect\citeauthoryear{{Fender} \& {Gallo}}{{Fender} \&
  {Gallo}}{2014}]{Fender2014}
{Fender} R.,  {Gallo} E.,  2014, \mn@doi [\ssr] {10.1007/s11214-014-0069-z},
  \href {http://adsabs.harvard.edu/abs/2014SSRv..183..323F} {183, 323}

\bibitem[\protect\citeauthoryear{{Fernandes} et~al.,}{{Fernandes}
  et~al.}{2015}]{Fernandes2015}
{Fernandes} C.~A.~C.,  et~al., 2015, \mn@doi [\mnras] {10.1093/mnras/stu2517},
  \href {https://ui.adsabs.harvard.edu/abs/2015MNRAS.447.1184F} {447, 1184}

\bibitem[\protect\citeauthoryear{{Gaspari}, {Ruszkowski}  \& {Oh}}{{Gaspari}
  et~al.}{2013}]{Gaspari2013}
{Gaspari} M.,  {Ruszkowski} M.,   {Oh} S.~P.,  2013, \mn@doi [\mnras]
  {10.1093/mnras/stt692}, \href
  {https://ui.adsabs.harvard.edu/abs/2013MNRAS.432.3401G} {432, 3401}

\bibitem[\protect\citeauthoryear{{Gaspari}, {Brighenti}  \& {Temi}}{{Gaspari}
  et~al.}{2015}]{Gaspari2015}
{Gaspari} M.,  {Brighenti} F.,   {Temi} P.,  2015, \mn@doi [\aap]
  {10.1051/0004-6361/201526151}, \href
  {https://ui.adsabs.harvard.edu/abs/2015A&A...579A..62G} {579, A62}

\bibitem[\protect\citeauthoryear{{Gendre}, {Best}  \& {Wall}}{{Gendre}
  et~al.}{2010}]{Gendre2010}
{Gendre} M.~A.,  {Best} P.~N.,   {Wall} J.~V.,  2010, \mn@doi [\mnras]
  {10.1111/j.1365-2966.2010.16413.x}, \href
  {https://ui.adsabs.harvard.edu/abs/2010MNRAS.404.1719G} {404, 1719}

\bibitem[\protect\citeauthoryear{{Gendre}, {Best}, {Wall}  \& {Ker}}{{Gendre}
  et~al.}{2013}]{Gendre2013}
{Gendre} M.~A.,  {Best} P.~N.,  {Wall} J.~V.,   {Ker} L.~M.,  2013, \mn@doi
  [\mnras] {10.1093/mnras/stt116}, \href
  {http://adsabs.harvard.edu/abs/2013MNRAS.430.3086G} {430, 3086}

\bibitem[\protect\citeauthoryear{{Grandi}, {Torresi}, {Macconi}, {Boccardi}  \&
  {Capetti}}{{Grandi} et~al.}{2021}]{Grandi2021}
{Grandi} P.,  {Torresi} E.,  {Macconi} D.,  {Boccardi} B.,   {Capetti} A.,
  2021, \mn@doi [\apj] {10.3847/1538-4357/abe776}, \href
  {https://ui.adsabs.harvard.edu/abs/2021ApJ...911...17G} {911, 17}

\bibitem[\protect\citeauthoryear{{G{\"u}rkan}, {Hardcastle}  \&
  {Jarvis}}{{G{\"u}rkan} et~al.}{2014}]{Gurkan2014}
{G{\"u}rkan} G.,  {Hardcastle} M.~J.,   {Jarvis} M.~J.,  2014, \mn@doi [\mnras]
  {10.1093/mnras/stt2264}, \href
  {http://cdsads.u-strasbg.fr/abs/2014MNRAS.438.1149G} {438, 1149}

\bibitem[\protect\citeauthoryear{{G{\"u}rkan} et~al.,}{{G{\"u}rkan}
  et~al.}{2018}]{Gurkan2018a}
{G{\"u}rkan} G.,  et~al., 2018, \mn@doi [\mnras] {10.1093/mnras/sty016}, \href
  {http://adsabs.harvard.edu/abs/2018MNRAS.475.3010G} {475, 3010}

\bibitem[\protect\citeauthoryear{{G{\"u}rkan} et~al.,}{{G{\"u}rkan}
  et~al.}{2019}]{Gurkan2019}
{G{\"u}rkan} G.,  et~al., 2019, \mn@doi [\aap] {10.1051/0004-6361/201833892},
  \href {http://adsabs.harvard.edu/abs/2019A%26A...622A..11G} {622, A11}

\bibitem[\protect\citeauthoryear{{Hardcastle}}{{Hardcastle}}{2018a}]{Hardcastle2018b}
{Hardcastle} M.,  2018a, \mn@doi [Nature Astronomy]
  {10.1038/s41550-018-0424-1}, \href
  {http://adsabs.harvard.edu/abs/2018NatAs...2..273H} {2, 273}

\bibitem[\protect\citeauthoryear{{Hardcastle}}{{Hardcastle}}{2018b}]{Hardcastle2018}
{Hardcastle} M.~J.,  2018b, \mn@doi [\mnras] {10.1093/mnras/stx3358}, \href
  {http://adsabs.harvard.edu/abs/2018MNRAS.475.2768H} {475, 2768}

\bibitem[\protect\citeauthoryear{{Hardcastle} \& {Croston}}{{Hardcastle} \&
  {Croston}}{2020}]{HardcastleCroston2020}
{Hardcastle} M.~J.,  {Croston} J.~H.,  2020, \mn@doi [\nar]
  {10.1016/j.newar.2020.101539}, \href
  {https://ui.adsabs.harvard.edu/abs/2020NewAR..8801539H} {88, 101539}

\bibitem[\protect\citeauthoryear{{Hardcastle} \& {Krause}}{{Hardcastle} \&
  {Krause}}{2013}]{Hardcastle2013}
{Hardcastle} M.~J.,  {Krause} M.~G.~H.,  2013, \mn@doi [\mnras]
  {10.1093/mnras/sts564}, \href
  {http://adsabs.harvard.edu/abs/2013MNRAS.430..174H} {430, 174}

\bibitem[\protect\citeauthoryear{{Hardcastle} \& {Krause}}{{Hardcastle} \&
  {Krause}}{2014}]{Hardcastle2014}
{Hardcastle} M.~J.,  {Krause} M.~G.~H.,  2014, \mn@doi [\mnras]
  {10.1093/mnras/stu1229}, \href
  {http://cdsads.u-strasbg.fr/abs/2014MNRAS.443.1482H} {443, 1482}

\bibitem[\protect\citeauthoryear{{Hardcastle} \& {Looney}}{{Hardcastle} \&
  {Looney}}{2008}]{Hardcastle2008}
{Hardcastle} M.~J.,  {Looney} L.~W.,  2008, \mn@doi [\mnras]
  {10.1111/j.1365-2966.2008.13370.x}, \href
  {http://adsabs.harvard.edu/abs/2008MNRAS.388..176H} {388, 176}

\bibitem[\protect\citeauthoryear{{Hardcastle}, {Harris}, {Worrall}  \&
  {Birkinshaw}}{{Hardcastle} et~al.}{2004}]{Hardcastle2004}
{Hardcastle} M.~J.,  {Harris} D.~E.,  {Worrall} D.~M.,   {Birkinshaw} M.,
  2004, \mn@doi [\apj] {10.1086/422808}, \href
  {http://adsabs.harvard.edu/abs/2004ApJ...612..729H} {612, 729}

\bibitem[\protect\citeauthoryear{{Hardcastle}, {Croston}  \&
  {Kraft}}{{Hardcastle} et~al.}{2007}]{Hardcastle2007}
{Hardcastle} M.~J.,  {Croston} J.~H.,   {Kraft} R.~P.,  2007, \mn@doi [\apj]
  {10.1086/521696}, \href {http://cdsads.u-strasbg.fr/abs/2007ApJ...669..893H}
  {669, 893}

\bibitem[\protect\citeauthoryear{{Hardcastle}, {Evans}  \&
  {Croston}}{{Hardcastle} et~al.}{2009}]{Hardcastle2009}
{Hardcastle} M.~J.,  {Evans} D.~A.,   {Croston} J.~H.,  2009, \mn@doi [\mnras]
  {10.1111/j.1365-2966.2009.14887.x}, \href
  {http://cdsads.u-strasbg.fr/abs/2009MNRAS.396.1929H} {396, 1929}

\bibitem[\protect\citeauthoryear{{Hardcastle} et~al.,}{{Hardcastle}
  et~al.}{2013}]{Hardcastle2013b}
{Hardcastle} M.~J.,  et~al., 2013, \mn@doi [\mnras] {10.1093/mnras/sts510},
  \href {http://cdsads.u-strasbg.fr/abs/2013MNRAS.429.2407H} {429, 2407}

\bibitem[\protect\citeauthoryear{{Hardcastle} et~al.,}{{Hardcastle}
  et~al.}{2019}]{Hardcastle2019}
{Hardcastle} M.~J.,  et~al., 2019, \mn@doi [\aap]
  {10.1051/0004-6361/201833893}, \href
  {http://adsabs.harvard.edu/abs/2019A%26A...622A..12H} {622, A12}

\bibitem[\protect\citeauthoryear{Harris et~al.,}{Harris et~al.}{2020}]{NumPy}
Harris C.~R.,  et~al., 2020, \mn@doi [Nature] {10.1038/s41586-020-2649-2}, 585,
  357

\bibitem[\protect\citeauthoryear{{Harwood}, {Vernstrom}  \& {Stroe}}{{Harwood}
  et~al.}{2020}]{Harwood2020}
{Harwood} J.~J.,  {Vernstrom} T.,   {Stroe} A.,  2020, \mn@doi [\mnras]
  {10.1093/mnras/stz3069}, \href
  {https://ui.adsabs.harvard.edu/abs/2020MNRAS.491..803H} {491, 803}

\bibitem[\protect\citeauthoryear{{Helfand}, {White}  \& {Becker}}{{Helfand}
  et~al.}{2015}]{FIRST2015}
{Helfand} D.~J.,  {White} R.~L.,   {Becker} R.~H.,  2015, \mn@doi [\apj]
  {10.1088/0004-637X/801/1/26}, \href
  {http://adsabs.harvard.edu/abs/2015ApJ...801...26H} {801, 26}

\bibitem[\protect\citeauthoryear{{Hine} \& {Longair}}{{Hine} \&
  {Longair}}{1979}]{Hine1979}
{Hine} R.~G.,  {Longair} M.~S.,  1979, \mnras, \href
  {http://cdsads.u-strasbg.fr/abs/1979MNRAS.188..111H} {188, 111}

\bibitem[\protect\citeauthoryear{{Hunter}}{{Hunter}}{2007}]{Matplotlib}
{Hunter} J.~D.,  2007, \mn@doi [Computing in Science and Engineering]
  {10.1109/MCSE.2007.55}, \href
  {http://adsabs.harvard.edu/abs/2007CSE.....9...90H} {9, 90}

\bibitem[\protect\citeauthoryear{{Hurley} et~al.,}{{Hurley}
  et~al.}{2017}]{Hurley2017}
{Hurley} P.~D.,  et~al., 2017, \mn@doi [\mnras] {10.1093/mnras/stw2375}, \href
  {https://ui.adsabs.harvard.edu/abs/2017MNRAS.464..885H} {464, 885}

\bibitem[\protect\citeauthoryear{{Ilbert} et~al.,}{{Ilbert}
  et~al.}{2015}]{Ilbert2015}
{Ilbert} O.,  et~al., 2015, \mn@doi [\aap] {10.1051/0004-6361/201425176}, \href
  {https://ui.adsabs.harvard.edu/abs/2015A&A...579A...2I} {579, A2}

\bibitem[\protect\citeauthoryear{{Ineson}, {Croston}, {Hardcastle}, {Kraft},
  {Evans}  \& {Jarvis}}{{Ineson} et~al.}{2015}]{Ineson2015}
{Ineson} J.,  {Croston} J.~H.,  {Hardcastle} M.~J.,  {Kraft} R.~P.,  {Evans}
  D.~A.,   {Jarvis} M.,  2015, \mn@doi [\mnras] {10.1093/mnras/stv1807}, \href
  {http://cdsads.u-strasbg.fr/abs/2015MNRAS.453.2682I} {453, 2682}

\bibitem[\protect\citeauthoryear{{Ineson}, {Croston}, {Hardcastle}  \&
  {Mingo}}{{Ineson} et~al.}{2017}]{Ineson2017}
{Ineson} J.,  {Croston} J.~H.,  {Hardcastle} M.~J.,   {Mingo} B.,  2017,
  \mn@doi [\mnras] {10.1093/mnras/stx189}, \href
  {http://adsabs.harvard.edu/abs/2017MNRAS.467.1586I} {467, 1586}

\bibitem[\protect\citeauthoryear{{Ishwara-Chandra} \&
  {Saikia}}{{Ishwara-Chandra} \& {Saikia}}{1999}]{Ishwara-Chandra1999}
{Ishwara-Chandra} C.~H.,  {Saikia} D.~J.,  1999, \mn@doi [\mnras]
  {10.1046/j.1365-8711.1999.02835.x}, \href
  {https://ui.adsabs.harvard.edu/abs/1999MNRAS.309..100I} {309, 100}

\bibitem[\protect\citeauthoryear{{Jackson} \& {Rawlings}}{{Jackson} \&
  {Rawlings}}{1997}]{Jackson1997}
{Jackson} N.,  {Rawlings} S.,  1997, \mn@doi [\mnras]
  {10.1093/mnras/286.1.241}, \href
  {https://ui.adsabs.harvard.edu/abs/1997MNRAS.286..241J} {286, 241}

\bibitem[\protect\citeauthoryear{{Jannuzi} \& {Dey}}{{Jannuzi} \&
  {Dey}}{1999}]{Jannuzi1999}
{Jannuzi} B.~T.,  {Dey} A.,  1999, in {Weymann} R.,  {Storrie-Lombardi} L.,
  {Sawicki} M.,   {Brunner} R.,  eds,  Astronomical Society of the Pacific
  Conference Series Vol. 191, Photometric Redshifts and the Detection of High
  Redshift Galaxies. p.~111

\bibitem[\protect\citeauthoryear{{Janssen}, {R{\"o}ttgering}, {Best}  \&
  {Brinchmann}}{{Janssen} et~al.}{2012}]{Janssen2012}
{Janssen} R.~M.~J.,  {R{\"o}ttgering} H.~J.~A.,  {Best} P.~N.,   {Brinchmann}
  J.,  2012, \mn@doi [\aap] {10.1051/0004-6361/201219052}, \href
  {http://cdsads.u-strasbg.fr/abs/2012A%26A...541A..62J} {541, A62}

\bibitem[\protect\citeauthoryear{{Jarvis} et~al.,}{{Jarvis}
  et~al.}{2016}]{MIGHTEE2016}
{Jarvis} M.,  et~al., 2016, in Proceedings of MeerKAT Science: On the Pathway
  to the SKA. 25-27 May. p.~6 (\mn@eprint {arXiv} {1709.01901})

\bibitem[\protect\citeauthoryear{{Jarvis} et~al.,}{{Jarvis}
  et~al.}{2019}]{Jarvis2019}
{Jarvis} M.~E.,  et~al., 2019, \mn@doi [\mnras] {10.1093/mnras/stz556}, \href
  {https://ui.adsabs.harvard.edu/abs/2019MNRAS.485.2710J} {485, 2710}

\bibitem[\protect\citeauthoryear{{Jimenez-Gallardo} et~al.,}{{Jimenez-Gallardo}
  et~al.}{2019}]{Jimenez-Gallardo2019}
{Jimenez-Gallardo} A.,  et~al., 2019, \mn@doi [\aap]
  {10.1051/0004-6361/201935104}, \href
  {https://ui.adsabs.harvard.edu/abs/2019A&A...627A.108J} {627, A108}

\bibitem[\protect\citeauthoryear{{Jurlin} et~al.,}{{Jurlin}
  et~al.}{2020}]{Jurlin2020}
{Jurlin} N.,  et~al., 2020, \mn@doi [\aap] {10.1051/0004-6361/201936955}, \href
  {https://ui.adsabs.harvard.edu/abs/2020A&A...638A..34J} {638, A34}

\bibitem[\protect\citeauthoryear{{Kaiser} \& {Best}}{{Kaiser} \&
  {Best}}{2007}]{Kaiser2007}
{Kaiser} C.~R.,  {Best} P.~N.,  2007, \mn@doi [\mnras]
  {10.1111/j.1365-2966.2007.12350.x}, \href
  {https://ui.adsabs.harvard.edu/abs/2007MNRAS.381.1548K} {381, 1548}

\bibitem[\protect\citeauthoryear{{Kaiser} et~al.,}{{Kaiser}
  et~al.}{2010}]{PanSTARRS2010}
{Kaiser} N.,  et~al., 2010, in {Stepp} L.~M.,  {Gilmozzi} R.,   {Hall} H.~J.,
  eds,  Society of Photo-Optical Instrumentation Engineers (SPIE) Conference
  Series Vol. 7733, Ground-based and Airborne Telescopes III. p. 77330E,
  \mn@doi{10.1117/12.859188}

\bibitem[\protect\citeauthoryear{{Kapi{\'n}ska}, {Uttley}  \&
  {Kaiser}}{{Kapi{\'n}ska} et~al.}{2012}]{Kapinska2012}
{Kapi{\'n}ska} A.~D.,  {Uttley} P.,   {Kaiser} C.~R.,  2012, \mn@doi [\mnras]
  {10.1111/j.1365-2966.2012.21351.x}, \href
  {http://adsabs.harvard.edu/abs/2012MNRAS.424.2028K} {424, 2028}

\bibitem[\protect\citeauthoryear{{Karouzos} et~al.,}{{Karouzos}
  et~al.}{2014}]{Karouzos2014}
{Karouzos} M.,  et~al., 2014, \mn@doi [\apj] {10.1088/0004-637X/784/2/137},
  \href {https://ui.adsabs.harvard.edu/abs/2014ApJ...784..137K} {784, 137}

\bibitem[\protect\citeauthoryear{{Keenan}, {Meyer}, {Georganopoulos}, {Reddy}
  \& {French}}{{Keenan} et~al.}{2021}]{Keenan2021}
{Keenan} M.,  {Meyer} E.~T.,  {Georganopoulos} M.,  {Reddy} K.,   {French}
  O.~J.,  2021, \mn@doi [\mnras] {10.1093/mnras/stab1182}, \href
  {https://ui.adsabs.harvard.edu/abs/2021MNRAS.505.4726K} {505, 4726}

\bibitem[\protect\citeauthoryear{{Kim} \& {Fabbiano}}{{Kim} \&
  {Fabbiano}}{2013}]{KimFabbiano2013}
{Kim} D.-W.,  {Fabbiano} G.,  2013, \mn@doi [\apj]
  {10.1088/0004-637X/776/2/116}, \href
  {https://ui.adsabs.harvard.edu/abs/2013ApJ...776..116K} {776, 116}

\bibitem[\protect\citeauthoryear{{Kondapally} et~al.,}{{Kondapally}
  et~al.}{2021}]{Kondapally2021}
{Kondapally} R.,  et~al., 2021, \mn@doi [\aap] {10.1051/0004-6361/202038813},
  \href {https://ui.adsabs.harvard.edu/abs/2021A&A...648A...3K} {648, A3}

\bibitem[\protect\citeauthoryear{{Laing} \& {Bridle}}{{Laing} \&
  {Bridle}}{2002}]{Laing2002}
{Laing} R.~A.,  {Bridle} A.~H.,  2002, \mn@doi [\mnras]
  {10.1046/j.1365-8711.2002.05873.x}, \href
  {http://adsabs.harvard.edu/abs/2002MNRAS.336.1161L} {336, 1161}

\bibitem[\protect\citeauthoryear{{Laing} \& {Bridle}}{{Laing} \&
  {Bridle}}{2014}]{LaingBridle2014}
{Laing} R.~A.,  {Bridle} A.~H.,  2014, \mn@doi [\mnras]
  {10.1093/mnras/stt2138}, \href
  {http://adsabs.harvard.edu/abs/2014MNRAS.437.3405L} {437, 3405}

\bibitem[\protect\citeauthoryear{{Laing}, {Riley}  \& {Longair}}{{Laing}
  et~al.}{1983}]{Laing1983}
{Laing} R.~A.,  {Riley} J.~M.,   {Longair} M.~S.,  1983, \mnras, \href
  {http://cdsads.u-strasbg.fr/abs/1983MNRAS.204..151L} {204, 151}

\bibitem[\protect\citeauthoryear{{Lake}, {Wright}, {Petty}, {Assef}, {Jarrett},
  {Stanford}, {Stern}  \& {Tsai}}{{Lake} et~al.}{2012}]{Lake2012}
{Lake} S.~E.,  {Wright} E.~L.,  {Petty} S.,  {Assef} R.~J.,  {Jarrett} T.~H.,
  {Stanford} S.~A.,  {Stern} D.,   {Tsai} C.-W.,  2012, \mn@doi [\aj]
  {10.1088/0004-6256/143/1/7}, \href
  {http://cdsads.u-strasbg.fr/abs/2012AJ....143....7L} {143, 7}

\bibitem[\protect\citeauthoryear{{Lawrence} et~al.,}{{Lawrence}
  et~al.}{2007}]{UKIDSS2007}
{Lawrence} A.,  et~al., 2007, \mn@doi [\mnras]
  {10.1111/j.1365-2966.2007.12040.x}, \href
  {https://ui.adsabs.harvard.edu/abs/2007MNRAS.379.1599L} {379, 1599}

\bibitem[\protect\citeauthoryear{{Ledlow} \& {Owen}}{{Ledlow} \&
  {Owen}}{1996}]{Ledlow1996}
{Ledlow} M.~J.,  {Owen} F.~N.,  1996, \mn@doi [\aj] {10.1086/117985}, \href
  {http://adsabs.harvard.edu/abs/1996AJ....112....9L} {112, 9}

\bibitem[\protect\citeauthoryear{{Lockman}, {Jahoda}  \& {McCammon}}{{Lockman}
  et~al.}{1986}]{Lockman1986}
{Lockman} F.~J.,  {Jahoda} K.,   {McCammon} D.,  1986, \mn@doi [\apj]
  {10.1086/164002}, \href
  {https://ui.adsabs.harvard.edu/abs/1986ApJ...302..432L} {302, 432}

\bibitem[\protect\citeauthoryear{{Macconi}, {Torresi}, {Grandi}, {Boccardi}  \&
  {Vignali}}{{Macconi} et~al.}{2020}]{Macconi2020}
{Macconi} D.,  {Torresi} E.,  {Grandi} P.,  {Boccardi} B.,   {Vignali} C.,
  2020, \mn@doi [\mnras] {10.1093/mnras/staa560}, \href
  {https://ui.adsabs.harvard.edu/abs/2020MNRAS.493.4355M} {493, 4355}

\bibitem[\protect\citeauthoryear{{Machalski}, {Jamrozy}  \& {Zola}}{{Machalski}
  et~al.}{2001}]{Machalski2001}
{Machalski} J.,  {Jamrozy} M.,   {Zola} S.,  2001, \mn@doi [\aap]
  {10.1051/0004-6361:20010352}, \href
  {https://ui.adsabs.harvard.edu/abs/2001A%26A...371..445M} {371, 445}

\bibitem[\protect\citeauthoryear{{Machalski}, {Kozie{\l}-Wierzbowska},
  {Jamrozy}  \& {Saikia}}{{Machalski} et~al.}{2008}]{Machalski2008}
{Machalski} J.,  {Kozie{\l}-Wierzbowska} D.,  {Jamrozy} M.,   {Saikia} D.~J.,
  2008, \mn@doi [\apj] {10.1086/586703}, \href
  {https://ui.adsabs.harvard.edu/abs/2008ApJ...679..149M} {679, 149}

\bibitem[\protect\citeauthoryear{{Magorrian} et~al.,}{{Magorrian}
  et~al.}{1998}]{Magorrian1998}
{Magorrian} J.,  et~al., 1998, \mn@doi [\aj] {10.1086/300353}, \href
  {http://cdsads.u-strasbg.fr/abs/1998AJ....115.2285M} {115, 2285}

\bibitem[\protect\citeauthoryear{{Mahatma} et~al.,}{{Mahatma}
  et~al.}{2019}]{Mahatma2019}
{Mahatma} V.~H.,  et~al., 2019, \mn@doi [\aap] {10.1051/0004-6361/201833973},
  \href {http://adsabs.harvard.edu/abs/2019A%26A...622A..13M} {622, A13}

\bibitem[\protect\citeauthoryear{{Mahatma}, {Hardcastle}, {Harwood},
  {O'Sullivan}, {Heald}, {Horellou}  \& {Smith}}{{Mahatma}
  et~al.}{2021}]{Mahatma2021}
{Mahatma} V.~H.,  {Hardcastle} M.~J.,  {Harwood} J.,  {O'Sullivan} S.~P.,
  {Heald} G.,  {Horellou} C.,   {Smith} D.~J.~B.,  2021, \mn@doi [\mnras]
  {10.1093/mnras/staa3980}, \href
  {https://ui.adsabs.harvard.edu/abs/2021MNRAS.502..273M} {502, 273}

\bibitem[\protect\citeauthoryear{{Mainzer} et~al.,}{{Mainzer}
  et~al.}{2011}]{WISE2011}
{Mainzer} A.,  et~al., 2011, \mn@doi [\apj] {10.1088/0004-637X/731/1/53}, \href
  {http://adsabs.harvard.edu/abs/2011ApJ...731...53M} {731, 53}

\bibitem[\protect\citeauthoryear{{Massaro}, {{\'A}lvarez-Crespo}, {Capetti},
  {Baldi}, {Pillitteri}, {Campana}  \& {Paggi}}{{Massaro}
  et~al.}{2019}]{Massaro2019}
{Massaro} F.,  {{\'A}lvarez-Crespo} N.,  {Capetti} A.,  {Baldi} R.~D.,
  {Pillitteri} I.,  {Campana} R.,   {Paggi} A.,  2019, \mn@doi [\apjs]
  {10.3847/1538-4365/aaf1c7}, \href
  {https://ui.adsabs.harvard.edu/abs/2019ApJS..240...20M} {240, 20}

\bibitem[\protect\citeauthoryear{{Mateos} et~al.,}{{Mateos}
  et~al.}{2012}]{Mateos2012}
{Mateos} S.,  et~al., 2012, \mn@doi [\mnras]
  {10.1111/j.1365-2966.2012.21843.x}, \href
  {http://cdsads.u-strasbg.fr/abs/2012MNRAS.426.3271M} {426, 3271}

\bibitem[\protect\citeauthoryear{{Mehdipour} \& {Costantini}}{{Mehdipour} \&
  {Costantini}}{2019}]{Mehdipour2019}
{Mehdipour} M.,  {Costantini} E.,  2019, \mn@doi [\aap]
  {10.1051/0004-6361/201935205}, \href
  {https://ui.adsabs.harvard.edu/abs/2019A&A...625A..25M} {625, A25}

\bibitem[\protect\citeauthoryear{{Meyer}, {Fossati}, {Georganopoulos}  \&
  {Lister}}{{Meyer} et~al.}{2011}]{Meyer2011}
{Meyer} E.~T.,  {Fossati} G.,  {Georganopoulos} M.,   {Lister} M.~L.,  2011,
  \mn@doi [\apj] {10.1088/0004-637X/740/2/98}, \href
  {http://cdsads.u-strasbg.fr/abs/2011ApJ...740...98M} {740, 98}

\bibitem[\protect\citeauthoryear{{Miley}}{{Miley}}{1980}]{Miley1980}
{Miley} G.,  1980, \mn@doi [\araa] {10.1146/annurev.aa.18.090180.001121}, \href
  {https://ui.adsabs.harvard.edu/abs/1980ARA&A..18..165M} {18, 165}

\bibitem[\protect\citeauthoryear{{Mingo}, {Hardcastle}, {Croston}, {Dicken},
  {Evans}, {Morganti}  \& {Tadhunter}}{{Mingo} et~al.}{2014}]{Mingo2014}
{Mingo} B.,  {Hardcastle} M.~J.,  {Croston} J.~H.,  {Dicken} D.,  {Evans}
  D.~A.,  {Morganti} R.,   {Tadhunter} C.,  2014, \mn@doi [\mnras]
  {10.1093/mnras/stu263}, \href
  {http://cdsads.u-strasbg.fr/abs/2014MNRAS.440..269M} {440, 269}

\bibitem[\protect\citeauthoryear{{Mingo} et~al.,}{{Mingo}
  et~al.}{2016}]{Mingo2016}
{Mingo} B.,  et~al., 2016, \mn@doi [\mnras] {10.1093/mnras/stw1826}, \href
  {http://adsabs.harvard.edu/abs/2016MNRAS.462.2631M} {462, 2631}

\bibitem[\protect\citeauthoryear{{Mingo} et~al.,}{{Mingo}
  et~al.}{2019}]{Mingo2019}
{Mingo} B.,  et~al., 2019, \mn@doi [\mnras] {10.1093/mnras/stz1901}, \href
  {https://ui.adsabs.harvard.edu/abs/2019MNRAS.488.2701M} {488, 2701}

\bibitem[\protect\citeauthoryear{{Miraghaei} \& {Best}}{{Miraghaei} \&
  {Best}}{2017}]{Miraghaei2017}
{Miraghaei} H.,  {Best} P.~N.,  2017, \mn@doi [\mnras] {10.1093/mnras/stx007},
  \href {http://adsabs.harvard.edu/abs/2017MNRAS.466.4346M} {466, 4346}

\bibitem[\protect\citeauthoryear{{Mohan} \& {Rafferty}}{{Mohan} \&
  {Rafferty}}{2015}]{PyBDSF2015}
{Mohan} N.,  {Rafferty} D.,  2015, {PyBDSF: Python Blob Detection and Source
  Finder}, Astrophysics Source Code Library (\mn@eprint {ascl} {1502.007})

\bibitem[\protect\citeauthoryear{{Morganti} \& {Oosterloo}}{{Morganti} \&
  {Oosterloo}}{2018}]{Morganti2018}
{Morganti} R.,  {Oosterloo} T.,  2018, \mn@doi [\aapr]
  {10.1007/s00159-018-0109-x}, \href
  {https://ui.adsabs.harvard.edu/abs/2018A&ARv..26....4M} {26, 4}

\bibitem[\protect\citeauthoryear{{Morrissey} et~al.,}{{Morrissey}
  et~al.}{2007}]{GALEX2007}
{Morrissey} P.,  et~al., 2007, \mn@doi [\apjs] {10.1086/520512}, \href
  {https://ui.adsabs.harvard.edu/abs/2007ApJS..173..682M} {173, 682}

\bibitem[\protect\citeauthoryear{{Mullaney} et~al.,}{{Mullaney}
  et~al.}{2012}]{Mullaney2012}
{Mullaney} J.~R.,  et~al., 2012, \mn@doi [\apjl] {10.1088/2041-8205/753/2/L30},
  \href {https://ui.adsabs.harvard.edu/abs/2012ApJ...753L..30M} {753, L30}

\bibitem[\protect\citeauthoryear{{Mullin} \& {Hardcastle}}{{Mullin} \&
  {Hardcastle}}{2009}]{MullinHardcastle2009}
{Mullin} L.~M.,  {Hardcastle} M.~J.,  2009, \mn@doi [\mnras]
  {10.1111/j.1365-2966.2009.15232.x}, \href
  {https://ui.adsabs.harvard.edu/abs/2009MNRAS.398.1989M} {398, 1989}

\bibitem[\protect\citeauthoryear{{Narayan} \& {Yi}}{{Narayan} \&
  {Yi}}{1995}]{Narayan1995}
{Narayan} R.,  {Yi} I.,  1995, \mn@doi [\apj] {10.1086/176343}, \href
  {http://cdsads.u-strasbg.fr/abs/1995ApJ...452..710N} {452, 710}

\bibitem[\protect\citeauthoryear{{O'Dea} \& {Owen}}{{O'Dea} \&
  {Owen}}{1985}]{ODea1985NAT}
{O'Dea} C.~P.,  {Owen} F.~N.,  1985, \mn@doi [\aj] {10.1086/113802}, \href
  {http://adsabs.harvard.edu/abs/1985AJ.....90..954O} {90, 954}

\bibitem[\protect\citeauthoryear{{O'Dea} \& {Saikia}}{{O'Dea} \&
  {Saikia}}{2021}]{ODea2021}
{O'Dea} C.~P.,  {Saikia} D.~J.,  2021, \mn@doi [\aapr]
  {10.1007/s00159-021-00131-w}, \href
  {https://ui.adsabs.harvard.edu/abs/2021A&ARv..29....3O} {29, 3}

\bibitem[\protect\citeauthoryear{{O'Sullivan}, {Purcell}, {Anderson}, {Farnes},
  {Sun}  \& {Gaensler}}{{O'Sullivan} et~al.}{2017}]{OSullivan2017}
{O'Sullivan} S.~P.,  {Purcell} C.~R.,  {Anderson} C.~S.,  {Farnes} J.~S.,
  {Sun} X.~H.,   {Gaensler} B.~M.,  2017, \mn@doi [\mnras]
  {10.1093/mnras/stx1133}, \href
  {https://ui.adsabs.harvard.edu/abs/2017MNRAS.469.4034O} {469, 4034}

\bibitem[\protect\citeauthoryear{{Oliver}, {Serjeant}, {Efstathiou},
  {Crockett}, {Gruppioni}, {La Franca}  \& {Elais Consortium}}{{Oliver}
  et~al.}{2000}]{Oliver2000}
{Oliver} S.,  {Serjeant} S.,  {Efstathiou} A.,  {Crockett} H.,  {Gruppioni} C.,
   {La Franca} F.,   {Elais Consortium} 2000, {The European Large Area ISO
  Survey (ELAIS): Latest Results}.
p.~28

\bibitem[\protect\citeauthoryear{{Owen} \& {Laing}}{{Owen} \&
  {Laing}}{1989}]{Owen1989}
{Owen} F.~N.,  {Laing} R.~A.,  1989, \mn@doi [\mnras]
  {10.1093/mnras/238.2.357}, \href
  {https://ui.adsabs.harvard.edu/abs/1989MNRAS.238..357O} {238, 357}

\bibitem[\protect\citeauthoryear{{Owen} \& {Rudnick}}{{Owen} \&
  {Rudnick}}{1976}]{Owen1976}
{Owen} F.~N.,  {Rudnick} L.,  1976, \mn@doi [\apjl] {10.1086/182077}, \href
  {https://ui.adsabs.harvard.edu/abs/1976ApJ...205L...1O} {205, L1}

\bibitem[\protect\citeauthoryear{{P{\^a}ris} et~al.,}{{P{\^a}ris}
  et~al.}{2017}]{Paris2017}
{P{\^a}ris} I.,  et~al., 2017, \mn@doi [\aap] {10.1051/0004-6361/201527999},
  \href {https://ui.adsabs.harvard.edu/abs/2017A&A...597A..79P} {597, A79}

\bibitem[\protect\citeauthoryear{{Patil} et~al.,}{{Patil}
  et~al.}{2020}]{Patil2020}
{Patil} P.,  et~al., 2020, \mn@doi [\apj] {10.3847/1538-4357/ab9011}, \href
  {https://ui.adsabs.harvard.edu/abs/2020ApJ...896...18P} {896, 18}

\bibitem[\protect\citeauthoryear{P\'erez \& Granger}{P\'erez \&
  Granger}{2007}]{IPython}
P\'erez F.,  Granger B.~E.,  2007, \mn@doi [Computing in Science and
  Engineering] {10.1109/MCSE.2007.53}, 9, 21

\bibitem[\protect\citeauthoryear{{Perucho} \& {Mart{\'\i}}}{{Perucho} \&
  {Mart{\'\i}}}{2007}]{Perucho2007}
{Perucho} M.,  {Mart{\'\i}} J.~M.,  2007, \mn@doi [\mnras]
  {10.1111/j.1365-2966.2007.12454.x}, \href
  {https://ui.adsabs.harvard.edu/abs/2007MNRAS.382..526P} {382, 526}

\bibitem[\protect\citeauthoryear{{Perucho}, {Mart{\'\i}}, {Laing}  \&
  {Hardee}}{{Perucho} et~al.}{2014}]{Perucho2014}
{Perucho} M.,  {Mart{\'\i}} J.~M.,  {Laing} R.~A.,   {Hardee} P.~E.,  2014,
  \mn@doi [\mnras] {10.1093/mnras/stu676}, \href
  {https://ui.adsabs.harvard.edu/abs/2014MNRAS.441.1488P} {441, 1488}

\bibitem[\protect\citeauthoryear{{Pierce}, {Tadhunter}, {Ramos Almeida},
  {Bessiere}  \& {Rose}}{{Pierce} et~al.}{2019}]{Pierce2019}
{Pierce} J.~C.~S.,  {Tadhunter} C.~N.,  {Ramos Almeida} C.,  {Bessiere} P.~S.,
   {Rose} M.,  2019, \mn@doi [\mnras] {10.1093/mnras/stz1253}, \href
  {https://ui.adsabs.harvard.edu/abs/2019MNRAS.487.5490P} {487, 5490}

\bibitem[\protect\citeauthoryear{{Pierce}, {Tadhunter}  \& {Morganti}}{{Pierce}
  et~al.}{2020}]{Pierce2020}
{Pierce} J.~C.~S.,  {Tadhunter} C.~N.,   {Morganti} R.,  2020, \mn@doi [\mnras]
  {10.1093/mnras/staa531}, \href
  {https://ui.adsabs.harvard.edu/abs/2020MNRAS.494.2053P} {494, 2053}

\bibitem[\protect\citeauthoryear{{Pilbratt} et~al.,}{{Pilbratt}
  et~al.}{2010}]{Herschel2010}
{Pilbratt} G.~L.,  et~al., 2010, \mn@doi [\aap] {10.1051/0004-6361/201014759},
  \href {https://ui.adsabs.harvard.edu/abs/2010A&A...518L...1P} {518, L1}

\bibitem[\protect\citeauthoryear{{Pracy} et~al.,}{{Pracy}
  et~al.}{2016}]{Pracy2016}
{Pracy} M.~B.,  et~al., 2016, \mn@doi [\mnras] {10.1093/mnras/stw910}, \href
  {https://ui.adsabs.harvard.edu/abs/2016MNRAS.460....2P} {460, 2}

\bibitem[\protect\citeauthoryear{{Prandoni} \& {Seymour}}{{Prandoni} \&
  {Seymour}}{2015}]{Prandoni2015}
{Prandoni} I.,  {Seymour} N.,  2015, in Advancing Astrophysics with the Square
  Kilometre Array (AASKA14). p.~67 (\mn@eprint {arXiv} {1412.6512})

\bibitem[\protect\citeauthoryear{{Read} et~al.,}{{Read}
  et~al.}{2018}]{Read2018}
{Read} S.~C.,  et~al., 2018, \mn@doi [\mnras] {10.1093/mnras/sty2198}, \href
  {https://ui.adsabs.harvard.edu/abs/2018MNRAS.480.5625R} {480, 5625}

\bibitem[\protect\citeauthoryear{{Reines} \& {Volonteri}}{{Reines} \&
  {Volonteri}}{2015}]{Reines2015}
{Reines} A.~E.,  {Volonteri} M.,  2015, \mn@doi [\apj]
  {10.1088/0004-637X/813/2/82}, \href
  {https://ui.adsabs.harvard.edu/abs/2015ApJ...813...82R} {813, 82}

\bibitem[\protect\citeauthoryear{{Retana-Montenegro} \&
  {R{\"o}ttgering}}{{Retana-Montenegro} \&
  {R{\"o}ttgering}}{2020}]{RetanaMontenegro2020}
{Retana-Montenegro} E.,  {R{\"o}ttgering} H.~J.~A.,  2020, \mn@doi [\aap]
  {10.1051/0004-6361/201936577}, \href
  {https://ui.adsabs.harvard.edu/abs/2020A&A...636A..12R} {636, A12}

\bibitem[\protect\citeauthoryear{{Rovilos} et~al.,}{{Rovilos}
  et~al.}{2014}]{Rovilos2014}
{Rovilos} E.,  et~al., 2014, \mn@doi [\mnras] {10.1093/mnras/stt2228}, \href
  {http://cdsads.u-strasbg.fr/abs/2014MNRAS.438..494R} {438, 494}

\bibitem[\protect\citeauthoryear{{Rudnick} \& {Owen}}{{Rudnick} \&
  {Owen}}{1976}]{Rudnick1976}
{Rudnick} L.,  {Owen} F.~N.,  1976, \mn@doi [\apjl] {10.1086/182030}, \href
  {https://ui.adsabs.harvard.edu/abs/1976ApJ...203L.107R} {203, L107}

\bibitem[\protect\citeauthoryear{{Sabater} et~al.,}{{Sabater}
  et~al.}{2019}]{Sabater2019}
{Sabater} J.,  et~al., 2019, \mn@doi [\aap] {10.1051/0004-6361/201833883},
  \href {http://adsabs.harvard.edu/abs/2019A%26A...622A..17S} {622, A17}

\bibitem[\protect\citeauthoryear{{Sabater} et~al.,}{{Sabater}
  et~al.}{2021}]{Sabater2021}
{Sabater} J.,  et~al., 2021, \mn@doi [\aap] {10.1051/0004-6361/202038828},
  \href {https://ui.adsabs.harvard.edu/abs/2021A&A...648A...2S} {648, A2}

\bibitem[\protect\citeauthoryear{{Schoenmakers}, {Mack}, {de Bruyn},
  {R{\"o}ttgering}, {Klein}  \& {van der Laan}}{{Schoenmakers}
  et~al.}{2000a}]{schoenmakers2000b}
{Schoenmakers} A.~P.,  {Mack} K.-H.,  {de Bruyn} A.~G.,  {R{\"o}ttgering}
  H.~J.~A.,  {Klein} U.,   {van der Laan} H.,  2000a, \mn@doi [\aaps]
  {10.1051/aas:2000267}, \href
  {https://ui.adsabs.harvard.edu/abs/2000A%26AS..146..293S} {146, 293}

\bibitem[\protect\citeauthoryear{{Schoenmakers}, {de Bruyn}, {R{\"o}ttgering},
  {van der Laan}  \& {Kaiser}}{{Schoenmakers} et~al.}{2000b}]{Schoenmakers2000}
{Schoenmakers} A.~P.,  {de Bruyn} A.~G.,  {R{\"o}ttgering} H.~J.~A.,  {van der
  Laan} H.,   {Kaiser} C.~R.,  2000b, \mn@doi [\mnras]
  {10.1046/j.1365-8711.2000.03430.x}, \href
  {http://adsabs.harvard.edu/abs/2000MNRAS.315..371S} {315, 371}

\bibitem[\protect\citeauthoryear{{Secrest}, {Dudik}, {Dorland}, {Zacharias},
  {Makarov}, {Fey}, {Frouard}  \& {Finch}}{{Secrest}
  et~al.}{2015}]{Secrest2015}
{Secrest} N.~J.,  {Dudik} R.~P.,  {Dorland} B.~N.,  {Zacharias} N.,  {Makarov}
  V.,  {Fey} A.,  {Frouard} J.,   {Finch} C.,  2015, \mn@doi [\apjs]
  {10.1088/0067-0049/221/1/12}, \href
  {http://cdsads.u-strasbg.fr/abs/2015ApJS..221...12S} {221, 12}

\bibitem[\protect\citeauthoryear{{Shabala}, {Jurlin}, {Morganti}, {Brienza},
  {Hardcastle}, {Godfrey}, {Krause}  \& {Turner}}{{Shabala}
  et~al.}{2020}]{Shabala2020}
{Shabala} S.~S.,  {Jurlin} N.,  {Morganti} R.,  {Brienza} M.,  {Hardcastle}
  M.~J.,  {Godfrey} L. E.~H.,  {Krause} M. G.~H.,   {Turner} R.~J.,  2020,
  \mn@doi [\mnras] {10.1093/mnras/staa1172}, \href
  {https://ui.adsabs.harvard.edu/abs/2020MNRAS.496.1706S} {496, 1706}

\bibitem[\protect\citeauthoryear{{Shakura} \& {Sunyaev}}{{Shakura} \&
  {Sunyaev}}{1973}]{Shakura1973}
{Shakura} N.~I.,  {Sunyaev} R.~A.,  1973, \aap, \href
  {http://cdsads.u-strasbg.fr/abs/1973A%26A....24..337S} {24, 337}

\bibitem[\protect\citeauthoryear{{Shimwell} et~al.,}{{Shimwell}
  et~al.}{2017}]{Shimwell2017}
{Shimwell} T.~W.,  et~al., 2017, \mn@doi [\aap] {10.1051/0004-6361/201629313},
  \href {http://adsabs.harvard.edu/abs/2017A%26A...598A.104S} {598, A104}

\bibitem[\protect\citeauthoryear{{Shimwell} et~al.,}{{Shimwell}
  et~al.}{2019}]{Shimwell2019}
{Shimwell} T.~W.,  et~al., 2019, \mn@doi [\aap] {10.1051/0004-6361/201833559},
  \href {http://adsabs.harvard.edu/abs/2019A%26A...622A...1S} {622, A1}

\bibitem[\protect\citeauthoryear{{Smith} et~al.,}{{Smith}
  et~al.}{2021}]{Smith2021}
{Smith} D.~J.~B.,  et~al., 2021, \mn@doi [\aap] {10.1051/0004-6361/202039343},
  \href {https://ui.adsabs.harvard.edu/abs/2021A&A...648A...6S} {648, A6}

\bibitem[\protect\citeauthoryear{{Smol{\v c}i{\'c}}}{{Smol{\v
  c}i{\'c}}}{2009}]{Smolcic2009}
{Smol{\v c}i{\'c}} V.,  2009, \mn@doi [\apjl] {10.1088/0004-637X/699/1/L43},
  \href {http://adsabs.harvard.edu/abs/2009ApJ...699L..43S} {699, L43}

\bibitem[\protect\citeauthoryear{{Stern} et~al.,}{{Stern}
  et~al.}{2005}]{Stern2005}
{Stern} D.,  et~al., 2005, \mn@doi [\apj] {10.1086/432523}, \href
  {https://ui.adsabs.harvard.edu/abs/2005ApJ...631..163S} {631, 163}

\bibitem[\protect\citeauthoryear{{Stern} et~al.,}{{Stern}
  et~al.}{2012}]{Stern2012}
{Stern} D.,  et~al., 2012, \mn@doi [\apj] {10.1088/0004-637X/753/1/30}, \href
  {http://cdsads.u-strasbg.fr/abs/2012ApJ...753...30S} {753, 30}

\bibitem[\protect\citeauthoryear{{Tadhunter}}{{Tadhunter}}{2016}]{Tadhunter2016}
{Tadhunter} C.,  2016, \mn@doi [\aapr] {10.1007/s00159-016-0094-x}, \href
  {http://adsabs.harvard.edu/abs/2016A%26ARv..24...10T} {24, 10}

\bibitem[\protect\citeauthoryear{{Tasse}, {Best}, {R{\"o}ttgering}  \& {Le
  Borgne}}{{Tasse} et~al.}{2008}]{Tasse2008}
{Tasse} C.,  {Best} P.~N.,  {R{\"o}ttgering} H.,   {Le Borgne} D.,  2008,
  \mn@doi [\aap] {10.1051/0004-6361:20079299}, \href
  {https://ui.adsabs.harvard.edu/abs/2008A&A...490..893T} {490, 893}

\bibitem[\protect\citeauthoryear{{Tasse} et~al.,}{{Tasse}
  et~al.}{2021}]{Tasse2021}
{Tasse} C.,  et~al., 2021, \mn@doi [\aap] {10.1051/0004-6361/202038804}, \href
  {https://ui.adsabs.harvard.edu/abs/2021A&A...648A...1T} {648, A1}

\bibitem[\protect\citeauthoryear{{Taylor}}{{Taylor}}{2005}]{Taylor2005}
{Taylor} M.~B.,  2005, in {Shopbell} P.,  {Britton} M.,   {Ebert} R.,  eds,
  Astronomical Society of the Pacific Conference Series Vol. 347, Astronomical
  Data Analysis Software and Systems XIV. p.~29

\bibitem[\protect\citeauthoryear{{Turner} \& {Shabala}}{{Turner} \&
  {Shabala}}{2015}]{Turner2015}
{Turner} R.~J.,  {Shabala} S.~S.,  2015, \mn@doi [\apj]
  {10.1088/0004-637X/806/1/59}, \href
  {http://adsabs.harvard.edu/abs/2015ApJ...806...59T} {806, 59}

\bibitem[\protect\citeauthoryear{{Vardoulaki} et~al.,}{{Vardoulaki}
  et~al.}{2021}]{Vardoulaki2021}
{Vardoulaki} E.,  et~al., 2021, \mn@doi [\aap] {10.1051/0004-6361/202039488},
  \href {https://ui.adsabs.harvard.edu/abs/2021A&A...648A.102V} {648, A102}

\bibitem[\protect\citeauthoryear{{Wang} \& {Kaiser}}{{Wang} \&
  {Kaiser}}{2008}]{Wang2008}
{Wang} Y.,  {Kaiser} C.~R.,  2008, \mn@doi [\mnras]
  {10.1111/j.1365-2966.2008.13417.x}, \href
  {http://adsabs.harvard.edu/abs/2008MNRAS.388..677W} {388, 677}

\bibitem[\protect\citeauthoryear{{Webster} et~al.,}{{Webster}
  et~al.}{2021a}]{Webster2021}
{Webster} B.,  et~al., 2021a, \mn@doi [\mnras] {10.1093/mnras/staa3437}, \href
  {https://ui.adsabs.harvard.edu/abs/2021MNRAS.500.4921W} {500, 4921}

\bibitem[\protect\citeauthoryear{{Webster}, {Croston}, {Harwood}, {Baldi},
  {Hardcastle}, {Mingo}  \& {R{\"o}ttgering}}{{Webster}
  et~al.}{2021b}]{Webster2021b}
{Webster} B.,  {Croston} J.~H.,  {Harwood} J.~J.,  {Baldi} R.~D.,  {Hardcastle}
  M.~J.,  {Mingo} B.,   {R{\"o}ttgering} H.~J.~A.,  2021b, \mn@doi [\mnras]
  {10.1093/mnras/stab2939}, \href
  {https://ui.adsabs.harvard.edu/abs/2021MNRAS.508.5972W} {508, 5972}

\bibitem[\protect\citeauthoryear{{Weigel}, {Schawinski}, {Caplar}, {Wong},
  {Treister}  \& {Trakhtenbrot}}{{Weigel} et~al.}{2017}]{Weigel2017}
{Weigel} A.~K.,  {Schawinski} K.,  {Caplar} N.,  {Wong} O.~I.,  {Treister} E.,
   {Trakhtenbrot} B.,  2017, \mn@doi [\apj] {10.3847/1538-4357/aa803b}, \href
  {https://ui.adsabs.harvard.edu/\#abs/2017ApJ...845..134W} {845, 134}

\bibitem[\protect\citeauthoryear{{Werner} et~al.,}{{Werner}
  et~al.}{2004}]{Spitzer2004}
{Werner} M.~W.,  et~al., 2004, \mn@doi [\apjs] {10.1086/422992}, \href
  {https://ui.adsabs.harvard.edu/abs/2004ApJS..154....1W} {154, 1}

\bibitem[\protect\citeauthoryear{{Wild}, {Heckman}  \& {Charlot}}{{Wild}
  et~al.}{2010}]{Wild2010}
{Wild} V.,  {Heckman} T.,   {Charlot} S.,  2010, \mn@doi [\mnras]
  {10.1111/j.1365-2966.2010.16536.x}, \href
  {http://cdsads.u-strasbg.fr/abs/2010MNRAS.405..933W} {405, 933}

\bibitem[\protect\citeauthoryear{{Williams} \& {R{\"o}ttgering}}{{Williams} \&
  {R{\"o}ttgering}}{2015}]{Williams2015}
{Williams} W.~L.,  {R{\"o}ttgering} H.~J.~A.,  2015, \mn@doi [\mnras]
  {10.1093/mnras/stv692}, \href
  {http://cdsads.u-strasbg.fr/abs/2015MNRAS.450.1538W} {450, 1538}

\bibitem[\protect\citeauthoryear{{Williams} et~al.,}{{Williams}
  et~al.}{2018}]{Williams2018a}
{Williams} W.~L.,  et~al., 2018, \mn@doi [\mnras] {10.1093/mnras/sty026}, \href
  {http://adsabs.harvard.edu/abs/2018MNRAS.475.3429W} {475, 3429}

\bibitem[\protect\citeauthoryear{{Williams} et~al.,}{{Williams}
  et~al.}{2019}]{Williams2019}
{Williams} W.~L.,  et~al., 2019, \mn@doi [\aap] {10.1051/0004-6361/201833564},
  \href {http://adsabs.harvard.edu/abs/2019A%26A...622A...2W} {622, A2}

\bibitem[\protect\citeauthoryear{{Williams} et~al.,}{{Williams}
  et~al.}{2021}]{Williams2021}
{Williams} W.~L.,  et~al., 2021, \mn@doi [\aap] {10.1051/0004-6361/202141745},
  \href {https://ui.adsabs.harvard.edu/abs/2021A&A...655A..40W} {655, A40}

\bibitem[\protect\citeauthoryear{{Willott}, {Rawlings}, {Blundell}  \&
  {Lacy}}{{Willott} et~al.}{1999}]{Willott1999}
{Willott} C.~J.,  {Rawlings} S.,  {Blundell} K.~M.,   {Lacy} M.,  1999, \mn@doi
  [\mnras] {10.1046/j.1365-8711.1999.02907.x}, \href
  {http://cdsads.u-strasbg.fr/abs/1999MNRAS.309.1017W} {309, 1017}

\bibitem[\protect\citeauthoryear{{Willott}, {Rawlings}, {Blundell}, {Lacy}  \&
  {Eales}}{{Willott} et~al.}{2001}]{Willott2001}
{Willott} C.~J.,  {Rawlings} S.,  {Blundell} K.~M.,  {Lacy} M.,   {Eales}
  S.~A.,  2001, \mn@doi [\mnras] {10.1046/j.1365-8711.2001.04101.x}, \href
  {https://ui.adsabs.harvard.edu/\#abs/2001MNRAS.322..536W} {322, 536}

\bibitem[\protect\citeauthoryear{{Wright} et~al.,}{{Wright}
  et~al.}{2010}]{Wright2010}
{Wright} E.~L.,  et~al., 2010, \mn@doi [\aj] {10.1088/0004-6256/140/6/1868},
  \href {http://adsabs.harvard.edu/abs/2010AJ....140.1868W} {140, 1868}

\bibitem[\protect\citeauthoryear{{Yang} et~al.,}{{Yang}
  et~al.}{2020}]{cigale2020}
{Yang} G.,  et~al., 2020, \mn@doi [\mnras] {10.1093/mnras/stz3001}, \href
  {https://ui.adsabs.harvard.edu/abs/2020MNRAS.491..740Y} {491, 740}

\bibitem[\protect\citeauthoryear{{Ziparo} et~al.,}{{Ziparo}
  et~al.}{2016}]{Ziparo2016}
{Ziparo} F.,  et~al., 2016, \mn@doi [\aap] {10.1051/0004-6361/201526792}, \href
  {https://ui.adsabs.harvard.edu/abs/2016A&A...592A...9Z} {592, A9}

\bibitem[\protect\citeauthoryear{{da Cunha}, {Charlot}  \& {Elbaz}}{{da Cunha}
  et~al.}{2008}]{magphys2008}
{da Cunha} E.,  {Charlot} S.,   {Elbaz} D.,  2008, \mn@doi [\mnras]
  {10.1111/j.1365-2966.2008.13535.x}, \href
  {https://ui.adsabs.harvard.edu/abs/2008MNRAS.388.1595D} {388, 1595}

\bibitem[\protect\citeauthoryear{{de Gasperin} et~al.,}{{de Gasperin}
  et~al.}{2021}]{DeGasperin2021}
{de Gasperin} F.,  et~al., 2021, \mn@doi [\aap] {10.1051/0004-6361/202140316},
  \href {https://ui.adsabs.harvard.edu/abs/2021A&A...648A.104D} {648, A104}

\bibitem[\protect\citeauthoryear{{van Haarlem} et~al.,}{{van Haarlem}
  et~al.}{2013}]{vanHaarlem2013}
{van Haarlem} M.~P.,  et~al., 2013, \mn@doi [\aap]
  {10.1051/0004-6361/201220873}, \href
  {https://ui.adsabs.harvard.edu/abs/2013A&A...556A...2V} {556, A2}

\bibitem[\protect\citeauthoryear{{van der Walt}, {Colbert}  \&
  {Varoquaux}}{{van der Walt} et~al.}{2011}]{numpyArray}
{van der Walt} S.,  {Colbert} S.~C.,   {Varoquaux} G.,  2011, \mn@doi
  [Computing in Science Engineering] {10.1109/MCSE.2011.37}, 13, 22

\makeatother
\end{thebibliography}






\bsp	
\label{lastpage}
\end{document}